\documentclass{scrartcl}

\usepackage[breaklinks=true]{hyperref}
\usepackage{import}

\usepackage{amsmath,MnSymbol}
\usepackage{amsthm}
\usepackage{algorithm}
\usepackage[noend]{algpseudocode}
\usepackage{thmtools}
\usepackage{multicol}
\usepackage{colonequals}
\usepackage{graphicx}
\usepackage{xcolor}

\renewcommand{\paragraph}[1]{\smallskip \noindent \textbf{#1.}}

\newcommand{\var}{\mathcal{X}}
\newcommand{\captures}{\textsf{C}}

\newcommand{\spanc}[2]{[#1,#2\rangle}

\newcommand{\cA}{\mathcal{A}}

\newcommand{\cF}{\mathcal{F}}
\newcommand{\cI}{\mathcal{I}}
\newcommand{\cO}{\mathcal{O}}
\newcommand{\cS}{\mathcal{S}}
\newcommand{\eps}{\varepsilon}
\renewcommand{\epsilon}{\varepsilon}

\newcommand{\shape}{\mathsf{shape}}

\newcommand{\opS}{\Sigma^{\texttt{<}}}
\newcommand{\clS}{\Sigma^{\texttt{>}}}
\newcommand{\noS}{\Sigma^{\texttt{|}}}
\newcommand{\wnS}{\Sigma^{\texttt{<*>}}}
\newcommand{\wnSann}{(\Sigma\cup\Sigma\times\Omega)^{\texttt{<*>}}}

\newcommand{\sem}[1]{{\lsem{}{#1}\rsem}}
\newcommand{\trans}[2][]{\raisebox{-1pt}[10pt][0pt]{$\overset{#2}{\underset{^{#1}}{\raisebox{0pt}[3pt][0pt]{$\relbar\mspace{-8mu}\longrightarrow$}}}$}}

\newcommand{\cP}{\mathcal{P}}

\newcommand{\cG}{\mathcal{G}}
\newcommand{\cH}{\mathcal{H}}

\newcommand{\str}{\mathsf{str}}
\newcommand{\ann}{\mathsf{ann}}
\newcommand{\der}[1]{\Rightarrow_{#1}}
\newcommand{\ders}[1]{\Rightarrow_{#1}^*}

\newcommand{\tableI}{\mathbb{I}}

\newcommand{\dsD}{\mathcal{D}}

\renewcommand{\O}{\mathcal{O}}

\newcommand{\cT}{\mathcal{T}}
\newcommand{\outf}{\textsf{out}}

\DeclareMathOperator{\oout}{%
	\ooalign{\raisebox{0ex}{$o$}\cr\hidewidth\raisebox{.5ex}{\scalebox{.5}{$\ \,  \boldsymbol{\smallsmile}$}}\hidewidth}}%

\renewcommand{\prod}{\mathsf{prod}}
\newcommand{\union}{\mathsf{union}}
\newcommand{\empt}{\mathsf{empty}}

\newtheorem{lemma}{Lemma}

\newtheorem{theorem}{Theorem}
\newtheorem{corollary}{Corollary}
\newtheorem{principle}{Principle}

\newcommand{\proc}{\mathsf{proc}}
\newcommand{\sfi}{\mathsf{in}}
\newcommand{\sfo}{\mathsf{out}}
\newcommand{\sfl}{\mathsf{left}}
\newcommand{\sfm}{\mathsf{mid}}
\newcommand{\sfr}{\mathsf{right}}

\newcommand{\card}[1]{\left|#1\right|}

\newcommand{\Set}{\mathrm{S}}

\newcommand{\singleton}{\textsf{singleton}}

\newcommand{\D}{\mathrm{D}}

\usepackage[textwidth=2cm,textsize=small]{todonotes}

\usepackage{amsfonts}
\usepackage{microtype}

\hypersetup{
    colorlinks,
    linkcolor={red!50!black},
    citecolor={blue!50!black},
    urlcolor={blue!30!black}
}

\title{Efficient enumeration algorithms for annotated grammars}

\author{
\begin{tabular}[t]{c}
Antoine Amarilli \\
{\normalfont LTCI, Télécom Paris, Institut polytechnique de Paris} \\
{\normalfont antoine.amarilli@telecom-paris.fr} \\[0.3em]
Louis Jachiet \\
{\normalfont LTCI, Télécom Paris, Institut polytechnique de Paris} \\
{\normalfont louis.jachiet@telecom-paris.fr} \\[0.3em]
Mart\'in Mu\~noz \\
{\normalfont PUC \& IMFD} \\
{\normalfont mmunos@uc.cl} \\[0.3em]
Cristian Riveros \\
{\normalfont PUC \& IMFD} \\
{\normalfont cristian.riveros@uc.cl} \\[0.3em]
\end{tabular}
}

\date{}

\makeatletter
\DeclareOldFontCommand{\rm}{\normalfont\rmfamily}{\mathrm}
\DeclareOldFontCommand{\sf}{\normalfont\sffamily}{\mathsf}
\DeclareOldFontCommand{\tt}{\normalfont\ttfamily}{\mathtt}
\DeclareOldFontCommand{\bf}{\normalfont\bfseries}{\mathbf}
\DeclareOldFontCommand{\it}{\normalfont\itshape}{\mathit}
\DeclareOldFontCommand{\sl}{\normalfont\slshape}{\@nomath\sl}
\DeclareOldFontCommand{\sc}{\normalfont\scshape}{\@nomath\sc}
\makeatother

\newtheorem{definition}[theorem]{Definition}
\newtheorem{proposition}[theorem]{Proposition}
\newtheorem{example}[theorem]{Example}

\begin{document}
\maketitle
\begin{abstract}
We introduce annotated grammars, an extension of context-free grammars which allows annotations on terminals. Our model extends the standard notion of regular spanners, and 
is more expressive than the extraction grammars recently introduced by Peterfreund. We study the enumeration problem for annotated grammars: fixing a grammar, and given a string as input, enumerate all annotations of the string that form a word derivable from the grammar. Our first result is an algorithm for unambiguous annotated grammars, which preprocesses the input string in cubic time and enumerates all annotations with output-linear delay. This improves over Peterfreund's result, which needs quintic time preprocessing to achieve this delay bound. We then study how we can reduce the preprocessing time while keeping the same delay bound, by making additional assumptions on the grammar. Specifically, we present a class of grammars which only have one derivation shape for all outputs, for which we can enumerate with quadratic time preprocessing. We also give classes that generalize regular spanners for which linear time preprocessing suffices.

\end{abstract}

\section{Introduction}

Arguably the most fundamental problem in database research is query
evaluation: given as input a query and data, we must find the results of the query over the data.
Database theory research has studied the complexity of such problems for 
decades.
However, in some contexts, for instance over large datasets, the usual
complexity measures are not well-suited to this study.
Indeed, the number of query results might be so large that it is unreasonable in
practice to produce all of them.
Further, in theoretical terms, the complexity of an algorithm may be dominated by the cost of writing
the full output, hiding the actual complexity of the
computation.
For this reason, a significant line of research on query evaluation has adopted the perspective of \emph{enumeration algorithms}.
Instead of explicitly producing all results, the task is to \emph{enumerate} them, in any order and without repetition.
The cost of the algorithm is then measured across two dimensions: the
\emph{preprocessing time}, which is the time needed to read the input and
prepare an enumeration data structure; and the \emph{delay}, the worst-case time
elapsed between any two solutions while enumerating using the data structure.

This study of enumeration algorithms has managed in some cases to achieve a
\emph{constant-delay} guarantee. In this case, once the algorithm has
preprocessed its input, the delay between any two outputs is \emph{constant},
i.e., it is independent from the input. Of course, the challenge is to achieve
this strong guarantee after a preprocessing phase that runs in a limited amount
of time -- in particular, one that does not explicitly materialize all solutions.
Starting with the work of Durand and Grandjean~\cite{durand2007first}, researchers have designed
constant-delay algorithms for several query evaluation problems, e.g., the evaluation of some queries over relational databases~\cite{bagan2007acyclic,kara2020trade}, query evaluation over dynamic data~\cite{berkholz2017answering,idris2017dynamic}, query evaluation over graph data~\cite{hartig2018semantics,kroll2016complexity}, among others~\cite{Segoufin13}. 

One area where enumeration algorithms have been especially successful is the problem of \emph{information extraction}, studied through the lens of \emph{document spanners}~\cite{FaginKRV15}. In this data management task, the data is a textual document (i.e., a string), and the query is a declarative specification of information to extract from the text, formalized as a \emph{spanner}. The spanner describes \emph{mappings}, which are possible choices of how to map variables to substrings of the document (called spans). The enumeration problem is then to enumerate all mappings of a spanner on an input document, i.e., to enumerate efficiently all possible results for the information extraction task.
The work by Florenzano et al.~\cite{FlorenzanoRUVV18} showed that the task could be solved with preprocessing linear in the document and polynomial in a finite deterministic automaton describing the spanner, improving on a theoretical result by Bagan~\cite{bagan2006mso}; and 
this was extended in~\cite{amarilli2019constant} to spanners described using nondeterministic automata or regular expressions.

However, while regular spanners are natural, they do not capture all
possible information extraction tasks. More
expressiveness is needed for extraction over structured data (e.g., XML,
or JSON documents), over the source code of programs, or possibly over
natural language texts. We believe that a natural way to address such
tasks is to move from finite automata to \emph{context-free grammars}
(CFGs). Context-free grammars are a well-known formalism: they extend regular
expressions and are commonly used, e.g., in programming language
design. Common verification tasks on textual representations of tree
documents can be expressed using CFGs, and so can parsing tasks, e.g.,
to extract subexpressions from source code data. However, CFGs do not
describe \emph{captures}, i.e., they do not specify how to extract the
parts of interest of an input document, and thus cannot be used
directly for information extraction.

This question of information extraction with grammars was studied by Peterfreund in very recent work~\cite{Peterfreund21}. This paper proposed a formalism of \emph{extraction grammars}, which are CFGs extended via special terminals that describe the endpoints of spans.
Further, it presents an algorithm to enumerate the mappings captured by
\emph{unambiguous} extraction grammars on an input document.
However, while the algorithm achieves constant-delay, the preprocessing bound is significantly worse than in the case of regular spanners: it is quintic in the document, and exponential in the number of variables of the grammar. This complexity is also worse than CFG parsing, e.g., the standard CYK parsing algorithm runs in cubic time in the input string.

Our goal in this paper is to study the enumeration problem for CFGs while
achieving better complexities. Our algorithms ensure a constant-delay
guarantee when outputs have constant size, and more generally ensure
\emph{output-linear} delay when this is not the case: the delay is linear in the
size of each produced solution. Within this delay bound, the preprocessing
time has lower complexity: it is at worse cubic in the
input document, and improves to quadratic or even linear time for restricted
classes. We achieve these results by proposing a new formalism to extend CFGs, called
\emph{annotated grammars}, on which we impose an unambiguity restriction similar
to that of~\cite{Peterfreund21}. Let us present our specific contributions.

    \paragraph{Contributions}
    Our first contribution is to introduce \emph{annotated grammars} (Section~\ref{sec:models}).
    They are a natural extension of CFGs, where terminals are optionally annotated by the information that we wish to extract.
    We then study the problem, given an annotated grammar $\cG$ and document~$s$, of
    enumerating all annotations of~$s$ that are derived by~$\cG$.
This captures the enumeration problems for regular spanners~\cite{FlorenzanoRUVV18,amarilli2020constant}, nested words~\cite{icdt2020nested}, and
    even the extraction grammars of~\cite{Peterfreund21} (we explain this in Section~\ref{sec:spanners}). 
    As we explain, we aim for \emph{output-linear delay}, which is the best
    possible delay in our setting where the solutions to output may have
    non-constant size. 

    Our second contribution is to study the enumeration problem for \emph{unambiguous} annotated grammars (Section~\ref{sec:cubic}),
    that do not produce multiple times the same annotation of an input string. This is a natural restriction to avoid duplicate results, %
    which is also made in~\cite{Peterfreund21}. 
  For such grammars, 
    we present an algorithm to enumerate the annotations produced by a grammar $\cG$ on a string~$s$ with output-linear delay (independent from~$\cG$ or~$s$), after a preprocessing time of $\cO(|\cG| \cdot |s|^3)$, i.e., cubic time in~$s$, and linear time in~$\cG$. 
    This improves over the result of~\cite{Peterfreund21} whose preprocessing is quintic. 
    Our algorithm has a modular design: it follows a standard design of a CFG
    parsing algorithm, but uses the abstract data structure
    of~\cite{icdt2020nested}
    to represent
    the sets of annotations and combine them with operators in a way that allows for output-linear enumeration. 
    We further show a conditional lower bound on the best preprocessing time that can achieve output-linear delay, by reducing from the standard task of checking membership to a CFG, and using the lower bound of~\cite{AbboudBW18}. We show that the preprocessing time must be $\Omega(|s|^{\omega-c})$ for every $c > 0$, where $\omega$ is
    the Boolean matrix multiplication exponent.

    Our third contribution is to improve the preprocessing time by imposing a different requirement on grammars. Thus, we introduce \emph{rigid} annotated grammars (Section~\ref{sec:quadratic}) where, for every input string, all annotations on the string are intuitively produced by parse trees that have the same shape. In contrast with general annotated grammars, we show that rigid annotated grammars can always be made unambiguous, so that our algorithm applies to them. But we also show that, under this restriction, the data complexity of our algorithm goes down from cubic to quadratic time.
    Further, achieving sub-quadratic preprocessing time would imply a sub-quadratic algorithm to test membership to an unambiguous CFG, which is an open problem.

    Our last contribution shows how we can, in certain cases, achieve linear-time preprocessing complexity and output-linear delay (Section~\ref{sec:linear}). This is the complexity of enumeration for regular spanners, and is by definition the best possible. We show that the same complexity can be achieved, beyond regular spanners, for a subclass of rigid grammars, intuitively defined by a determinism requirement. We define it via the formalism of \emph{pushdown annotators} (PDAnn for short), which are the analogue of pushdown automata for CFGs, or the extraction pushdown automata of~\cite{Peterfreund21}. We show that PDAnn are equally expressive to annotated grammars, and that rigid CFGs correspond to a natural class of PDAnns where all runs have the same sequence of stack heights. Moreover, 
    we show that we can enumerate with linear-time preprocessing and output-linear delay in the case of \emph{profiled-deterministic} PDAnn, where the sequence of stack heights can be computed deterministically over the run: this generalizes regular spanners and visibly-pushdown automata.

    This paper is the extended version of the work published at
    PODS'22~\cite{amarilli2022efficientPODS}. It includes complete proofs of
    all results in the appendix.

\paragraph{Related work}
We have explained how our work is set in the context of document spanners~\cite{FaginKRV15}, and in particular of enumeration results for regular spanners~\cite{FlorenzanoRUVV18,amarilli2019constant}. A recent survey of much of this literature can be found in Peterfreund's PhD thesis~\cite{peterfreund2019complexity}. The most related work to ours is the more recent introduction of extraction grammars by Peterfreund~\cite{Peterfreund21}, which we already discussed.
Another related work is~\cite{icdt2020nested}, by some authors of the present
paper. This paper studies enumeration for spanners over nested documents,
defined as visibly pushdown transducers. In the present paper, we re-use the
enumeration data structure of~\cite{icdt2020nested}, and we consider a transducer model in
Section~\ref{sec:linear} that recaptures some of the results of that work. However, in our
problem, we do not require visibility guarantees. This poses new technical
challenges: the underlying tree structure (i.e., the parse tree) is not known in
advance and generally not unique.

There are also some other extensions of regular spanners that are reminiscent of CFGs, e.g., core spanners (featuring equality) or generalized core spanners (with difference) already introduced in~\cite{FaginKRV15}, or Datalog evaluated over regular spanners as in~\cite{peterfreund2019recursive}. However, to our knowledge, there are no known constant-delay enumeration algorithms in these contexts.

Our study of enumeration for annotated grammars is also reminiscent of
enumeration results for queries over trees expressed as tree automata. An
algorithm for this was given by Bagan~\cite{bagan2006mso} with linear-time
preprocessing and constant-delay in data complexity, for deterministic tree
automata, and this was extended in~\cite{amarilli2019enumeration} to
nondeterministic automata. However, this is again more restricted:
evaluating a tree automaton on a tree amounts to evaluating a \emph{visibly
pushdown} automaton over a string representation of the tree, which is again
more restrictive than general context-free grammars.

\section{Grammars and Annotators}\label{sec:models}

\paragraph{Strings and annotations} Let $\Sigma$ be a finite alphabet.
We write $\Sigma^*$ for the set of strings
over $\Sigma$. The \emph{length} of a string $w = w_1 \cdots w_n \in
\Sigma^*$ is $|w| \colonequals n$. The string of length 0 is written
$\epsilon$. We write $u \cdot v$ or $uv$ for the \emph{concatenation} of $u, v \in \Sigma^*$.

Let $\Omega$ be a finite set of annotations. An
\emph{annotated string} is a string $\hat{w} \in  (\Sigma \cup \Sigma \times \Omega)^*$. 
We denote strings by~$w$ and annotated strings by~$\hat{w}$ when this avoids confusion. 
Intuitively, if $\hat{w} = \hat{w}_1 \cdots \hat{w}_n$, then $\hat{w}_i = (a, \oout) \in \Sigma \times \Omega$ means that the letter $a$ at position~$i$ is annotated with $\oout$ (called an \emph{annotated letter}) and $\hat{w}_i \in \Sigma$ means that there is no annotation at position~$i$. 
Given an annotated string $\hat{w} = \hat{w}_1 \cdots \hat{w}_n$, we denote by
$\str(\hat{w}) = \str(\hat{w}_1) \cdot
\cdots \cdot \str(\hat{w}_n)$ the \emph{unannotated string} of~$\hat{w}$, i.e., $\str((a,
\oout)) \colonequals a$ and $\str(a) \colonequals a$, and we denote by
$\ann(\hat{w}) = \ann(\hat{w}_1, 1) \cdot \cdots \cdot \ann( \hat{w}_n, n)$
the \emph{annotations} of~$\hat{w}$, i.e.,
$\ann((a, \oout), i) \colonequals (\oout, i)$ and $\ann(a, i) \colonequals
\epsilon$.
Note that $\card{\str(\hat{w})} = \card{w}$, but the length
$\card{\ann(\hat{w})}$ of $\ann(\hat{w})$ can be
much less than~$\card{w}$.

\paragraph{Annotated grammars} 
A \emph{context-free grammar} (CFG) over~$\Sigma$ is a tuple $G = (V, \Sigma, P,
S)$, where $V$ is a set of \emph{nonterminals}, $\Sigma$ is the alphabet (whose
letters are called \emph{terminals}),
$S \in V$ is the \emph{start symbol}, and $P$ is a finite set of \emph{rules} of the form $X \to \alpha$
where $X \in V$ and $\alpha \in (V \cup \Sigma)^*$. We assume that $V$ and
$\Sigma$ are disjoint. In this paper, we extend this definition to an
\emph{annotated (context-free) grammar} $\cG = (V, \Sigma, \Omega, P, S)$, which
is simply the CFG $(V, \Sigma \cup \Sigma \times \Omega, P, S)$.
We use $G$ to denote a CFG and $\cG$ to denote an annotated grammar.
The \emph{terminals} of~$\cG$ are letters $a \in \Sigma$ and annotated letters $(a, \oout) \in \Sigma \times \Omega$. 

We recall the semantics of a CFG $G = (V, \Sigma, P, S)$.
Given a string $u \in \Sigma^*$, two strings $\gamma, \delta \in (V \cup
\Sigma)^*$, and $X \in V$, we say that $u X \delta$ \emph{produces} $u \gamma
\delta$, denoted by $u X \delta \der{G}u \gamma \delta$, if $P$ contains the rule
$X \rightarrow \gamma$. We then say that $\alpha \in (V \cup \Sigma)^*$
\emph{derives} $\beta \in (V \cup \Sigma)^*$, denoted by $\alpha \ders{\cG}
\beta$ or just $\alpha \ders{} \beta$, if there is a sequence of strings $\alpha_1, \ldots, \alpha_m$ 
with $m\geq 1$
such that $\alpha = \alpha_1 \der{} \alpha_2 \der{} \ldots \der{} \alpha_m =
\beta$. We say that $G$ \emph{derives} $\alpha \in (V \cup \Sigma)^*$ if $S
\ders{} \alpha$, and define the \emph{language} $L(G)$ of~$G$ as the set of
strings $\{w \in \Sigma^* \mid S \ders{} w\}$. Note that our derivations
are leftmost derivations, which is standard for the unambiguity notions that we
introduce afterwards.
The \emph{language of an annotated grammar} $\cG$ is that of the underlying CFG on the
alphabet of terminals $\Sigma \cup \Sigma \times \Omega$. In particular,
$L(\cG)$ is a set of annotated strings.

The purpose of annotated grammars is to consider all possible
annotations of an input unannotated string $w \in \Sigma^*$. 
Specifically, the \emph{semantics} of an annotated grammar~$\cG$ is the
function $\sem{\cG}$ mapping each string $w \in \Sigma^*$ to the following
(possibly empty) set of annotations:
$  \sem{\cG}(w) \ := \ \{\ann(\hat{w}) \mid \hat{w} \in
L(\cG) \wedge \str(\hat{w}) = w\}$.

An \emph{output} of evaluating $\cG$ over $w$ is just an element $\mu \in
\sem{\cG}(w)$.
Note that, in the case when $\Omega = \emptyset$, for all $w \in \Sigma^*$ we have $\sem{\cG}(w) =
\emptyset$ if $w \notin L(\cG)$ and $\sem{\cG}(w) = \{\epsilon\}$ if $w \in
L(\cG)$. So, annotated grammars subsume CFGs. In Section~\ref{sec:spanners}, we show that they also subsume the extraction grammars of~\cite{Peterfreund21}, which implies that annotated grammars are more expressive than regular spanners~\cite{FlorenzanoRUVV18,amarilli2020constant}, or even visibly
pushdown transducers from~\cite{icdt2020nested}.

Towards ensuring tractability, we call a CFG $G$ \emph{unambiguous} if for every
$w \in L(G)$ there is a unique derivation of $w$ by~$G$. We call an annotated
grammar
$\cG$ {\em unambiguous} if the underlying CFG over $\Sigma
\cup \Sigma\times\Omega$ is unambiguous.
Intuitively, this means that each output $\mu \in \sem{\cG}(w)$ can be produced
in only one way.
Remember that there are CFGs $G$ with no unambiguous CFG $G'$ 
\emph{equivalent} to $G$ (i.e., such that $L(G') = L(G)$), and it is undecidable to check
whether an input CFG is unambiguous, or has an equivalent unambiguous CFG. The same
is immediately true for annotated grammars.

\paragraph{Problem statement} 
The goal of this paper is to study how to efficiently enumerate the
annotations of an annotated grammar:
\smallskip

\begin{center}
	\framebox{
		\begin{tabular}{rl}
			\textbf{Input:} & An annotated grammar $\cG$
                        and a string $s\in \Sigma^*$ \\
			\textbf{Output:} & Enumerate the outputs of $\sem{\cG}(s)$
		\end{tabular}
	}
\end{center}
\smallskip
We work in the standard computational model of Random Access Machines (RAM)
with logarithmic word size and
uniform cost measure, having addition and subtraction as basic
operations~\cite{AhoHU74}. The size of~$\cG$ is measured as the sum of rule lengths.

As the set of outputs $\sem{\cG}(w)$ can be large, we
work in the framework of \emph{enumeration algorithms}.
Such algorithms consist of two \emph{phases}.
First, in the \emph{preprocessing phase},
the algorithm receives the input annotated grammar $\cG$ and string~$w$, and produces some
\emph{index} structure $D$.
The \emph{preprocessing time} is the worst-case running time of this
preprocessing phase, measured as a function of the input, i.e., in terms of~$w$
and~$\cG$ when studying \emph{combined complexity}, and in terms of~$w$ only
when studying \emph{data complexity}.

Second, in the \emph{enumeration phase}, the algorithm can use $\cG$, $w$, and~$D$,
and must produce all outputs of $\sem{\cG}(w)$
one after the other and without
repetitions. The \emph{delay} of this phase is the worst-case time to
produce any of the outputs, i.e., for $N$ the number of outputs,
if we call $\mathsf{time}_0$ the moment the preprocessing ends, 
$\mathsf{time}_i$ the moment the algorithm finishes producing the $i$-th output
with $1 \leq i \leq N$, and 
$\mathsf{time}_{N+1}$ the moment when the algorithm terminates,
then the delay is the maximum of the values $(\mathsf{time}_i - \mathsf{time}_{i-1})$ for any 
$0 < i \leq N+1$.
We aim for \emph{output-linear} delay~\cite{FlorenzanoRUVV20} (also called linear delay~\cite{Courcelle09}, or constant delay~\cite{Segoufin13}
for constant-sized outputs), where the delay is
linear in the size of each produced output, and is independent from the input
(i.e., from~$w$ and~$\cG$).
The \emph{memory usage} of the algorithm is the maximum memory
used across both phases, including the size of~$D$.

The ultimate goal of this paper is to find enumeration algorithms to enumerate
the outputs of annotated grammars
with linear preprocessing and
output-linear delay.  However, as we will see, this goal
is not always realistic, so we will initially settle for a higher
processing time, i.e., quadratic or cubic, before presenting classes with linear
preprocessing in data complexity.  We present our first results
towards this goal in this next section.

\section{Unambiguous Grammars} \label{sec:cubic}

In this section we start presenting our results and show a first
algorithm to enumerate the outputs of an annotated grammar on an input string.
The algorithm applies to any unambiguous annotated grammar, and ensures cubic-time preprocessing and
output-linear delay in data complexity; in terms of combined complexity, the
preprocessing is linear on the grammar.
This improves the result by
Peterfreund~\cite{Peterfreund21}, which had quintic-time preprocessing in data complexity.

The section is structured as follows.
We present a general-purpose enumeration data structure called
\emph{enumerable sets}, which is the basis of 
our enumeration algorithms. 
We then introduce the \emph{arity-two normal form} for annotated grammars,
designed to ensure efficient enumeration, and which can be
enforced in linear time.
After this, we present
our algorithm and state our main result (Theorem~\ref{thm:cubic}).
Last, we state a conditional
data complexity lower bound.

\paragraph{Enumerable sets}
The preprocessing phase of our enumeration algorithm builds data structures
representing the set of outputs to enumerate. For this, we essentially re-use
the $\epsilon$-ECS data structure of~\cite{icdt2020nested}, but for convenience
we present them in a self-contained way for our context, and
name them \emph{enumerable sets}.
We now define them and state that we can enumerate their contents
with output-linear delay (Theorem~\ref{thm:enum}).
The enumeration phase of our algorithm simply enumerates the outputs of 
an enumerable set using this delay guarantee.

An \emph{enumerable set}  is a representation of a set of
strings over some alphabet $\O$.  For our case, we want strings of
$\O^*$ to describe outputs, so $\O$ consists of pairs of
annotations
with positions of the input
string~$w$, i.e.,
$\O \colonequals \Omega \times \{1, \ldots, |w|\}$.

The basic enumerable sets are:
\begin{itemize}
  \item $\textsf{empty}$, the empty set;
  \item $\singleton(\epsilon)$, the singleton set containing the empty string;
  \item $\singleton(x)$ for
$x \in \O$, the singleton set with the
    single-character string~$x$.
\end{itemize}

Enumerable sets can be combined using operators to form more complex
enumerable sets. The operators that we consider all take constant-time
and are {\em
fully-persistent}~\cite{driscoll1986making}. Specifically, given
enumerable sets $\dsD_1$ and $\dsD_2$, combining them creates an
enumerable set without modifying $\dsD_1$ and $\dsD_2$ (i.e., they can still be
used in other operator applications). To make this
possible, enumerable sets can share some components, e.g., some parts
of the arguments $\dsD_1$ and $\dsD_2$ can be shared in memory, and
the result can also have some parts that are shared with
$\dsD_1$ and $\dsD_2$. This is similar, e.g., to persistent lists,
where we can only extend a list by adding an element to its head: this
does not modify the original list, and returns a new
list sharing some memory with the original list.

The two operators to combine enumerable sets are:

\begin{itemize}
  \item The \emph{union} operator $\textsf{union}(\dsD_1, \dsD_2)$ can
    be applied if the sets represented by $\dsD_1$ and $\dsD_2$ are
    disjoint, intuitively to avoid duplicates. It returns an enumerable set representing the
    union of these sets,

  \item The \emph{product} operator $\textsf{prod}(\dsD_1, \dsD_2)$
    can be applied if there are no common letters in the strings of
    the sets represented by $\dsD_1$ and $\dsD_2$, i.e., if $\dsD_1$
    (resp. $\dsD_2$) represents $S_1$ (resp. $S_2$) then the sets
    $\{x_1 \in \O \mid x_1 \text{~occurs~in~some~} w_1 \in S_1\}$ and
    $\{x_2 \in \O \mid x_2 \text{~occurs~in~some~} w_2 \in S_2\}$ are
    disjoint.  Then, the operation returns an enumerable set $\dsD$
    which represents the concatenations of the strings in~$\dsD_1$ and
    in~$\dsD_2$: formally, it represents $S_1 \cdot S_2 = \{w_1 \cdot w_2 \mid
    w_1 \in S_1, w_2 \in S_2\}$.
\end{itemize}

It is known that enumerable sets can be enumerated efficiently:

\begin{theorem}[\cite{amarilli2017circuit,icdt2020nested}]
  \label{thm:enum}
  We can implement enumerable sets such that:
\begin{itemize} \item The enumerable
        sets $\empt$, $\singleton(\epsilon)$, and $\singleton(x)$ for $x \in \O$ can be built in
  constant time;
\item The union and product
  operations can be implemented in constant time and in a fully
  persistent way;
\item Given an enumerable set,  we can enumerate the strings
      it represents with output-linear delay and memory usage linear
  in the number of instructions used to build it.
\end{itemize}
\end{theorem}

This was shown in~\cite[Theorem 9]{icdt2020nested} with the data structure of
\emph{Enumerable Compact Sets} ($\epsilon$-ECS). It also follows from the work
in~\cite{amarilli2017circuit} on \emph{zero-suppressed d-DNNF circuits}. For the
reader's convenience, and to derive the bound on memory usage, we give a
self-contained proof in Appendix~\ref{apx:enum}.

\paragraph{Arity-two normal form}
Having presented enumerable sets, we now present the normal form to
enforce on annotated grammars. Our results could be shown using the
commonly known Chomsky normal form (CNF), but we cannot always obtain
an equivalent CNF of linear size from a grammar. For this
reason, we use a variant of CNF, the
\emph{arity-two normal
form} (2NF)~\cite{lange2009cnf}, which is intuitively like CNF but without
disallowing rules of the form $X \rightarrow Y$ or $X \rightarrow \epsilon$. Formally,
  we say that an annotated grammar $(V, \Sigma, \Omega, P, S)$ is in \emph{arity-two normal
  form} (2NF) if the following hold:
  \begin{itemize}
\item Every nonterminal $X$ can derive some
string, i.e., there exists $\hat{w}\in (\Sigma \cup \Sigma\times \Omega)^*$
such that $X \ders{\cG} \hat{w}$.
\item Every nonterminal $X$ can be
reached from the start symbol $S$, i.e., there exists $\alpha,\beta \in  (V \cup
\Sigma \cup \Sigma \times \Omega)^*$ such that $S \ders{\cG}
\alpha X \beta$.
     \item For every rule $X \rightarrow \alpha$ in~$P$, we have $|\alpha| \leq
       2$, and if $\alpha = 2$ then it consists of two nonterminals.
\end{itemize}

We can easily translate annotated grammars to 2NF, as in~\cite{lange2009cnf}:

\begin{proposition}[\cite{lange2009cnf}]
  \label{prp:2nf}
  Given any annotated grammar\/ $\mathcal{G}$, we can compute in linear time an annotated
  grammar $\mathcal{G}'$ in 2NF such that $\mathcal{G}$ and $\mathcal{G}'$ are
  equivalent. Furthermore, if $\mathcal{G}$ is unambiguous then $\mathcal{G}'$ is unambiguous as well.
\end{proposition}

The proof is straightforward and shown in Appendix~\ref{apx:arity2}.
By Proposition~\ref{prp:2nf}, we assume that the 
input grammar~$\cG$ is in 2NF.

We also compute in linear time some more information about $\cG$. First, we precompute which
nonterminals are \emph{nullable}, i.e., are such that $X \ders{\cG} \epsilon$: if we have a rule
$X \rightarrow \epsilon$ then $X$ is nullable, and if we have a
rule $X \rightarrow Y$ where $Y$ is nullable or $X \rightarrow YZ$ where $Y$ and
$Z$ are nullable then $X$ is also nullable. From this information, we further
compute for each nonterminal $Y$ a set $\D[Y]$ of all
nonterminals $X$ such that one of the following rules exist: a rule
$X\rightarrow Y$, a rule $X\rightarrow YZ$ where $Z$ is nullable, or a rule $X\rightarrow ZY$ where $Z$ is
nullable. We can clearly compute all of this in linear time (note that each rule contributes at most two
entries to~$\D$).

Second, as the grammar is assumed to be unambiguous, it also contains
no \emph{cycles}, %
i.e., there are no sequence of nonterminals
$X_1 \dots X_n$ such that for $1\leq i < n$, $X_i\in \D[X_{i+1}]$ and
$X_1\in \D[X_n]$. Indeed, otherwise 
there would be infinitely many possible derivations of some string
starting at~$X_1$, contradicting the unambiguity of~$\cG$.  Thus, we can sort the nonterminals of $\cG$
in \emph{topological order}, by which we mean that when $Y\in \D[X]$
then $Y$ is enumerated after $X$. Intuitively, when we consider a nonterminal
$X$, we want to be done with processing the nonterminals $Y$ such that $X \rightarrow
Y$ or $X \rightarrow YZ$ or $X \rightarrow ZY$ with $Z$ nullable. This order can also be computed in linear time.
%


\paragraph{Enumeration algorithm}
We now present the preprocessing phase of the enumeration algorithm, formalized as
Algorithm~\ref{alg:cubic} where the  input string $w=a_1\ldots a_n$ is
assumed nonempty. 

The principle of the algorithm is the following:
\begin{principle}
  \label{pri:main}
    For every triple
of the form $(i,j,X)$ with $1\leq i < j \leq n+1$ and $X\in N$, the
table cell $\tableI[i][j][X]$ will contain an enumerable set
representing the annotations of the string $a_i \cdots a_{j-1}$ that
can be derived from symbol $X$ in the grammar.
\end{principle}
These sets are initialized to be empty.
In lines~\ref{ln:initstart}--\ref{ln:initend} of the algorithm, the cells $\tableI[i][j][X]$ with
$j-i=1$ are initialized
to consider derivations via ``simple rules'' of the form $X \rightarrow a$ or
$X \rightarrow (a, \oout)$.
(For now, ignore the role of the
$\mathsf{endIn}$ table.) Note that the rules of the form $X \rightarrow
\epsilon$ are considered when defining~$\D$ and not further examined by the algorithm. At the end, line~\ref{ln:last} returns the
enumerable set for the annotations of the entire string derivable from
the start symbol, i.e., the outputs of~$\cG$ on~$w$.

The main part of the algorithm consists in satisfying Principle~\ref{pri:main} by adding
the annotations corresponding to ``complex'' rules (i.e., of the form
$X\rightarrow YZ$ or $X\rightarrow Y$). At the beginning of the
algorithm the cells of the table $\tableI$ might lack some annotations
corresponding to complex rules, but each cell will be considered
complete at some point during the execution, at which point it will
satisfy Principle~\ref{pri:main} and will not be modified anymore.
We define the order in which the cells are considered complete as follows:
$(i,j,X)<(i',j',X')$ when $j<j'$ or
$(j=j' \land i>i')$ or $(j=j' \land i=i' \land X < X')$ where we order
nonterminals $X$ and $X'$ following the topological order from $\D$.

Consider the \emph{complex derivations} starting from $X$ of the string $a_i \cdots
a_{j-1}$, i.e., those that begin with a complex rule. We will see here how to
reflect them in $\tableI[i][j][X]$. There are two kinds of complex
derivations. The first kind is the derivations where we first rewrite $X$
to another nonterminal $Z$ with a rule $X \rightarrow Z$, or by
rewriting $X$ to $YZ$ or $ZY$ but where $Y$ is nullable and will be
rewritten to~$\epsilon$. In these three cases, we have
$X \in \D[Z]$. Thus, we fill the index $\tableI[i][j][X]$ with the
contents of $\tableI[i][j][Z]$, which is already complete, for
$X\in \D[Z]$ (lines \ref{ln:directProdBegin}-\ref{ln:directProdEnd}).

The second kind of complex derivation begins with a complex rule $X
\rightarrow YZ$ where neither $Y$ nor $Z$ will be rewritten to $\epsilon$. In this case, the set of
annotations to add into $\tableI[i][j][X]$ using this rule is the union of
products of all the $\tableI[i][k][Y]$ and $\tableI[k][j][Z]$ where $i < k <
j$. We have $(i,k,Y)<(k,j,Z)<(i,j,X)$, so we can fill $\tableI[i][j][X]$ with the product of the contents of $\tableI[i][k][Y]$ and $\tableI[k][j][Z]$,
at the moment where $\tableI[k][j][Z]$ is considered complete.

To summarize, from line~\ref{ln:calc} onwards, the algorithm
considers the positions $j$ in ascending order, and 
populates all cells $\tableI[i][j][X]$ so that they are complete. To do
so, we consider the triples $(k,j,Z)$ by increasing order in our
sorting criterion, i.e., by decreasing $k$, then increasing~$Z$ in the
order of the topological sort. Whenever we consider a cell, it is
complete, and we consider its contributions to cells of the form
$\tableI[i][j][X]$ with $i=k$ using complex rules of the first kind 
(lines \ref{ln:directProdBegin}-\ref{ln:directProdEnd}), and if it is non-empty
we consider how to combine it with
a neighboring cell (which is also complete and non-empty) as we explained previously,
adding the results to a cell $\tableI[i][j][X]$ with $i < k$ which is not yet complete (lines
\ref{ln:kind2begin}--\ref{ln:kind2end}).

We now explain the optimization involving the set $\mathsf{endIn}$. It
is not necessary to achieve the cubic running time of
this section, but is required for the quadratic bound in
Section~\ref{sec:quadratic}.  The optimization is that, when processing the
triple $(k,j,Z)$ and the rule $X\rightarrow YZ$, we do not test all the
possible cells $\tableI[i][k][Y]$, but only those that are
non-empty. Indeed, if $\tableI[i][k][Y]$ is empty, then the concatenation of
$\tableI[i][k][Y]$ with $\tableI[k][j][Z]$ is also empty. Thus, we maintain
the list $\mathsf{endIn}[k][Y]$ of all the~$i$'s to consider
with $i<k$, i.e., those such that $\tableI[i][k][Y]$ is non-empty. We
initialize this list to be empty, add $i$ to $\mathsf{endIn}[k][Y]$ whenever
$\tableI[i][k][Y]$ becomes non-empty (at line~\ref{ln:endset1} in the base
case, or at line~\ref{ln:endset2} before adding to an empty cell for
the first time). Then, we only consider the indices
$i$ of this list to combine $\tableI[i][k][Y]$ with another cell.

We now argue that our algorithm is correct, and in particular
that 
\textbf{(i)} 
we satisfy Principle~\ref{pri:main};
that
\textbf{(ii)} all the unions are disjoint, and that \textbf{(iii)} all the
products involve enumerable sets on disjoint alphabets. 
One can establish \textbf{(i)} by showing by induction over cells that the invariant is
correct when each cell is considered complete by our algorithm (and the
cell is not changed afterwards). Knowing \textbf{(i)}, the first violation of \textbf{(ii)} would witness that
the same annotation of some factor $a_i \cdots a_{j-1}$ can be derived in two
different ways from a nonterminal $X$,
contradicting unambiguity, so there are no
violations of \textbf{(ii)}. For
\textbf{(iii)}, we simply observe that, by \textbf{(i)}, $\tableI[i][j][X]$
only contains pairs of the form $(\oout, k)$ for some $i \leq k < j$, so we can
indeed perform the product of $\tableI[i][k][Y]$ and $\tableI[k][j][Z]$.

This establishes that the algorithm is correct. Now, the running time
of the preprocessing phase of the algorithm is clearly in $\cO(n^3 |\cG|)$, because (1) the $\mathsf{endIn}$ lists
are of size $\cO(n)$ at most, and (2) the consideration of all $Z \in N$ and $X \in \D[Z]$
is in $\cO(|\cG|)$: every $X \in \D[Z]$ corresponds to a rule, so the consideration of all $Z \in N$ and
rules in CRule$[Z]$ is in $\cO(|\cG|)$.
The enumeration phase is then simply that of Theorem~\ref{thm:enum}. Hence, we have
shown that enumeration for unambiguous annotated grammars can be achieved
with cubic time preprocessing and output-linear delay:

\begin{theorem}
  \label{thm:cubic}
	Given an unambiguous annotated grammar $\cG$ and an input string
        $w$, we can enumerate $\sem{\cG}(w)$ with preprocessing in $\cO(|w|^3
        \cdot
        |\cG|)$ (hence cubic in data complexity), and output-linear delay
        (independent from~$w$ or~$\cG$). The memory usage is in $\cO(|w|^3 \cdot
        |\cG|)$.
\end{theorem}

\newcommand{\To}{\textbf{to}}
\newcommand{\Downto}{\textbf{downto}}

\begin{algorithm}[t]
	\caption{Preprocessing phase: given a 2NF
        unambiguous annotated grammar $\cG = (N, \Sigma, \Omega, P, S)$ and
        a non-empty string $w = a_1\cdots a_n$, compute an enumerable set
        representing $\sem{\cA}(w)$.}\label{alg:preprocessing}
        \label{alg:cubic}
		\begin{algorithmic}[1]
                  \State I $\gets$ an array $(n+1)\times (n+1)\times N$ initialized with $\empt$
                  \State  $\mathsf{endIn}\gets$ an array $(n+1)\times N$
                  initialized with empty lists
                  \State CRule $\gets$ an array such that CRule$[Z]=\{X\rightarrow YZ \in P \}$
                  \State $\D \gets $ an array as described in the presentation of 2NF

                  \For{$1 \leq i \leq n$} \label{ln:initstart} 
                    \If{rule $(X \rightarrow a_i)$ in $P$}
                      \State $\tableI[i][i+1][X] \gets \union(\tableI[i][i+1][X], \singleton(\epsilon))$
                    \EndIf
                    \For{rule $(X \rightarrow (a_i, \oout))$ in $P$}
                      \State $\tableI[i][i+1][X] \gets \union(\tableI[i][i+1][X], \singleton((\oout, i)))$
                    \EndFor
                    \If{$\tableI[i][i+1][X] \neq \empt$}
                      \State $\mathsf{endIn}[i+1][X]$.append($i$)
                      \label{ln:endset1}
                    \EndIf
                  \EndFor \label{ln:initend}

                  \For{$j=1$ \To{} $n+1$} \label{ln:calc} 
                    \For{$k=j-1$ \Downto{} $1$}
                      \For{nonterminal $Z \in N$ in topological order}
                          \For{nonterminal $X\in \D[Z]$} \label{ln:directProdBegin}                             
                            \State $\tableI[k][j][X] \gets \union(\tableI[k][j][X],
                            \tableI[k][j][Z])$ \label{ln:directProdEnd}
                          \EndFor
                        \If{$\tableI[k][j][Z] \neq \empt $}
                        \label{ln:kind2begin}                             
                            \For{rule $(X \rightarrow YZ)$ in CRule$[Z]$}
                              \For{$i \in\mathsf{endIn}[k][Y]$}
                                  \label{ln:innermost}
                                \If{$\tableI[i][j][X]=\empt$}
                                  \label{ln:innermost3}
                                  \State $\mathsf{endIn}[j][X]$.append($i$)
                                  \label{ln:endset2}
                                \EndIf
                                  \label{ln:innermost2}
                                \State $\tableI[i][j][X] \gets \union(\tableI[i][j][X],$
                                \State $\mbox{~}\hspace{3em} \prod(\tableI[i][k][Y], \tableI[k][j][Z]))$
                               \EndFor
                            \EndFor
                        \label{ln:kind2end}                             
                         \EndIf
                    \EndFor
                  \EndFor
                  \EndFor
                  \State\Return $\tableI[1][n+1][S]$ \label{ln:last}
\end{algorithmic}%
\end{algorithm}%


\paragraph{Lower bounds}
We cannot show a lower bound that matches the complexity of our algorithm, but
we can prove that we cannot achieve a preprocessing time better than the time to
test whether a string is accepted by a CFG, which is essentially
the complexity of Boolean matrix multiplication~\cite{AbboudBW18}:

\begin{proposition}
  \label{prp:cubiclowerbound}
  Let $\omega$ be the
  smallest value
  such that we can multiply two Boolean $n\times
  n$ matrices in time $\cO(n^{\omega+o(1)})$. Then for any $c>0$ there is an unambiguous
  annotated grammar $\cG$ such that, given an input string~$w$, enumerating 
  $\sem{\cG}(w)$ with output-linear delay requires a preprocessing time of
  $\Omega(|w|^{\omega-c})$.
\end{proposition}

\section{Rigid Grammars} \label{sec:quadratic}

In the previous section, we have shown how to enumerate the output of
unambiguous annotated grammars on strings, with output-linear delay and cubic
preprocessing in the input string. This algorithm has two drawbacks: it requires
us to impose that the grammar is unambiguous, and the cubic preprocessing 
may be expensive.

In this section, we introduce a new class of annotated grammars, called
\emph{rigid grammars}. Rigid grammars do not need to be
unambiguous, but as we will show a rigid grammar can always be converted to
additionally impose unambiguity. The point of rigid grammars is that we can show
a quadratic bound on Algorithm~\ref{alg:cubic} for them.

We first define rigid grammars in this section. We then state that we can impose
unambiguity for rigid grammars, and derive some consequences about their
expressiveness and the complexity of recognizing
them. Last, we show a quadratic bound on the
preprocessing time for output-linear enumeration for such grammars, and explain
why a better bound would be challenging to achieve.

%
%
%
%
%
%

\label{sec:profile}

\paragraph{Rigid grammars}
We first define the restricted notion of grammars that we study.
Consider an annotated grammar $\cG = (N, \Sigma, \Omega, P, S)$, and a string
$\gamma\in (\Sigma \cup (\Sigma \times \Omega)\cup N)^*$ of nonterminals and
of terminals which may carry an annotation in~$\Omega$. We will be interested in
the \emph{shape}
of~$\gamma$, written $\shape(\gamma)$: it is the string over $\{0, 1\}$
obtained by replacing every nonterminal of~$N$ in~$\gamma$ by~$1$ and replacing
all terminals (annotated or not) by~$0$: note that $|\shape(\gamma)| = |\gamma|$.

We then say that an annotated grammar $\cG$ is \emph{rigid}
        if for every string
        $w \in \Sigma^*$, all derivations from the start symbol~$S$ of~$\cG$ to
        an annotated string $\hat{w}$ of~$w$ have the same sequence of shapes
        (remember that we only consider leftmost derivations).
        Formally, there exists a sequence $s_1, \ldots, s_k \in \{0,
        1\}^*$ depending only on~$w$ such that for any derivation 
        $S = \alpha_1 \der{\cG} \alpha_2 \der{\cG} \ldots \der{\cG} \alpha_m =
        \hat{w}$ with $\str(\hat{w}) = w$, we have $m = k$ and $\mathrm{shape}(\alpha_i) = s_i$ for all $1
        \leq i \leq k$.

Intuitively, the sequence of shapes of a derivation describes the
skeleton of the corresponding derivation tree. Thus, a rigid annotated
grammar is one where, for each unannotated string, all derivation trees for all
annotations of the string are isomorphic (ignoring the labels of 
nonterminals and the annotation of terminals).

\paragraph{Rigidity vs unambiguity}
Unambiguity and rigidity for annotated grammars seem
incomparable: unambiguity imposes that every annotation is produced by only one
derivation, whereas rigidity imposes that all derivations across all annotations
have the same shape (but the same annotation may be obtained multiple times).

However, it turns out that, on rigid grammars, we can impose unambiguity without
loss of generality: all rigid grammars can be converted to equivalent rigid and
unambiguous grammars.

\begin{theorem}
  \label{thm:profileu-iou} For any rigid grammar
	$\cG$ we can build an equivalent rigid and
	unambiguous grammar $\cG'$. The transformation runs in exponential time,
        i.e., time $\cO(2^{|\cG|^c})$ for some $c> 0$.
\end{theorem}

The transformation to impose unambiguity goes via a notion of annotated pushdown
automata (introduced in the next section), and is inspired by the
determinization procedure for visibly pushdown automata~\cite{alur2004visibly}, even though rigid
grammars generally do not define visibly pushdown languages. The transformation
comes at a cost, as it will generally blow up the size of the grammar
exponentially.

\paragraph{Expressiveness of rigid grammars}
Armed with Theorem~\ref{thm:profileu-iou}, we study what is the expressive power
of rigid grammars. For this, let us first go back to the setting without
annotations. Theorem~\ref{thm:profileu-iou} tells us that for (unannotated) CFGs the rigidity
requirement is equivalent to the usual unambiguity requirement: each accepted
word has a unique derivation.
Now, for the case of an annotated grammar $\cG$,
rigidity additionally imposes the requirement
that all annotations of an input string have the same parse tree.
In particular, the language of the strings where $\cG$ accepts some
annotation must be recognizable by a rigid (unannotated) CFG, hence an unambiguous CFG (by
Theorem~\ref{thm:profileu-iou}). Formally:

\begin{proposition}
  \label{prp:baseunambig}
   For a rigid grammar $\cG$, let $L'$ be the set of strings with nonempty output, i.e., $L' = \{w \mid \sem{\cG}(w) \not = \emptyset\} $. Then $L'$ is
  recognized by an unambiguous CFG.
\end{proposition}

This yields concrete examples of languages (on the empty annotation alphabet) that cannot be recognized by a
rigid annotated grammar, e.g., inherently ambiguous context-free languages
such as $L_\mathrm{a} = \{a^i b^j c^k \mid i, j, k \geq 1 \land (i = j \lor j =
k)\}$ on~$\{a, b, c\}^*$~\cite{maurer1969direct}.
Proposition~\ref{prp:baseunambig} also implies that we cannot decide if the language
of an annotated grammar can be expressed instead by a rigid grammar,
or if an annotated grammar is rigid:

\begin{proposition}
  \label{prp:unambundec2}
  Given an unannotated grammar $\cG$, it is undecidable to determine whether $\cG$
  is rigid, and it is undecidable to determine whether there
  is some equivalent rigid grammar $\cG'$.
\end{proposition}

These undecidability results make 
rigid grammars less appealing, but note that our enumeration algorithm for such
grammars applies in particular to decidable grammar classes which are
designed to ensure rigidity. For instance, this would be the case
of grammars arising from visibly pushdown automata, which we discuss in more
detail in the next section.

\paragraph{Enumeration algorithm}
We now give our algorithm with quadratic preprocessing time for
rigid grammars. Given a rigid grammar, we first make it unambiguous if
necessary, using Theorem~\ref{thm:profileu-iou}, in exponential time in the
input grammar. The result is a rigid and unambiguous annotated grammar. Now, we
transform it in 2NF like in Section~\ref{sec:quadratic}: this
takes linear time, preserves unambiguity, and one can check that it also
preserves rigidity.

Armed with our rigid and unambiguous grammar $\cG$ in 2NF, we can simply use 
Algorithm~\ref{alg:cubic} to construct a data structure allowing us to enumerate
the outputs with output-linear delay. But we now claim that
Algorithm~\ref{alg:cubic} runs in time $\cO(|\cG| \cdot |w|^2)$ because $\cG$
is~rigid.

For this, we study for every nonterminal $X$ and pair $1 \leq i \leq j \leq n+1$ how
many times we can consider the cell $\tableI[i][j][X]$ in lines
\ref{ln:innermost3}--\ref{ln:kind2end}. Whenever
we consider it, we witness the existence of a complex rule $X \rightarrow Y Z$
and a value~$k$ such that $\tableI[i][k][Y]$ and $\tableI[k][j][Z]$ are nonempty (the first
is because $i \in $ endIn$[k][Z]$). Thus, we witness a derivation from $X$ of
some annotation of the string $a_i \cdots a_{j-1}$ that starts with a rule $X
\rightarrow YZ$ where $Y$ derives some annotation of the string $a_i \cdots
a_{k-1}$ and $Z$ derives some annotation of the string $a_k \cdots a_{j-1}$. We
now claim that, for $(i, j, X)$, the rigidity of the grammar ensures that there is
only one such value~$k$. Indeed, assume by contradiction that we have two
rules $X \rightarrow YZ$ and $X \rightarrow Y'Z'$ and two values $i \leq k < k'
\leq j$ such that $Y$ and $Y'$ respectively derive some annotation of the
strings
$a_i \cdots a_{k-1}$ and $a_i \cdots a_{k'-1}$, and $Z$ and $Z'$ respectively
derive some annotation of the strings
$a_k \cdots a_{j-1}$ and $a_{k'} \cdots a_{j-1}$. Then once we are done
rewriting $Y$ and all the nonterminals that it generates in the first derivation, we
obtain a different shape from what we obtain after rewriting $Y'$ and all the
nonterminals it generates in the second derivation, contradicting the rigidity
of the grammar.

Thus, whenever we consider the cell $\tableI[i][j][X]$ in lines~\ref{ln:innermost2}--\ref{ln:kind2end}, it is
for one value of~$k$ which is unique for $(i,j,X)$, and we thus consider the
cell once
at most for every complex rule of the grammar with $X$ as left-hand-side. Thus,
we consider the cells of $\tableI[i][j]$ at most $|\cG|$ times in total. As
there are $\cO(n^2)$ pairs $(i,j)$, this ensures that the total running time of the
innermost for loop (lines~\ref{ln:innermost}--\ref{ln:kind2end}), and that of the entire algorithm, is indeed
in $\cO(|\cG| \cdot |w|^2)$:

\begin{theorem}
\label{thm:quadratic}
  Given a rigid annotated grammar $\cG$ and an input string $w$, we can enumerate
  $\sem{\cG}(w)$ with preprocessing in $\cO(|w|^2)$ data complexity and
  output-linear delay (independent from~$w$ or~$\cG$). The combined complexity
  of the preprocessing is $\cO(2^{|\cG|^c} \cdot |w|^2)$ for some $c >0$, or
  $\cO(|\cG| \cdot
  |w|^2)$ if $\cG$ is additionally assumed to be unambiguous.
\end{theorem}

\paragraph{Optimality}
We now turn to the question of whether the quadratic preprocessing time for
rigid grammars is optimal. For this, we notice that the parsing of
(unannotated) unambiguous grammars can be performed in quadratic time, but the
question of finding a better algorithm was open as of
2012~\cite{schmitz2012can}. Now, this is a special case of our problem, because
an unannotated unambiguous grammar is in particular a rigid and unambiguous
annotated grammar, and enumerating the outputs of an unannotated grammar just
means deciding in constant time after the preprocessing whether the input
unannotated string is accepted or not. Thus:

\begin{proposition}\label{prp:redunann}
  Any algorithm to enumerate the accepted outputs of a rigid
  annotated grammar can be used to test if an input string is accepted by an
  unambiguous unannotated grammar, with same complexity as that of the
  preprocessing phase.
\end{proposition}

For this reason, we leave open the question of whether a better than quadratic
preprocessing time can be achieved in this case.

\section{Pushdown Annotators} \label{sec:linear}

We have presented an enumeration algorithm for annotated grammars that achieves
quadratic-time preprocessing and output linear delay on rigid annotated
grammars. We now study whether the bound can be improved even further to achieve
linear-time preprocessing and output-linear delay, which is the best possible
data complexity bound in our model.

To achieve this, it is natural to look for a class of grammars having some ``deterministic'' behavior.
Unfortunately, grammars are not convenient for this purpose, and so we move to
the equivalent model of pushdown automata.
We thus introduce \emph{pushdown annotators} and show that they are equally expressive to annotated grammars.
We present syntactic restrictions on pushdown annotators that ensure
quadratic-time preprocessing, similarly to rigid grammars.
Then, we propose additional deterministic conditions on pushdown annotators that allow for linear-time preprocessing.

\paragraph{Pushdown annotators} A \emph{pushdown annotator} (PDAnn) is a tuple
$\cP = (Q, \Sigma, \Omega, \Gamma, \Delta, \allowbreak q_0, F)$ where $Q$ is a finite set of \emph{states},
$\Sigma$ is the alphabet, $\Omega$ is a finite set of annotations, $\Gamma$ is a finite set of \emph{stack symbols}, $q_0 \in Q$ is the \emph{initial state}, and $F \subseteq Q$ are the \emph{final states}. We assume that the set $\Gamma$ of stack symbols is disjoint from $(\Sigma \cup \Sigma\times \Omega)$. Finally, $\Delta$ is a finite set of \emph{transitions} that are of the following kinds:
\begin{itemize}
	\item {\em Read-write transitions} of the form $(p, (a, \oout), q) \in Q\times(\Sigma\times \Omega)\times Q$, meaning that, if the next letter of the string is~$a$, the annotator can go from states~$p$ to~$q$ while reading that letter and writing the annotation~$\oout$;
	\item {\em Read-only transitions} of the form $(p, a, q) \in
          Q\times\Sigma\times Q$, meaning that the annotator can go from $p$ to
          $q$ while reading~$a$;
	\item {\em Push transitions} of the form $(p, q, \gamma) \in Q\times (Q\times\Gamma)$, meaning that the annotator can go from $p$ to $q$ while pushing the symbol~$\gamma$ on the stack;
	\item {\em Pop transitions} of the form $(p, \gamma, q) \in (Q\times \Gamma)\times Q$, meaning that, if the topmost symbol of the stack is~$\gamma$, the annotator can go from $p$ to $q$ while removing this topmost symbol~$\gamma$.
\end{itemize}

We now give the semantics of PDAnns.  Fix a string $w = w_1 \cdots
w_n \in \Sigma^*$. A \emph{configuration} of $\cP$ over $w$ is a pair
$C = (q, i) \in Q\times [0,n]$ of the current state and position in
$w$. An \emph{instantaneous description} of $\cP$ is a pair
$(C, \alpha)$ where $C$ is a configuration and $\alpha\in \Gamma^*$
describes the stack. A \emph{run} of $\cP$ over $w$ is a sequence:
\begin{equation} \label{eq:run}\tag{\!\dagger}
	\rho \ \colonequals \ (C_0, \alpha_0) \trans{t_1} (C_1, \alpha_1) \trans{t_2} \, \ldots \, \trans{t_m} (C_m, \alpha_m)
\end{equation}
such that $C_0 = (q_0, 0)$ and $\alpha_0 = \eps$, each $t_k$ is a transition in $\Delta$, and for each $k\in[1,m]$ the following hold: 
\begin{itemize}
	\item if $t_k$ is a read-write transition $(p, (a, \oout), q)$ or a read-only transition $(p, a, q)$, then $\alpha_k = \alpha_{k-1}$, $C_{k-1} = (p, i-1)$, $C_{k} = (q, i)$ and $a = a_i$ for some $i\in [1,n]$;
	\item if $t_k$ is a push transition $(p, q, \gamma)$, then $\alpha_k = \alpha_{k-1}\gamma$ and for some $i\in [1,n]$, $C_{k-1} = (p,i)$, $C_k = (q,i)$; and
	\item if $t_k$ is a pop transition $(p,\gamma,q)$, then  $\alpha_{k-1} = \alpha_k \gamma$ and for some $i\in [1,n]$, $C_{k-1} = (p,i)$, $C_k = (q,i)$.
\end{itemize}
We say that $\rho$ is \emph{accepting} if $(C_m, \alpha_m) = ((q_f, n), \epsilon)$ for
some $q_f\in F$.
We define the \emph{annotation} of $\rho$ as $\ann(\rho) = \ann(t_1, C_1) \cdot \cdots \cdot \ann(t_m, C_m)$ such that $\ann(t, C) = \epsilon$ if $t$ is a push, pop, or a  read-only transition $(p, a, q)$, and $\ann(t, C) = (i, \oout)$ if $t$ is a read-write transition $(p, (a, \oout), q)$ and $C = (q, i)$.
Finally, we define the function $\sem{\cP}$ that maps any $w \in \Sigma^*$ to its set of outputs:
\[
\sem{\cP}(w) =
\{\ann(\rho) \mid \text{$\rho$ is an accepting run of $\cP$ over $w$}\}.
\]
Similarly to annotated grammars, we say that $\cP$ is \emph{unambiguous} if for every $w\in \Sigma^*$ and output~$\mu$, there exists at most one accepting run $\rho$ of $\cP$ over $w$ such that $\ann(\rho) = \mu$. 

One can alternatively see a PDAnn as a \emph{pushdown transducer}~\cite{berstel2013transductions}, which is the standard way to extend automata to have an output. However, an important difference is that a PDAnn concisely represents outputs by only writing the annotations and their positions: this can be much smaller than the input string, and cannot easily be encoded as a transducer on a finite alphabet. For instance, where a PDAnn can produce an output such as $(2,\oout), (5, \oout')$, a transducer would either write $\oout \oout'$ (losing the position information) or \textvisiblespace~$\oout$~\textvisiblespace~\textvisiblespace~$\oout'$ (whose length is always linear in the input) for a special symbol~\textvisiblespace.

\paragraph{Profiled PDAnns and annotated grammars} To define the analogue
of rigid annotated grammars on PDAnn, we will study the
\emph{stack profile} (or simply \emph{profile}) of PDAnn runs, which is
informally the sequence of all stack heights.
Formally, let $\cP$ be a PDAnn, $w$ be a string, and consider a run $\rho$ of $\cT$
over~$w$ like in ($\dagger$).
The \emph{profile}~$\pi$ of $\rho$ is the sequence $\pi:= \card{\alpha_0}, \ldots,
\card{\alpha_m}$.
We then introduce \emph{profiled} PDAnns by requiring that all
accepting runs of the PDAnn on an input string have the same profile (no matter
their output). Formally, we say that a PDAnn $\cP$ is \emph{profiled} if, for every string $w$, all accepting runs of $\cT$ over~$w$ have the same profile.

As usual for context-free grammars and pushdown automata, the formalisms of annotated grammars and PDAnn have the same expressive power.
We call two annotated grammars
$\cG$ and $\cG'$ \emph{equivalent} if they define the same functions, i.e.,
$\sem{\cG} = \sem{\cG'}$, and extend this notion to PDAnn in the expected way.
We then have:
\begin{proposition}\label{prop:grammar-pdann}
	Annotated grammars and PDAnn are equally expressive. Specifically, for any annotated grammar $\cG$, we can build an equivalent PDAnn $\cP$ in polynomial time, and vice versa.
	Further, $\cG$ is unambiguous (resp., rigid) iff $\cP$ is unambiguous (resp., profiled).
\end{proposition}

Let us now study the enumeration for PDAnns. We know that the problem
for unambiguous PDAnns can be solved via Proposition~\ref{prop:grammar-pdann}
with cubic-time preprocessing in data complexity and output-linear delay (with
Theorem~\ref{thm:cubic}). We know that
profiled PDAnns can be made unambiguous (via Proposition~\ref{prop:grammar-pdann} and Theorem~\ref{thm:profileu-iou}) and so that
we can solve enumeration for them in
quadratic-time preprocessing in data complexity and output-linear delay (using
Theorem~\ref{thm:quadratic}).
We now show that, if we are given a profile of an 
unambiguous PDAnn $\cP$ on an input string~$w$, we can use it as a guide to enumerate with linear
preprocessing and output-linear delay the set $\sem{\cP}_{\pi}(w)$ of
annotations for that profile, i.e., all $\ann(\rho)$ such that $\rho$ is an
accepting run with profile $\pi$ of $\cP$ over $w$. Formally:

\begin{lemma}\label{lem:vpaconv}
	Given an unambiguous PDAnn $\cP$, there exists an enumeration algorithm
        that receives as input a string $w$ and a profile $\pi$ of $\cP$ over
        $w$, and enumerates $\sem{\cP}_{\pi}(w)$ with output-linear delay after
        linear-time preprocessing in data complexity.
\end{lemma}

This result implies that we could achieve linear-time enumeration over profiled
PDAnn if we could easily discover their (unique) profile.
We achieve this with \emph{profiled-deterministic PDAnns}.

\paragraph{Profiled-deterministic PDAnn}
Let $\cP = (Q, \Sigma, \Omega, \Gamma, \Delta, q_0, F)$ be a PDAnn. We
say that a PDAnn $\cP$ is \emph{profiled-deterministic} if, for any
string $w \in \Sigma^*$, for any two partial runs $\rho$ and
$\rho'$ of $\cP$ over $w$ with the same length, $\rho$ and $\rho'$ have the same profile.

The relationship between profiled-deterministic PDAnns and 
deterministic pushdown automata (formally defined in Appendix~\ref{apx:linear}) is similar to the relationship between 
profiled PDAnns and unambiguous pushdown automata (the latter relationship was
stated as
Proposition~\ref{prp:baseunambig} in the context of grammars). Specifically:

\begin{proposition}
  \label{prp:basedeterm}
  For a profiled-deterministic PDAnn $\cP$, let $L'$ be the set of strings with nonempty output, i.e., $L' = \{w \mid \sem{\cP}(w) \not = \emptyset\} $. Then $L'$ is
  recognized by a deterministic pushdown automaton.
\end{proposition}

This result gives a concrete picture of the expressive power of 
profiled-deterministic PDAnn $\cA$, i.e., as acceptors they are more powerful
than the class of
\emph{visibly pushdown
automata}~\cite{alur2004visibly}, where each alphabet letter must have a
specific effect on the profile. Profiled-deterministic PDAnn are also
reminiscent of 
the \emph{height-determinism} notion introduced for pushdown automata~\cite{nowotka2007height}, but
extend this with the support of annotations.

Profiled-deterministic PDAnn are designed to ensure that they have only one
profile (i.e., they are profiled), and further that their unique
profile can be constructed in linear time:

\begin{proposition}\label{prp:detlinear}
  A profiled-deterministic PDAnn $\cP$ is always profiled, and given a string
  $w$, the unique profile of accepting runs of $\cP$ over~$w$ can be computed in
  linear time in~$w$.
\end{proposition}

Together with Lemma~\ref{lem:vpaconv}, this yields:

\begin{corollary}\label{cor:linear-time-pdann}
	Let $\cP$ be a profiled-deterministic PDAnn. Then for every string $w$
        the set $\sem{\cP}(w)$ can be enumerated with output-linear delay after
        linear-time preprocessing in data complexity.
\end{corollary}

\section{Application: Document Spanners} \label{sec:spanners}

We have presented our enumeration results for annotated grammars and
pushdown annotators. We conclude the paper by applying them to the
standard context of \emph{document spanners}~\cite{FaginKRV15} and to the
\emph{extraction grammars} recently introduced in~\cite{Peterfreund21}.

\paragraph{Mappings, spanners, extraction grammars}
Our paper studies strings~$s \in \Sigma^*$ with letters
\emph{annotated} by an annotation from a finite set, but document spanners work
with \emph{mappings} selecting so-called \emph{spans} of~$s$. Formally, 
a \emph{span} of~$s$ is a pair 
$\spanc{i}{j}$ with $1 \leq i \leq j \leq |s|+1$ describing a
possibly empty factor of~$s$.
For a finite set of \emph{variables} $\var$,
a \emph{document spanner} (or just \emph{spanner}) is a
function that maps every string $s \in \Sigma^*$ to a set of \emph{mappings},
where a mapping (intuitively denoting an extraction result)
assigns each variable of~$\var$ to a span of~$s$.

\newcommand{\splain}{\textsf{plain}}
Peterfreund~\cite{Peterfreund21} specifies spanners via \emph{extraction
grammars}. An \emph{extraction grammar} $\cH$ is simply a
CFG over the alphabet $\Sigma$ extended with \emph{variable operations} of the form
$\vdash_x$ and $\dashv_x$ for $x \in \var$, intuitively denoting the beginning and end of a
span for~$x$. Any such grammar $\cH$ denotes a \emph{language} $L(\cH)$ of words with
variable operations, called \emph{ref-words}, and gives a function
$\sem{\cH}$ associating every string $w \in \Sigma^*$ to the set of
mappings $\sem{\cH}(w)$ defined as follows. We consider every ref-word $\hat{w}$ of~$L(\cH)$ whose restriction to~$\Sigma$
is~$w$ and which is \emph{valid}, i.e., 
for every $x \in \var$ there is one occurrence of~$\vdash_x$ followed by
one occurrence of~$\dashv_x$ in~$\hat{w}$. Every valid ref-word
$\hat{w}$ defines a mapping that intuitively associates each variable $x \in
\var$ to the span of the characters of~$w$ between 
$\vdash_x$ and $\dashv_x$ in~$\hat{w}$.

Thus, extraction grammars are like annotated grammars but with variable
operations that
describe span endpoints
(whereas our annotations are arbitrary), and that are expressed as separate
variable operation characters (not annotations of existing letters).

\paragraph{Comparing both formalisms}
Given an extraction grammar $\cH$ on alphabet $\Sigma$ and with variables
$\var$, an annotated  grammar $\cG$ is \emph{equivalent} to~$\cH$ if it is over the
alphabet $\Sigma \cup \{\#\}$ 
(with a special end-of-word character~$\#$ to account for indexing differences),
if its annotation set $\Omega$
is the powerset of the set $\bigcup_{x \in \var} \{\vdash_x, \dashv_x\}$ of
markers symbols, and if the following holds:
for any $w \in \Sigma^*$, the outputs $\sem{\cG}(w \#)$ are in
one-to-one correspondence with the mappings of~$\sem{\cH}(w)$ in the expected
way (the formal definition is in Appendix~\ref{apx:formalextraction}).
We show that every extraction grammar has an equivalent annotation grammar
in this sense, and the translation further preserves unambiguity:

\begin{proposition}\label{prop:extrac-grammars-general}
  Given any extraction grammar $\cH$ with $k$ variables, we can build an
  equivalent annotated grammar $\cG$ in time $\cO(9^{3k}\cdot|\cH|^2)$. Moreover, if $\cH$ is unambiguous then so is $\cG$.
\end{proposition}

Hence, our formalism of annotated grammars captures that of extraction grammars.
Unfortunately, the translation is exponential, 
intuitively because $\Omega$ must cover all possible sets of variable operations: we
explain in Appendix~\ref{apx:spanexa} why we believe this to be unavoidable.
We note that, in exchange for this, annotated grammars
are strictly more \emph{expressive}: each output can annotate an arbitrary
number of positions in the string (e.g., every other character), unlike
extraction grammars whose mappings have a fixed number of variables.

\paragraph{Enumeration for extraction grammars}
As extraction grammars can be rewritten to annotated grammars in an unambiguity-preserving way (Proposition~\ref{prop:extrac-grammars-general}), we can derive from
Theorem~\ref{thm:cubic} an enumeration result for unambiguous extraction grammars with cubic preprocessing time in data complexity.

\begin{theorem}
  Given an unambiguous extraction grammar $\cH$ with $k$ variables and a string $s$, we can
  enumerate the mappings of~$\sem{\cH}(s)$ with preprocessing time 
  $\cO(9^{3k} \cdot |\cH|^2 \cdot |s|^3)$ (hence, cubic in~$|s|$), and with output-linear delay
  (independent from~$s$, $k$, or~$\cH$).
\end{theorem}

In data complexity, this improves over the result of~\cite{Peterfreund21}
for unambiguous extraction grammars, whose preprocessing time is
$\cO(9^{2k} \cdot |\cH|^2 \cdot |s|^5)$, i.e.,
our data complexity is cubic instead of quintic.
We leave to future work a study of enumeration results for
restricted classes of extraction grammars via %
Theorems~\ref{thm:cubic} and~\ref{thm:quadratic}.

\section{Conclusions and Future Work}

We have presented our formalism of annotated grammars and our results on the
efficient enumeration of all annotations of an input string. Our results achieve
output-linear delay, and cubic-time preprocessing if the grammar is unambiguous,
quadratic-time if it is rigid, and linear-time for profiled-deterministic PDAnns.

The main question left open by our work is that of the precise complexity of
this task, depending on the grammar formalism. For instance, can we improve the 
$\cO(n^3)$ algorithm to match the complexity of Valiant's parser? For which grammar
classes can we extend the linear-time preprocessing approach? We believe,
however, that a complete classification is out of reach, given that classifying
the fine-grained complexity of parsing is still open to a large extent even in
the case of unannotated CFGs.

\paragraph{Acknowledgements}
Amarilli was partially supported by the ANR project EQUUS
ANR-19-CE48-0019 and by the Deutsche Forschungsgemeinschaft (DFG, German
Research Foundation) – 431183758. Muñoz and Riveros were funded by ANID - Millennium Science Initiative Program - Code \verb|ICN17_002|.

\bibliographystyle{abbrv}
\bibliography{main_arxiv}

\newpage
\appendix
\section{Proofs of Section~\ref{sec:cubic}}

\subsection{Proof of Theorem~\ref{thm:enum}}
\label{apx:enum}
We give here a self-contained argument for the proof of
Theorem~\ref{thm:enum}. A more detailed presentation can be found
in~\cite{amarilli2017circuit} or~\cite{icdt2020nested}.

We will first show how to build a structure that only represents
sets that are non-empty and that cannot contain the empty string
($\singleton(\epsilon)$). Then we will explain how we can
extend this structure to support the empty set, and to support
$\singleton(\epsilon)$.

\paragraph{The DAG structure of enumerable sets}
As enumerable sets are a fully-persistent structure, we can represent
our data structure as a global directed acyclic graph storing all
enumerable sets constructed so far, to which we can only add new nodes
pointing to existing nodes. In particular, for any node that is built, its
set of descendants is immutable.
Each node $n$ will be associated to a set
$\Set(n)$. 
In the global DAG, there are three 
kinds of nodes:

\begin{itemize}
  \item \emph{Singleton nodes}, a singleton node $n$ carries a value
  $x \in \O$ and as its name suggests, $\Set(n)=\{x\}$, i.e. $n$ represents
  the set containing a single value which is the single-character string $x$.
  
  \item \emph{Product nodes}, a product node has two outgoing
    edges. Intuitively, a product node $n$ is such that $\Set(n)
    = \Set(n_1) \otimes \Set(n_2)$, where $\otimes$ denotes the
    product operation on sets defined on the main text, and $n_1$
    and $n_2$ are the two inputs.
    
    \item \emph{Union nodes}, a union node has two or three outgoing
    edges. The first outgoing edge necessarily points to a node which
    is a singleton or product node. Intuitively, a union node $n$ is
    such that $\Set(n)$ is $\cup_{c}\Set(c)$ where $c$ ranges over the nodes to
    which $n$ has an outgoing edge.
    In our construction we will make sure that
    union nodes with three children are never used directly to
    represent an enumerable sets but only used internally, in particular we
    never start the enumeration on such nodes.
\end{itemize}

\paragraph{Implementing the various construction operations}
We now present how to implement the various operations for enumerable
sets. For $\singleton$ and $\prod$ the operation is easy: for
$\singleton(x)$, we create a leaf node carrying the value
$\singleton(x)$ and return it; for $\prod(n_1,n_2)$ we
create a product node with $n_1$ and $n_2$ as outgoing edges and
return it.

The union operation is the most complicated operation by far.  Let $n_1$ and
$n_2$ be the arguments
Remember that $n_1$ and $n_2$ cannot be union nodes that have three outgoing
edges, because $n_1$ and $n_2$ are nodes that represent enumerable sets.
We implement the union operator on~$n_1$ and $n_2$ 
in the following way
\begin{itemize}
  \item If one of the arguments is a singleton or product node, then
    the result is a union node with two children, having the singleton
    or product node as first child and the other argument as second
    child.
    \item If both arguments are union nodes $n$ and $n'$, let
    $n_1$ and $n_2$ be the children of $n$ with $n_1$ a singleton or
    product node, and define likewise $n_1'$ and $n_2'$. The result is
    a union node $n''$ having as children $n_1$ and a fresh 
    union node $n'''$ having as children $n_1'$, $n_2$, and $n_2'$.
    Notice the returned node, $n''$ has only two outgoing edges satisfying the
    condition that union nodes with three children are never used directly.
\end{itemize}

One can verify that the construction satisfies the invariants, i.e.,
that $\Set(n)$ is as prescribed. This is clear for the product
operator, and for the union operator the only difficult case is the
union of two union nodes. Then, given that we had $\Set(n)
= \Set(n_1) \cup \Set(n_2)$, and $\Set(n') = \Set(n_1') \cup
\Set(n_2')$, the result is such that $\Set(n'') = \Set(n_1) \cup \Set(n_1') \cup \Set(n_2)
\cup \Set(n_2')$ which is correct. Furthermore, it is clear that we never return
a union node with three children as the result of an operator, and
that the unions that we create have the right number of children and
that their first child is always a singleton or product node.

What is more, all operators are indeed in constant-time.

\paragraph{Reduction to the enumeration of enumerable sets starting
with a product node} For the sake of simplicity, we will suppose in
the rest of this section that the node to be enumerated is always a
product node. If it is not we can easily create the product with a
singleton node \$, enumerate the strings represented by this new node
and remove from each string produced the last character (which will be
the \$).

\paragraph{Auxiliary procedure: enumerating the exits of union nodes}
Given a node $n$, we say that an \emph{exit} of $n$ is a product or a
singleton node reachable from $n$ going only through union nodes.  For
our enumeration scheme, we will need a structure capable of
enumerating the exits of a node with constant-delay between the
any two exits, such that the structure can be initialized in constant time.

The set of exits for a node $n$ can be expressed recursively: it is
either $n$ when $n$ itself is a singleton or product node or it is the
union of exits of the children of $n$ otherwise. Our enumeration thus
performs a depth first search but with a constant delay guarantee.

To do this, we will rely (as in a normal depth first search) on a
stack $S$ of nodes ``to be enumerated''. At the beginning of the
enumeration, $S$ contains the node $n$, and the enumeration ends
whenever $S$ becomes empty. When looking for the next exit, we pop one
node $v$ from $S$. If $v$ is a singleton or product node, the next
exit is $v$. If $v$ is a union node, we push back the children of $v$
on the stack while making sure that the first child of $v$ is on
top. Then, we reiterate the procedure to find the next exit.

It is clear that this scheme enumerates all the exits and only once.
To show that this procedure is indeed constant-delay, notice that, by
design, the first child of a union node is never a
union node itself.  Therefore if we pop a union node, the next pop
cannot be a union node again. Therefore once every two pops we produce
at least one exit and therefore the procedure outputs the exits
with constant delay.

 \paragraph{Enumeration state}
 In the design of our enumeration algorithm, we must explain what is the
 \emph{state} stored by the enumeration between any two outputs. This state
 is an \emph{enumeration tree}. In
 an enumeration tree, the nodes are of one of the following two
 types: \emph{concatenation} nodes (similar to the product nodes in our DAG)
 and \emph{pointer} nodes, pointing to either a product or a singleton
 node in the DAG. For each concatenation node~$n$ we will store multiple
 pieces of information: two children, two auxiliary data structures for
 exit enumeration, and 
 as a pointer to the product node that $n$ corresponds to. For
 a pointer node we will simply keep a pointer to the node in the DAG of
 enumerable sets, which is either a product node or a union node.
 
 An enumeration-tree is said to be \emph{output-ready} when no pointer nodes are
 pointing to product nodes. This means that the children
 of concatenation nodes are either concatenation nodes themselves or
 pointer nodes pointing to singleton nodes.
 
 \paragraph{Overall structure of the enumeration}
 Each step of the enumeration is decomposed into three sub-phases:
 (1.) an \emph{unfolding sub-phase} where we ``unfold'' the pointer nodes
 pointing to product nodes into concatenation nodes (resulting in an
 output-ready enumeration tree), then (2.) an \emph{output sub-phase} where we
 compute the output, and finally (3.) a \emph{pruning sub-phase} where we change
 the tree to point to the next solution by removing nodes where there
 are no more outputs. Each sub-phase will take a time proportional to the
 output currently being produced. After the pruning sub-phase we return the computed
 output. Thus, we ensure output-linear delay.
 
 At the beginning of the enumeration for a enumerable set, remembering our
 assumption that the enumerable set corresponds 
 to a product node $n$, the state of the enumeration is a pointer node
 pointing to $n$. The enumeration ends whenever $n$ has no next
 solution (we explain below how this can be computed).
 
 Note that we need to separate the pruning sub-phase of the current output
 from the next unfolding sub-phase, because the running time of the pruning sub-phase 
 depends on the size of the current output, while the running time of the
 unfolding sub-phase of
 the next output depends on the size of the next output. Thus, without running
 the pruning sub-phase with the current output, we could violate the output-linear
 delay guarantee, because two successive outputs may have very different sizes.

 \paragraph{Unfolding sub-phase}
 During the unfolding sub-phase, we explore the enumeration tree to look for
 a \emph{pointer} node. Each time we find a pointer node $p$ pointing to a
 product node $n$ with children $n_1$ and $n_2$, we replace $p$ by a new
 concatenation node $c$. The two exit-enumeration structures $S_1$ and
 $S_2$ of the new concatenation node $c$ enumerate the exits
 of the inputs $n_1$ and $n_2$ to the product node~$n$. We use these
 structures to retrieve the first exits $e_1$ and $e_2$ of $n_1$ and
 $n_2$, and the children $c_1$ and $c_2$ of $c$ are pointer nodes
 pointing to these $e_1$ and $e_2$. We recursively call the unfolding sub-phase
 into the two newly created nodes $c_1$ and $c_2$. Clearly this
 unfolding sub-phase overall takes a time proportional to the size of the final
 tree. Further, at the end of this sub-phase, the enumeration tree is
 output-ready.
 
 \paragraph{Output sub-phase}
 For the output sub-phase, we have a output buffer and we simply traverse
 our enumeration tree. Each time we encounter a pointer node, it points
 to a singleton node carrying the value $x$, and we append $x$ to our
 buffer. This sub-phase clearly takes a time proportional to the size of
 the tree. Notice that the output produced in the buffer is also of size proportional to the
 tree: this is because the number of concatenation node corresponds to the number
 of pointer nodes (all of which point to singletons), minus one.
 
 \paragraph{Pruning sub-phase}
 For the pruning sub-phase, we want to prepare for the next output. For
 this we will require to compute the Boolean information of whether a node has a
 next solution. This can computed from the output-ready tree
 recursively: for a pointer node, there is no next solution (because it
 points to a singleton), for a concatenation node there exists a next
 solution if either one of its children has a next solution or if one
 the two exit-enumeration structures contain an exit to be enumerated.
 
 To understand how we compute the next solution, note that for a
 product node $n$ of $n_1$ and $n_2$ we will compute the elements of $\Set(n)$ by
 computing first the concatenation of all strings of $\Set(n_1)$ with 
 the first string of $\Set(n_2)$, then with the second string of $\Set(n_2)$,
 etc., until we have covered all strings in~$\Set(n_2)$.
 
 Now, we can explain our recursive computation of the enumeration tree
 corresponding to the next solution: given a concatenation node $c$
 with children $c_1$ and $c_2$ and exit-enumeration structure $S_1$ and
 $S_2$ pointing to a product node $n$ with children $n_1$ and $n_2$
 there are several cases:
 \begin{itemize}
 \item When $c_1$ has a next solution, we simply recurse
       into $c_1$ to find the next solution.
 
 \item When $c_1$ has no next solution but $S_1$ has a next exit  $e_1$
       to enumerate, we replace $c_1$ with a pointer node to $e_1$.
 
 \item In the other cases, it means that we are done doing the enumeration
       of $n_1$ for the current solution of~$n_2$. We thus need to restart
       the enumeration of solutions to $n_1$ which is done by resetting
       $S_1$ to enumerate the exits of $n_1$ and replacing $c_1$ by the
       pointer node corresponding to the first exit enumerated
       by $S_1$.  We also need to move the enumeration of $n_2$ to its
       next solution. If $c_2$ has a next solution we compute it
       recursively otherwise we replace $c_2$ with the next exit in
       $S_2$.
  \item If $c_2$ has no next solution and $S_2$ has no next exit, then there is
    no next solution.
 \end{itemize}
 
 \paragraph{Memory usage of the enumeration phase}
 In Theorem~\ref{thm:enum}, we claimed that our enumeration scheme uses a memory
 linear in the size of the input string. As we already explained, the
 number of nodes that we keep in an enumeration tree is at most linear
 in the size of the current output, but we also need to prove that the
 memory used by the auxiliary structures for exit enumeration stays linear in the
 size of the DAG of enumerable sets. For this, notice that any node $n$
 of the DAG cannot be stored twice. It cannot be stored within the same
 exit enumeration data structure as it would mean that the same solution will
 be enumerated twice, which is impossible by our assumption that unions are
 disjoint. Further, it cannot be stored in two different
 exit enumeration schemes, as otherwise there would be a product node where both
 inputs contain the letters of some element in $\Set(n)$, which is again
 forbidden.

\paragraph{Adding the $\emptyset$ operation}
To extend the data structure to support the $\emptyset$ operator, we add a new node type to have a node representing the empty
set. There will be only one node of this type.
The $\emptyset$ basic operation is simply to create the $\emptyset$ node if it
does not exist and return it, otherwise return the one that exists.
We change the definition of the product operator to do the following
on arguments $n_1$ and $n_2$: if one of $n_1$ and $n_2$ is the
$\emptyset$ node, then return the $\emptyset$ node; otherwise do as
before.
We change the definition of the union operator to do the following on arguments
$n_1$ and $n_2$: if one of $n_1$ and $n_2$ is the $\emptyset$ node, then return
the other argument; otherwise do as before.
It is clear that these alternative definitions do not alter the semantics of
nodes, and they guarantee that whenever we apply the definitions from before
then the arguments are not $\emptyset$ nodes.

For the enumeration phase, we add a new base case: if the node to enumerate is
$\emptyset$, we immediately halt. Otherwise, it is clear that the~$\emptyset$
node can never be visited in the enumeration, because in our construction the
$\emptyset$ node never has an incoming edge.

\paragraph{Adding the $\singleton(\epsilon)$ operation}
Making a data structure that supports $\singleton(\epsilon)$ operator
is done by storing, for each enumerable set, a pair of an enumerable set without support for empty string
solutions (as defined above), and a Boolean indicating whether the empty string
$\epsilon$ is captured. We now present the new operations on
enumerable sets. We distinguish between the original functions and the new
functions described here by adding an epsilon index to the new functions, e.g.,
$\empt_\epsilon$.

For $\empt_\epsilon$, we return $(\emptyset, \text{false})$.

For $\singleton_\epsilon(x)$ we return the pair
$(\singleton(x),\text{false})$ if $x\neq \epsilon$ and
$(\emptyset,\text{true})$ otherwise. 

For $\union_\epsilon( (e_1,b_1),
(e_2,b_2))$ we return $(\union(e_1,e_2),b_1\lor b_2)$. 

For
$\prod_\epsilon( (e_1,b_1), (e_2,b_2))$, the Boolean that we return is $b_1
\land b_2$, and the enumerable set that we return depends on the value of
$b_1$ and $b_2$ we return $\prod(e_1,e_2)$ when $\lnot b_1\land \lnot
b_2$, $\union(\prod(e_1,e_2),e_1)$ when $b_1\land \lnot b_2$,
$\union(\prod(e_1,e_2),e_2)$ when $b_2 \land \lnot b_1$ and
$\union(\prod(e_1,e_2),\union(e_1,e_2))$ when $b_1\land b_2$. 

To run the enumeration phase on a pair $(e,b)$, we start by outputting $\epsilon$ if $b$ is
set, and then we enumerate $e$ as explained previously.

\subsection{Proof of Proposition~\ref{prp:2nf}}
\label{apx:arity2}
We show Proposition~\ref{prp:2nf}. The result is shown as
in~\cite{lange2009cnf}, though we give the proof in a self-contained fashion for
convenience. We additionally show here that the transformation
preserves rigidity (defined in Section~\ref{sec:quadratic}) as this property is
used in that section.

\paragraph{Conditions 1 and 2: removing useless nonterminals}
We first perform a linear-time exploration from the terminals to mark the
nonterminals $X$ that can derive some string of terminals. The base case is if a nonterminal
$X$ has a rule $X \rightarrow \alpha$ where $\alpha$ only consists of terminals
(in particular $\alpha = \epsilon$), then we mark it.
The induction is that whenever a nonterminal $X$ has a rule $X
\rightarrow \alpha$ where $\alpha$ only consists of terminals and of
marked nonterminals, then we mark~$X$. At the end of this process, it is clear that any nonterminal
that is not known to derive a string of terminals indeed does not derive any
string, because any derivation of a string of terminals from a nonterminal $X$
would witness that all nonterminals in this derivation, including $X$, should
have been marked, which is impossible. Hence, we can remove the nonterminals
that are not marked without changing the language or
successful derivations of the grammar, and satisfy condition 1 in linear time.

Second, we perform a linear-time exploration from the start symbol $S$ to mark the
nonterminals $X$ that can be reached in a derivation from $S$. The base case is
that $S$ is marked. The induction is that whenever a nonterminal $Y$ occurs in
the right-hand side of a rule having $X$ as its left-hand side, and $X$ is
marked, then we mark $Y$. At the end of the process, if a nonterminal $X$ is
not marked, then indeed there is no derivation from $S$ that produces a string
featuring $X$, as otherwise it would witness that $X$ is marked, which is
impossible. Hence, we can again remove the nonterminals that are not marked,
the grammar and successful derivations are again unchanged, and we
satisfy condition 2 in linear time.

As the transformations here only remove nonterminals and rules that cannot
appear in a derivation, they clearly preserve unambiguity as well as rigidity.

\paragraph{Condition 3: shape of rules}
We first ensure that every right-hand side of a rule is of size $\leq 2$.
Given the annotated grammar $\mathcal{G}$, for every rule $X \rightarrow \alpha$ where
$|\alpha| > 2$, letting $\alpha = \alpha_1\cdots\alpha_n$,
we introduce $n-2$ fresh nonterminals $X_{\alpha,1},
\ldots, X_{\alpha,n-2}$, and replace the rule by the following: $X
\rightarrow \alpha_1 X_{\alpha,1}$, $X_{\alpha,1} \rightarrow \alpha_2
X_{\alpha,2}$, ..., $X_{\alpha,n-2} \rightarrow \alpha_{n-1} \alpha_{n}$.

We make sure that the right-hand side of rules of size 2 consist only of
nonterminals by introducing fresh intermediate nonterminals whenever
necessary, which rewrite to the requisite terminal.

It is then clear that the result satisfies condition 3, and that there is
a one-to-one correspondence between derivations in the original grammar and derivations in the rewritten grammar. To see this, note that there is an obvious one-to-one function which maps derivations from the original grammar into derivations in the new grammar, and that there is a slightly more involved function which receives a derivation in the new grammar, and builds a derivation in the original grammar by following the steps detailed above (and using the fact that each fresh nonterminal is associated to exactly one rule), which is also one-to-one. We conclude that the original grammar is unambiguous if and only if the new grammar is unambiguous.

The last point to check is that the arity-2 transformation preserves rigidity,
i.e., if the original annotated grammar is rigid then so is the image of the
transformation. Let $X$ be some symbol of the original grammar $\cG$, and
$w \in \Sigma^*$ be a string. Let us show that all derivations from the
corresponding
symbol $X'$ of the rewritten grammar $\cG'$ have same shape. We do so by
induction on the length of~$w$ and then on the topological order on
nonterminals. The base case of $w$ of length $0$ is clear: the possible
derivations are sequences of applications of rules of the form $Y \rightarrow Z$
in a sequence of some fixed length, followed by a rule of the form $Y
\rightarrow \epsilon$, and what can happen in the rewritten grammar is the same.

For the inductive case, as $\cG$ is rigid, we know that there must be one
fixed profile $\pi \in \{0, 1\}^k$ such that all derivations of~$w$ from~$X$ start by the
application of a rule $X \rightarrow \alpha$ where $\alpha$ corresponds to
profile~$\pi$, i.e., it has length~$k$ and its $i$-th character is a nonterminal
or terminal according to the value of the $i$-th bit of~$\pi$.
Otherwise the existence of two different right-hand-side profiles would
contradict rigidity. Furthermore, by considering the possible sub-derivations
from~$\alpha_1$ (including the empty derivation if $\alpha_1$ is a terminal), we
know that $\alpha_1$ derives some fixed prefix of~$w$ and that all
such derivations have the same sequence of profiles; otherwise we would witness
a contradiction to rigidity. By applying the same argument successively to
$\alpha_2,
\ldots, \alpha_k$, we deduce
that there must be a partition of $w =
w_1 \cdots w_k$ such that, in all derivations of~$w$ from~$X$, the derivation
applies a rule with right-hand having profile $\pi$ to produce some string $\alpha_1 \cdots
\alpha_k$, and then each $\alpha_i$ derives an annotation of~$w_i$ and
for each $i$ all possible derivations of some annotation of~$w_i$ by some $i$-th
element in the right-hand size of such a rule has the same sequence of
profiles.

As the string is nonempty we know that $k > 0$. Further, if $k = 1$ then $X$ and
the productions involving $X$ were not rewritten so we immediately conclude
either with the case of a rule $X \rightarrow \tau$ for a terminal $\tau$ or by
induction hypothesis on the nonterminals in the topological order for the case
of a rule of the form $X \rightarrow Y$. Hence, we assume that $k \geq 2$.

We know by induction that, in the rewritten grammar, the derivation from $X$
will start by rewriting $X$ to $Y_1 X_{\alpha, 1}$, the $X_{\alpha, 1}$ being
itself rewritten to $Y_2 X_{\alpha, 2}$, and so on, for some right-hand size
$\alpha$ of a rule $X \rightarrow \alpha$ having profile $\pi$. Clearly each
$Y_i$ will have to derive an annotation of the $w_i$ in the partitioning of~$w$, as
a derivation following a different partitioning would witness a derivation in
the original grammar that contradicts rigidity.
Now, the profile
$\pi$ indicates if each $Y_i$ is a nonterminal of the initial grammar or a fresh
nonterminal introduced to rewrite to a terminal. In the latter case, there is no
possible deviation in profiles. In the former case, we conclude by induction
hypothesis that each $Y_i$ derives annotations of its~$w_i$ that all have the
same profile, and we conclude that all derivations in the rewritten grammar
indeed have the same profile, concluding the proof.

\subsection{Proof of Proposition~\ref{prp:cubiclowerbound}}
\label{apx:cubiclowerbound}

We know from~\cite{AbboudBW18} that for any $c>0$, there exists a fixed
grammar $\cG$ such that determining whether a string $w$ is
derived by $\cG$, cannot be solved in time $\cO(|w|^{\omega-c})$, unless the conjecture in graph algorithms mentioned in~\cite{AbboudBW18} is false.

We will see that this conditional lower bound translates to unambiguous
annotated grammars. Indeed, we will show that for each grammar
$\cG$ there exists an unambiguous annotated grammar
$\mathcal{G}'$ such that $w$ is derived by $\cG$ if and only
if $\sem{\mathcal{G}'}(w)$ is non-empty. Therefore after the
preprocessing of $w$ for $\mathcal{G}'$, we know in constant time
whether $w$ is derived by $\cG$ which proves that the
preprocessing of $\mathcal{G}'$ on $w$ requires $\cO(|w|^{\omega-c})$
time, assuming the conjecture is true.

Now let us show how to translate a grammar $\cG$ into an
unambiguous annotated grammar $\mathcal{G}'$. This can be challenging, because
$\cG$ is not necessarily unambiguous: for this reason we need to define
$\mathcal{G}'$ intuitively by adding annotations that disambiguate the various
possible derivations of $\cG$, to guarantee that the result is
unambiguous. As this is cumbersome to do on grammars, we use the
correspondence between annotated grammars and pushdown annotators
(Proposition~\ref{prop:grammar-pdann}), shown later in the article. 

In this proof, we will use the notion of \emph{pushdown automata} (PDA); see
Definition~\ref{def:pda} for the formal definition.
Let us consider a PDA $\mathcal{P}$ 
which is equivalent to $\cG$.
As is standard with PDAs, we can change the given definition to suppose without
loss of generality that no transition in $\mathcal{P}$ is an
$\epsilon$-transition. Specifically, we consider PDAs in a slightly different
model where transitions 
are of the form
$(q_1,a,s_1,q_2,s_2)\in Q\times \Sigma \times \Gamma^+ \times
Q \times \Gamma^+$: such a transition means that in state $q_1$, when the top
stack symbols are $s_1$ and the next letter to read is~$a$, the automaton can
read the letter, move to state
$q_2$ and replace $s_1$ by $s_2$ on the stack. We create our
unambiguous PDAnn $\mathcal{P}'$ from $\mathcal{P}$ by replacing
each transition $t=(q_1,a,s_1,q_2,s_2)$ to a set of transitions that
first pop the symbols of $s_1$ from the stack, then reads $a$, then
pushes the symbols of $s_2$ onto the stack. The first state of this
transition is $q_1$, the last state is $q_2$ but we make sure that
each of the intermediate states are unique to $t$. Furthermore, the
transition that reads the letter $a$ outputs a symbol unique to the
transition $t$. Therefore, by construction there is a bijection
between runs of $\mathcal{P}$ and runs of $\mathcal{P}'$ and the 
PDAnn $\mathcal{P}'$ is unambiguous because the run used for each
output can be retrieved from that output.

We conclude by using Proposition~\ref{prop:grammar-pdann} to obtain an equivalent annotated
grammar $\cG'$, which is also unambiguous. Thus, we know that on any unannotated
string $w$, the set $\sem{\cG'}(w)$ is empty if $\cG$ does not derive
$w$, and non-empty if it does. Thus, we know that, if we assume the conjecture is true, we cannot determine
in $\cO(|w|^{\omega-c})$ whether $\sem{\cG'}(w)$ is empty or not. But if we have
an algorithm to enumerate $\sem{\cG'}(w)$  with output-linear delay, as any
output has size $\cO(|w|)$ in~$|w|$, we can do this with a complexity linear in~$|w|$
which is that of the preprocessing of the enumeration algorithm. Thus, we
conclude that the preprocessing conditionally requires $\Omega(|w|^{\omega-c})$ time.

\section{Proofs of Section~\ref{sec:quadratic}}

\subsection{Proof of Theorem~\ref{thm:profileu-iou}}

In this appendix, we will use the notion of PDAnn introduced in
Section~\ref{sec:linear}, and we will use 
Proposition~\ref{prop:grammar-pdann}, which is also stated in
Section~\ref{sec:linear} and proved in
Appendix~\ref{apx:grammar-pdann}.

To prove Theorem~\ref{thm:profileu-iou}, we introduce a general-purpose normal form on PDAnn, where,
intuitively, the only choices that can be made during a run are between the \emph{types}
of transition to apply. This is similar to Lemma~1 in~\cite{icdt2020nested}.

\begin{definition}
  A PDAnn $\cP$ is \emph{deterministic-modulo-profile} if it satisfies the following conditions:

  \begin{enumerate}
    \item for each state $p$ there is at most one push transition that
      starts on $p$, formally $\card{\{q, \gamma \in Q \times \Gamma \mid
      (p, q, \gamma) \in \Delta\}} \leq 1$
    \item for each state $p$ and stack symbol $\gamma$ there is at most
      one pop transition that starts on $p,\gamma$, formally $\card{\{q
      \in Q \mid (p, \gamma, q) \in \Delta\}} \leq 1$
    \item for each state $p$, letter $a$, and output $\oout \in \Omega$, there is at most one read-write transition that starts on
      $p,a,\oout$, formally, we have $\card{\{q \in Q \mid (p,(a,\oout),q) \in \Delta\}}
      \leq 1$.
      \item for each state $p$ and letter $a$, there is at most one read transition that starts on
      $p,a$, formally $\card{\{q \in Q \mid (p,a,q) \in \Delta\}}
      \leq 1$.
  \end{enumerate}
\end{definition}

\begin{lemma}\label{lem:pdtonechoice}
        Let $\cP$ be a PDAnn. We can build an equivalent PDAnn $\cP'$ which is
        deterministic-modulo-profile. The transformation takes exponential time,
        i.e., time $\cO(2^{|\cP|^c})$ for some $c > 0$.

        Further, on any string~$w$, there is an accepting run of~$\cP$ on~$w$ with
        profile $\pi$ iff there is an accepting run of~$\cP'$ on~$w$ with the same profile.
\end{lemma}

\begin{proof}
  The proof is similar to the determinization of visibly pushdown
  automata~\cite[Lemma~1]{icdt2020nested}.

  Given $\cP = (Q, \Sigma, \Omega, \Gamma, \Delta, q_0, F)$,
        we build $\cP' = (Q', \Sigma, \Omega, \Gamma', \Delta', S_I, F')$ as
        follows.
	We build $Q' = 2^{Q\times Q}$, intuitively denoting a set of pairs of
        states $(p, q)$ of $\cP$ such that $\cP$ can be at state $q$ at this
        point if it was at state $p$ when the topmost stack symbol was pushed. 
        We build $\Gamma' = 2^{Q\times\Gamma\times Q}$,
        intuitively specifying the sets of possible stack symbols and
        remembering the state just after the previous stack symbol was pushed
        and the state just after that symbol was pushed.
	We build $S_I = \{(q_0,q_0)\}$, meaning that initially we are at
        the initial state~$q_0$ and were here when the stack was initialized.
        We build
        $F' = \{S\mid (q_0,q)\in S\text{ for some }q\in F \}$,
        meaning that we accept when $\cP$ reaches a final
        state and we were at the initial state when the stack was initialized.
	Let $\Delta'$ be defined as follows:
	\begin{itemize}
          \item The (unique) push transition from a state $S \in Q'$ 
                  makes $\cP'$ push a stack symbol $S'$ and move to a state $T$,
                  intuitively defined as follows. For every pair $(p, p')$
                  of~$S$ and push transition $(p', q, \gamma) \in \Delta$ in the original PDAnn,
                  we can move to state $(q, q)$ and push on the stack the symbol
                  $(p, \gamma, q)$. The stack symbol $S'$ is the set of all
                  possible stack symbols that can be pushed in this way, and
                  $T$ is the set of all possible states that can be reached in
                  this way.

                  Formally, 
                  for every $S\in Q'$ we include $(S, S', T)$ in $\Delta'$, where:
		\begin{align*}
			T &= \{(p,\gamma,q)\mid (p,p')\in S \text{ and } (p',q,\gamma)\in\Delta\text{ for some }p,p',q\in Q, \gamma\in\Gamma \},\\
			S'&= \{(q,q)\mid(p,p')\in S\text{ and }(p',q,\gamma)\in\Delta\text{ for some }p,p',q\in Q,  \gamma\in\Gamma\}
		\end{align*}
              \item The (unique) pop transition from a state $S \in Q'$ and topmost
                stack symbol $T \in \Gamma'$ makes $\cP'$ move to a state $T'$
                intuitively defined as follows. For every pair $(p', q')$ of~$S$,
                we consider all triples $(p, \gamma, p')$ of the topmost stack
                symbol $T$, and if the original PDAnn had a pop transition $(q',
                \gamma, q) \in \Delta$, then we can pop the topmost stack symbol
                and go to the state $(p, q)$. The new state $T'$ is the set of
                all pairs $(p, q)$ that can be reached in this way.

                  Formally, 
                  for every $(S, T)\in Q'\times\Gamma'$ we include $(S, T, S')$ in $\Delta'$, where:
                  {\small
		\begin{align*}
			S' &= \{(p,q)\mid(p,\gamma,p')\in T\text{ and }(p',q')\in S\text{ and }(q',\gamma,q)\in\Delta \text{ for some }p,p',q,q'\in Q,\gamma\in\Gamma\},
		\end{align*}
                }

              \item The (unique) read-write transition from a state $S \in Q'$ on a letter
                  $a \in \Sigma$ and output $\oout \in \Omega$ makes
                  $\cP'$ move to a state $S'$ intuitively defined as follows: we
                  consider all pairs $(p, p')$ in $S$ and all transitions from
                  $p'$ with $a$ and $\oout$ in~$\cP$ to some state $q$, and
                  move to all possible pairs $(p', q)$.

                  Formally, for every $(S, a, \oout)\in Q'\times\Sigma\times\Omega$ we include $(S, (a, \oout), S')$ in $\Delta'$, where:
		\begin{align*}
			S' &= \{(p,q)\mid (p,p')\in S\text{ and }	(p',(a,\oout),q)\in\Delta\text{ for some }p,p',q\in Q\}.
		\end{align*}
	 \item The (unique) read transition from a state $S \in Q'$ on a letter
	$a \in \Sigma$ makes
	$\cP'$ move to a state $S'$ intuitively defined as follows: we
	consider all pairs $(p, p')$ in $S$ and all transitions from
	$p'$ with $a$ in~$\cP$ to some state $q$, and
	move to all possible pairs $(p', q)$.
	
	Formally, for every $(S, a)\in Q'\times\Sigma$ we include $(S, a, S')$ in $\Delta'$, where:
	\begin{align*}
		S' &= \{(p,q)\mid (p,p')\in S\text{ and }	(p',a,q)\in\Delta\text{ for some }p,p',q\in Q\}.
	\end{align*}
	\end{itemize}

        It is clear by definition that $\cP'$ is deterministic-modulo-profile,
        and it is clear that the running time of the construction satisfies the
        claimed time bound.

        We now show that $\cP$ and $\cP'$ are equivalent.

        Now, for the forward
        direction, let us first assume without loss of generality that whenever
        $\cP$ makes a push transition then the stack symbol that it pushes is
        annotated with the state reached just after the push. Then we will show
        that every instantaneous description that can be reached by $\cP$ can be
        reached by $\cP'$ by induction on the run. Specifically, we show by
        induction on the length of the run $\rho$ the following claim: if $\cP$
        has a run $\rho$ on a string $w$ that produces $\mu$ from an initial state $q_0 \in T$ to an instantaneous
        description $(q, i), \alpha$, with $\alpha =
        \gamma_0 p_0, \ldots, \gamma_m p_m$ being the sequence of the stack
        symbols and states annotating them, then $\cP'$ has a run $\rho'$ on $w$
        from $S_I$ to an instantaneous description $(S, i), \alpha'$ with
        $\alpha' = T_0 \ldots T_m$ such that $T_0$ contains $(q_0, \gamma_0,
        p_0)$, $T_1$ contains $(p_0, \gamma_1, p_1)$, ..., $T_m$ contains
        $(p_{m-1}, \gamma_m, p_m)$ and $S$ contains $(p_m, q)$; further $\rho$
        and $\rho'$ have the same profile.

        The base case of an empty run on a string is immediate: if $\cP$ has an
        empty run from an initial state $q_0$, then it reaches the instantaneous
        description with $(q_0, 0)$ and the empty stack, and then $\cP'$ then
        has an empty run reaching the instantaneous description $(S, 0)$ with the
        empty stack and $S$ indeed contains $(q_0, q_0)$. 

        For the induction case, assume that $\cP$ has a non-empty run $\rho_+$ on a string $w$ that produces $\mu$.
        First, write $\rho_+$ as a run $\rho$ followed by one single transition
        of $\cP$. We know $\cP$ has a run $\rho$ on $w$ which produces $\mu$ from an initial
        state $q_0$ to an instantaneous description $(q, i), \alpha$,
        with $\alpha = \gamma_0 p_0, \ldots, \gamma_m
        p_m$. By the induction hypothesis, we know that $\cP'$ has a run
        $\rho'$ on $w$ from $(q_0, q_0)$ to an instantaneous description $(S,
        i), \alpha'$ with $\alpha' = T_0 \ldots T_m$ such that $T_0$ contains
        $(q_0, \gamma_0, p_0)$, $T_1$ contains $(p_0, \gamma_1, p_1)$, ...,
        $T_m$ contains $(p_{m-1}, \gamma, p_m)$ and $S$ contains $(p_m, q)$; and
        $\rho'$ and $\rho$ have the same profile. We
        now distinguish on the type of the transition used to extend $\rho$ to
        $\rho_+$.

        If that transition is a read-write transition $(q, (a, \oout), q')$, we consider
        the read-write transition of~$\cP'$ labeled with $(a,\oout)$ from~$T$,
        and call $S'$ the state that $\cP'$ reaches.
        As $(p_m, q) \in S$ and $(q, (a, \oout), q') \in \Delta$, we know that
        $(p_m, q') \in S'$. Thus, $\cP'$ can read $(a,\oout)$ and
        reach a suitable state $S'$ and position $i+1$ and the stacks are
        unchanged so the claim is proven.
        
        If that transition is a read transition $(q, a, q')$, we follow an analogous reasoning.

        If that transition is a push transition $(q, q', \gamma)$, the position
        of $\cP$ is unchanged and the new stack is extended by $\gamma$
        annotated with state $q'$. Consider the push transition of $\cP'$
        from~$q$, and call $S'$ the state reached and $T = T_{m+1}$ the stack symbol that is
        pushed. As $(p_m, q) \in S$ and $(q, q', \gamma) \in \Delta$, we know
        that $T$ contains $(p_m, \gamma, q')$, and $S'$ contains $(q', q')$,
        which is what we needed to show.

        If that transition is a pop transition, $(q, \gamma_m, q')$, the position
        of $\cP$ is unchanged and the topmost stack symbol is removed. Consider
        the topmost stack symbol $T_m$ and the transition of $\cP'$ that pops
        it from $S$, and call $S'$ the state that we reach. We know that $S$
        contains $(p_m, q)$ and $T_m$ contains $(p_{m-1}, \gamma_m, p_m)$ and
        $(q, \gamma_m, q') \in \Delta$, so $S'$ contains $(p_{m-1}, q')$, which
        is what we needed to show.

        Note that, in all four cases, the profile of $\rho_+$ and $\rho_+'$ is
        the same, because this was true of $\rho$ and $\rho'$, and the type of
        transition done to extend $\rho'$ to~$\rho_+'$ is the same as the type
        of transition done to extend $\rho$ to~$\rho_+$.

        The inductive claim is therefore shown, and thus if $\cP$ has a run $\rho$ on some string $w$ that produces $\mu$ starting at some initial state $q_0$ and
        ending at state $q$, then $\cP'$ has a run $\rho'$
        on $w$ which produces $\mu$ and ending at a state of the form $(q_0,
        q)$ for $q_0$ and having same profile. Thus, if $\rho$ is accepting then $q$ is final for
        $\cP$ and $(q_0, q)$ is final for $\cP'$ so $\rho'$ is accepting.
        This concludes the forward implication.

        We now show the backward implication, and show it again by induction,
        again assuming that $\cP$ annotates the symbols of its stack with the
        state reached just after pushing them. We
        show by induction on the length of a run $\rho'$ the following claim: if
        $\cP'$ has a run $\rho'$ on a string $w$ that produces $\mu$from its initial state to an instantaneous
        description $(S, i), \alpha'$ with $\alpha' =
        T_0, \ldots, T_m$ being the sequence of the stack
        symbols, then for any choice of elements $(q_0, \gamma_0, p_0) \in T_0$,
        $(p_0, \gamma_1, p_1) \in T_1$, ..., $(p_{m-1}, \gamma_m, p_m) \in T_m$
        and $(p_m, q) \in S$ it holds that $\cP$ has a run $\rho$ on $w$ producing $\mu$
        from some initial state $q_0$ to the instantaneous description $(q, i), \alpha$ with
        $\alpha = \gamma_0 q_0, \ldots, \gamma_m q_m$ (writing next to each
        stack symbol the state that annotates it), and $\rho'$ and $\rho$ have
        the same profile.

        The base case of an empty run on a string is again immediate: if $\cP'$ has an
        empty run from its initial state, then it reaches the instantaneous
        description with $(S_I, 0)$ and empty stack, and then $\cP$
        has an empty run from any initial state $q_0$ to $q_0$ so that indeed
        $S_I$ contains $(q_0, q_0)$.

        For the induction case, assume that $\cP'$ has a non-empty run $\rho'_+$ on $w$ which produces $\mu$,
        We write again $\rho'_+$ as a run $\rho'$ followed by one single transition
        of $\cP'$. We know $\cP'$ has a run $\rho'$ on $w$ which produces $\mu$ from the initial
        state $S_I$ to an instantaneous description $(S, i), \alpha'$,
        with $\alpha = T_0 \ldots T_m$.
        By the induction hypothesis, we know that for any choice of elements
        $(q_0, \gamma_0, p_0) \in T_0$,
        $(p_0, \gamma_1, p_1) \in T_1$, ..., $(p_{m-1}, \gamma_m, p_m) \in T_m$
        and $(p_m, q) \in S$, 
        then $\cP$ has a run $\rho$ on $w$ which produces $\mu$
        from some initial state $q_0$ to the instantaneous description $(q, i), \alpha$ with
        $\alpha = \gamma_0 q_0, \ldots, \gamma_m q_m$, and $\rho$ and $\rho'$
        have the same profile.
        We now distinguish on the type of transition used to extend $\rho'$
        to~$\rho'_+$. 

        If the last transition is a read-write transition $(S, (a, \oout), S')$ with $S'$
        defined as in the construction, consider any choice of 
        $(q_0, \gamma_0, p_0) \in T_0$,
        $(p_0, \gamma_1, p_1) \in T_1$, ..., $(p_{m-1}, \gamma_m, p_m) \in T_m$
        and $(p_m, q') \in S'$, and then there must be some state $p''$ such that
        $(p'', (a, \oout), q) \in \Delta$ and $(p_m, p'') \in S$. Using the
        induction hypothesis but picking $(p_m, p'') \in S$, we obtain a run
        $\rho$ of~$\cP$ on $w$ which produces $\mu$, with the correct stack and ending at
        position $i$ on state $p''$, which we can extend by the read transition
        $(p'', (a, \oout), q)$ to reach state $q$ at position $i+1$ without
        touching the stack, proving the result.
        
        If the last transition is a read transition $(S, a, S')$ with $S'$
        defined as in the construction, we follow an analogous reasoning.

        If the last transition is a push transition $(S, S', T)$ with $T$ defined as
        in the construction, consider any choice of 
        $(q_0, \gamma_0, p_0) \in T_0$,
        $(p_0, \gamma_1, p_1) \in T_1$, ..., $(p_{m-1}, \gamma_m, p_m) \in T_m$,
        $(p_m, \gamma_{m+1}, p_{m+1}) \in T_{m+1}$ and $(p_{m+1}, q') \in S'$. We
        know that we must have $q' = p_{m+1}$, 
        and that there must be
        some state $p''$ and push transition $(p'', p_{m+1}, \gamma_{m+1})$ and
        pair $(p_m, p'')$ in $S$. Using the induction hypothesis but picking
        $(p_m, p'') \in S$, we obtain a run $\rho$ of~$\cP$ on $w$ which produces $\mu$ with topmost stack symbol $\gamma_m$, ending at
        state $p''$, which we can extend with the push transition $(p'',
        p_{m+1}, \gamma_{m+1})$ to obtain the desired stack and reach state
        $p_{m+1} = q'$, proving the result.

        If the last transition is a pop transition $(S, T, S')$ with $S'$ defined
        as in the construction, consider any choice of 
        $(q_0, \gamma_0, p_0) \in T_0$,
        $(p_0, \gamma_1, p_1) \in T_1$, ..., $(p_{m-2}, \gamma_{m-1}, p_{m-1})
        \in T_{m-1}$, and $(p_{m-1}, q) \in S'$. We know that there is a pair
        $(p', q') \in S$ and a triple $(p_{m-1}, \gamma_m, p')$ in $T_m$ and a
        pop transition $(q', \gamma_m, q)$ in $\Delta$. Applying the induction
        hypothesis, we get a run $\rho$ of $\cP$ on $w$ which produces $\mu$ and with topmost stack symbol $\gamma_m$ annotated with state
        $p'$ and ending at state $q'$. The pop transition $(q', \gamma_m, q)$
        allows us to extend this run to reach state~$q$ and remove the topmost
        stack symbol, while the rest of the stack is correct, proving the
        result.

        Again, we have ensured that $\rho$ is extended to $\rho_+$ with the same
        transition as the transition used to extend $\rho'$ to~$\rho_+'$,
        ensuring that $\rho_+$ and $\rho_+'$ have same profile.
        This concludes the proof of the backward induction, ensuring that
        if $\cP'$ has a run from $S_I$ to some final state $S$ reading a string $w$ and producing $\mu$, and having $(q_0, q_{\mathrm{f}})$
        with $q_{\mathrm{f}} \in
        F$ in $S$, then $\cP$ has a run reading $w$ which produces $\mu$ going from $q_0$ to the final state $q_{\mathrm{f}}$. This concludes the
        backward implication and completes the proof.
\end{proof}

We can now show Theorem~\ref{thm:profileu-iou} via
Proposition~\ref{prop:grammar-pdann}, using also the notion of \emph{profiled
PDAnn} defined in Section~\ref{sec:linear}:

\begin{proof}[Proof of Theorem~\ref{thm:profileu-iou}]
  Let $\cG$ be a rigid annotated grammar. Using Proposition~\ref{prop:grammar-pdann}, we
  transform it in polynomial time to a profiled PDAnn $\cP$.
  Using Lemma~\ref{lem:pdtonechoice}, we
  build in exponential time an equivalent PDAnn $\cP'$ satisfying the conditions of the lemma.

  We know that $\cP'$ is still profiled. Indeed, if we assume by
  contradiction that there is a string~$w$ on which $\cP'$ has two accepting runs
  with different profiles, then by the last condition of
  Lemma~\ref{lem:pdtonechoice}, the same is true of~$\cP$, contradicting the
  fact that $\cP$ is profiled.

  Now, we claim that $\cP'$ is necessarily also unambiguous. To see why,
  consider two accepting runs~$\rho$ and~$\rho'$ of~$\cP'$ on some string $w$. Since
  $\cP'$ is profiled, $\rho$ and $\rho'$ must have the same
  profile. But now, the conditions of Lemma~\ref{lem:pdtonechoice} ensure that,
  knowing the input string $w$ and profile, the runs $\rho$ and $\rho'$ are completely
  determined. Specifically, this is an immediate
  induction on the run. The base case is that there is only one initial state,
  so both $\rho$ and $\rho'$ must have the same initial state. Now, assuming by
  induction that the runs so far are identical and have the same stack, there
  are three cases. First, if the profile tells us that both runs make a push
  transition, the symbol pushed and state reached are determined by the last
  states of the runs so
  far, which are identical by inductive hypothesis. Second, if the profile tells us that both runs make
  a read-write transition (or read transition), the state reached is determined by the input and output
  symbols (or just the input symbol), and by the last states of the run so far, which are identical by
  inductive hypothesis. Third, if the profile tells
  us that both runs make a pop transition, the state reaches is determined by
  the last state of the run so far, and the topmost symbol of the stack, which
  are identical by inductive hypothesis. This concludes the inductive proof.

  Thus, for any two accepting runs $\rho$ and $\rho'$ on the string~$w$ which produce the same output, they must identical. Thus, $\cP'$ is unambiguous.
  We use Theorem~\ref{prop:grammar-pdann} to transform $\cP'$ back into an
  annotated grammar, which is still rigid and unambiguous, and equivalent to the
  original rigid annotated grammar $\cG$. The overall complexity of the
  transformation is in $\cO((2^{(|\cG|^c)^{c'}})^{c''})$ for some $c, c', c'' > 0$,
  so it is in $\cO(2^{|\cG|^d})$ for some $d>0$ overall, and the time complexity
  is as stated.
\end{proof}

\subsection{Proof of Proposition~\ref{prp:baseunambig}}

This proof is based on extending the definitions of unambiguity and rigidness of annotated grammars over unannotated context-free grammars. Indeed, an unambiguous annotated grammar with an empty output set is just an unambiguous CFG, and a rigid annotated grammar with an empty output set is a CFG for which every derivation of a given string $w\in \Sigma^*$ has the same shape.

  Consider the (unannotated) grammar $\cG'$ obtained from~$\cG$ by removing all
  annotations on terminals, and making $\Omega = \emptyset$. It can be seen that $L(\cG') = L'$ since for each string $w$, if $w\in L(\cG')$, then there is at least one $\hat{w}\in L(\cG)$ with $\str(\hat{w}) = w$ and vice versa. Now, we claim that
  $\cG'$ is \emph{rigid}, by extending the notion onto CFGs in the obvious way. To see this, consider a string $w \in L(\cG')$; all
  derivations of~$w$ by~$\cG'$ correspond to derivations by $\cG$ of some $\hat{w}$ such that $\str(\hat{w}) =w $. Because $\cG$ is rigid, all these derivations have the
  same shape. Now, using Theorem~\ref{thm:profileu-iou}, we can compute a rigid and
  unambiguous grammar $\cG''$ recognizing the same language over~$\Sigma^*$
  as~$\cG'$, i.e., $L'$. But as $L'$ is a language without output, the
  unambiguity of~$\cG''$ actually means that $\cG''$ is an unambiguous CFG. Hence, $L'$
  is recognized by an unambiguous grammar, concluding the proof.

\subsection{Proof of Proposition~\ref{prp:unambundec2}: Undecidability results
on rigid grammars}

We first show the undecidability of checking if an annotated grammar has an
equivalent rigid annotated grammar:

\begin{proposition}
  \label{prp:unambundec}
  Consider the problem, given an annotated grammar~$\cG$, of determining whether
  there exists some equivalent rigid annotated grammar equivalent to~$\cG$.
  This problem is undecidable.
\end{proposition}

\begin{proof}
  We reduce from the problem of deciding whether the language $L_2$ of an input
  (unannotated) context-free grammar $\cG_2$ can be recognized by an unambiguous
  context-free grammar: this task is known to be
  undecidable~\cite{ginsburg1966ambiguity}. Consider $\cG_2$ as an annotated
  grammar (with empty annotations). Let us show that $L_2$ can be recognized by
  a rigid annotated grammar iff it can be recognized by an
  unambiguous context-free grammar, which concludes. For the forward direction,
  if $L_2$ can be recognized
  by an unambiguous context-free grammar, then that grammar is in particular
  rigid. 
  For the backward direction, if $L_2$ can be recognized by a rigid
  grammar, then
  Proposition~\ref{prp:baseunambig} implies that $L_2$ can also be recognized by
  an unambiguous context-free grammar.
  Thus, we have showed that the (trivial) reduction is correct.
\end{proof}

We next show that it is undecidable to check if an input annotated grammar is
rigid:

\begin{proposition}
  \label{prp:unambundecb}
  Consider the problem, given an annotated grammar~$\cG$, of determining whether
  it is rigid. This problem is undecidable.
\end{proposition}

\begin{proof}
  We adapt the standard proof of undecidability~\cite[Ambiguity Theorem
  2]{chomsky1959algebraic} for the problem of deciding, given an input
  unannotated grammar~$\cG$, if it is unambiguous. The reduction is from the Post
  Correspondence Problem (PCP), which is undecidable: we are given as input sequences $\alpha_1, \ldots, \alpha_n$
  and $\beta_1, \ldots, \beta_n$ of strings over some alphabet~$\Sigma$, and we ask whether
  there is a non-empty sequence of indices $i_1, \ldots, i_m$ of integers in $[1, n]$ such that
  $\alpha_{i_1} \ldots \alpha_{i_m} = \beta_{i_1} \ldots \beta_{i_m}$. Given the
  input sequences $\alpha_1, \ldots, \alpha_n$ and $\beta_1, \ldots, \beta_n$ to
  the PCP, we
  consider the alphabet $\Sigma' = \Sigma \cup \{1, \ldots, n\}$, and we
  consider the CFG having nonterminals $S$, $S_1$, and $S_2$, start symbol $S$,
  and rules $S \rightarrow S_1$, $S \rightarrow S_2$, $S_1 \rightarrow
  \epsilon$, $S_2\rightarrow \epsilon$, and for each $1 \leq
  i \leq n$ the productions $S_1 \rightarrow \alpha_i S_1 i$ and $S_2
  \rightarrow \beta_i S_2 i$.

  We claim that this grammar is ambiguous iff there is a solution to the Post
  correspondence problem. Indeed, given any solution $\alpha_{i_1} \cdots
  \alpha_{i_m} = \beta_{i_1} \ldots \beta_{i_m}$, considering the string 
$\alpha_{i_1} \cdots
  \alpha_{i_m} i_m \cdots i_1 = \beta_{i_1} \ldots \beta_{i_m} i_m \cdots i_1$,
  we can parse it with one derivation  featuring $S_1$ and one derivation
  featuring $S_2$. Conversely, if we can parse a string $w \in \Sigma^*$ with two
  different derivations, we know that there cannot be two different derivations
  featuring $S_1$. Indeed, reading the string from right to left uniquely
  identifies the possible derivations from $S_1$. The same argument applies to
  derivations featuring $S_2$. Hence, if the grammar is ambiguous, then there is
  exactly one derivation featuring $S_1$ and exactly one derivation featuring
  $S_2$. These two derivations can be used to find a solution to the Post
  correspondence problem.

  We now adapt this proof to show the undecidability of rigidity. We
  say that an input to the PCP is \emph{trivial} if there is $i$ such that
  $\alpha_i = \beta_i$. We can clearly decide in linear time, given the input to
  the PCP, if it is trivial. Hence, the PCP is also undecidable in the case
  where the PCP is non-trivial. Now, when doing the reduction above on a PCP
  instance that is not trivial, we observe that two derivations of the same string
  can never have the same sequences of shapes. Indeed, if we have two derivations of the same
  string, then as we explained one must feature $S_1$ and the other must feature
  $S_2$, and they give a solution $\alpha_{i_1} \cdots \alpha_{i_m} =
  \beta_{i_1} \ldots \beta_{i_m}$ to the PCP. Assume by contradiction that both
  derivations have the same sequences of shapes. Then, 
  it means that we have $\card{\alpha_{i_j}} = \card{\beta_{i_j}}$ for every $1
  \leq j \leq m$. In particular we have $\card{\alpha_{i_1}} =
  \card{\beta_{i_1}}$, and so we know that $\alpha_{i_1} =
  \beta_{i_1}$ and the PCP instance was trivial, a contradiction.

  Hence, let us reduce from the PCP on non-trivial instances to the problem of
  deciding whether an input annotated grammar is not rigid. Given a
  non-trivial PCP instance, we construct $\cG$ as above, but seeing it as an
  annotated grammar with no outputs. Then $\cG$ is not rigid iff there is a
  string $w$ such that the empty annotation of $w$ has two derivations that do not
  have the same sequence of shapes. But this is equivalent to $\cG$ being
  unambiguous when seen as a CFG. Indeed, for the forward direction, if $\cG$ has
  two such derivations on a string~$w$ then clearly $w$ witnesses that $\cG$ is
  ambiguous when seen as a CFG. Conversely, if $\cG$ is ambiguous when seen as a
  CFG, we have explained in the previous paragraph that the two derivations must
  have different sequences of shapes, so $\cG$ is not rigid. Hence, we
  conclude that there is a solution to the input non-trivial PCP instance iff
  $\cG$ is not rigid. This establishes that the problem is undecidable
  and concludes the proof.
\end{proof}

\subsection{Proof of Proposition~\ref{prp:redunann}}

Assume we have such an algorithm $\cA$. Consider a procedure which receives an unambiguous unannotated CFG $\cG$ and an input string $w$, converts $\cG$ into an annotated grammar $\cG'$ with empty output set. Since $\cG'$ is unambiguous and rigid, we can run $\cA$ over $\cG'$ and $w$. If $w\in L(\cG)$, then $\sem{\cG'}(w) = \{\eps\}$, and if $w\not\in L(\cG)$, then $\sem{\cG'}(w) = \emptyset$. Thus, after the preprocessing phase of $\cA$ we need only to wait a constant amount of time to see if the string $\eps$ is given as output, or none is. We conclude that this procedure solves the problem with the same complexity as the preprocessing phase of $\cA$.

\section{Proofs of Section~\ref{sec:linear}}
\label{apx:linear}

In this appendix, we will need to use the standard notion of a \emph{pushdown
automaton} (PDA), whose definition was omitted from the main text of the paper.
We give it here:

\begin{definition}
  \label{def:pda}
  A \emph{pushdown automaton} (PDA) is a tuple $\cA = (Q, \Sigma, \Gamma, \Delta,
  q_0, F)$, where $Q$ is a finite set of \emph{states}, $\Sigma$ is the
  alphabet, $\Gamma$ is a finite alphabet of \emph{stack symbols}, $q_0 \in Q$
  is the \emph{initial state}, $F \subseteq Q$ are the \emph{final states}. We
  assume that $\Gamma$ is disjoint from~$\Sigma$. Further, $\Delta$ is a finite
  set of \emph{transitions} of the following kind:
  \begin{itemize}
    \item \emph{Read transitions} of the form $(p, a, q) \in Q \times \Sigma
      \times Q$, meaning that the automaton can go from state $p$ to
      state~$q$ while reading the letter~$a$;
    \item \emph{Push transitions} of the form $(p, q, \gamma) \in Q \times Q
      \times \Gamma$, meaning that the automaton can go from state $p$ to
      state~$q$ while pushing the symbol $\gamma$ on the stack;
    \item \emph{Pop transitions} of the form $(p, \gamma, q) \in Q \times \Gamma
      \times Q$, meaning that, if the topmost symbol of the stack is~$\gamma$,
      the automaton can go from~$p$ to~$q$ while removing this topmost
      symbol~$\gamma$.
  \end{itemize}
\end{definition}

We omit the definition of the semantics of PDAs, which are standard, and allow
us to define the \emph{language} $L(\cA)$ accepted by a PDA. It is also well-known
that CFGs and PDAs have the same expressive power, i.e., given a CFG $G$, we can
build in polynomial time a PDA $\cA$ which is \emph{equivalent} in the sense that
$L(G) = L(\cA)$, and vice-versa.

We will also need to use the standard notion of a \emph{deterministic} PDA (with
acceptance by final state). Formally:
\begin{definition}
  \label{def:dpda}
  Let $\cA = (Q, \Sigma, \Gamma, \Delta, q_0, F)$ be a PDA. For $p \in Q$, we define the \emph{next-transitions} of $p$ as the set $\Delta(p)$ of all transitions in $\Delta$ that start on $p$, i.e.,  $\Delta(p) = \{(p, x, y) \mid (p, x, y) \in \Delta\}$.
  We say that a PDA $\cA$ is \emph{deterministic} if
for every state $q \in Q$, one of the following conditions hold:
\begin{itemize}
	\item[(a)] $\Delta(q) \subseteq Q \times \Sigma \times Q$ and $|\{q'\mid
          (q, a, q')\in \Delta\}| \leq 1$ for each $a\in\Sigma$. Informally, all
          applicable transitions are read transitions, and there is at most one such
          applicable transition for each letter.
	\item[(b)] $\Delta(q) \subseteq Q \times (Q \times \Gamma)$, and
          $|\Delta(q)| \leq 1$. Informally, all applicable transitions are push
          transitions, and there is at most one such transition from $q$.
	\item[(c)] $\Delta(q) \subseteq (Q \times \Gamma) \times Q$ and
          $|\{q'\mid (q, \gamma, q')\}| \leq 1$ for each $\gamma \in \Gamma$.
          Informally, all applicable
          transitions from~$q$ are pop transitions, and there is at most one
          such applicable transition for each stack symbol.
\end{itemize}
\end{definition}

It is clear that the definition ensures that, on every input string~$w$, a
deterministic PDA $\cA$ has at most one run accepting~$w$, so that we can
check in linear time in $\cA$ and $w$ if $w \in L(\cA)$. Further, it is known that
deterministic PDAs are strictly less expressive than general PDAs.

\subsection{Proof of Proposition~\ref{prop:grammar-pdann}}
\label{apx:grammar-pdann}

Let us first give the formal definitions needed for the statement of the result.
We say that two annotated grammars
$\cG$ and $\cG'$ are \emph{equivalent} if they define the same functions, i.e.,
$\sem{\cG} = \sem{\cG'}$. We define equivalence in the same way for two PDAnns, or
for an annotated grammar and a PDAnn.

We first show one direction:

\begin{proposition}
	\label{prp:g2pdt}
	For any annotated grammar $\cG$, we can build an equivalent PDAnn $\cP$ in polynomial
	time. Further, if $\cG$ is unambiguous then so is $\cP$. Moreover, if $\cG$ is rigid, then $\cP$ is profiled.
\end{proposition}

\begin{proof}
	This is a standard transformation. Let $\cG = (V, \Sigma, \Omega, P, S)$. We build a PDAnn $\cP = (Q, \Sigma, \Omega, \Gamma, \Delta, q_0, F)$ as follows: For every rule $X \rightarrow \alpha$ of
	$\cG$ and position $0 \leq i \leq \card{\alpha}$, the PDAnn $\cP$ has a state $(X,
	\alpha, i)$ in $Q$, plus a special state $q_0$ which is the only initial and only
	final state. Also, $\Gamma = Q$. The set $\Delta$ has
	the following transitions:
	\begin{itemize}
		\item A push transition $(q_0, (S, \alpha, 0), q_0)$ and a pop transition $((S, \alpha, \card{\alpha}), q_0, q_0)$, for
		every rule $S \rightarrow \alpha$.
		\item For each state $(X, \alpha, \card{\alpha})$ for a production $X
		\rightarrow \alpha$, a pop transition reading a state from the stack and
		moving to that state.
		\item For each state $(X, \alpha, i)$ where the $(i+1)$-th element of $\alpha$
		(numbered from~$1$) is a nonterminal $Y$, for every rule $Y \rightarrow \beta$,
		a pop transition pushing $(X, \alpha, i+1)$, and moving to $(Y, \beta, 0)$.
		\item For each state $(X, \alpha, i)$ where the $(i+1)$-th element of $\alpha$
		(numbered from~$1$) is a terminal $\tau$, a
		read transition moving to $(X, \alpha, i+1)$ reading the symbol of $\tau$
		and outputting the annotation of~$\tau$ (if any).
	\end{itemize}

We will show a function that maps a given leftmost derivation $S\der{\cG} \gamma_1\der{\cG}\ldots\der{\cG}\gamma_m = \hat{w}$ into a run in $\cP$. To do this, we convert is sequence of productions into a sequence of strings which has the same size as the run (minus one). These strings serve as an intermediate representation of both the derivation and the run. The process is essentially to simulate the run in $\cP$.
\begin{itemize}
	\item First, we reduce the derivation into a sequence of productions $X_1\der{}\gamma_1, X_2\der{}\gamma_2,\ldots, X_m\der{}\gamma_m$ which uniquely defines the derivation. 
	\item The alphabet in which we represent strings that produce other strings include two special markers $\downarrow$ and $\uparrow$.
	\item We start on the string $\downarrow S\uparrow$.
	\item If the current string is $\hat{u} \downarrow X \beta$, and it is the $i$-th one that has reached a string of this form, then it must hold that $X = X_i$. We follow it by $\hat{u} \downarrow \gamma_i\uparrow\beta$.
	\item If the current string is $\hat{u}\downarrow \tau\beta$, for some terminal $\tau$, we follow it by $\hat{u} \tau \downarrow \beta$.
	\item If the current string is $\hat{u}\downarrow\uparrow\beta$, then we follow it by $\hat{u}\downarrow\beta$.
	\item If the current string is $\hat{u}\downarrow$, there is no follow up.
\end{itemize}
Interestingly, this function is completely reversible, since to obtain a sequence of productions from a sequence of strings in this model, all we need to do is to remove the markers $\downarrow$ and $\uparrow$ and eliminate the duplicate strings that appear. We will borrow the name $\splain$ to talk about the function which receives a string and returns one which deletes all markers. It is obvious that the resulting derivation is the original one. 

Furthermore, and more interestingly, we can extend the function $\shape$ to receive one of these strings and return a string in the alphabet $\{0, 1, \downarrow, \uparrow\}$. For two derivations that have the same shape, the resulting sequences have the same shape as well.

This sequence of strings represents a run in $\cP$ almost verbatim, and we only need to adapt it into a sequence of pushes, pops and reads: We make a run $\rho$ which starts on $q_0$, pushes $(X, \gamma_1, 0)$ to the stack, and moves to the state $(X, \gamma_1, 0)$. This pairs exactly to the strings $\downarrow S$ and $
\downarrow \gamma_1\uparrow$, which are the first two in the sequence. Then, we read the sequence of strings in order. If the current string is $\hat{u}\downarrow X\beta$, and this is the $i$-th time a string of this form is seen, then the current state must be $(Y, \alpha_1 X_i \alpha_2, k)$, where $|\alpha_1| = k$; we push $(Y, \alpha_1 X_i \alpha_2, k+1)$ onto the stack, and move to the state $(X_i, \gamma_i, 0)$. If the current string is $\hat{u}\downarrow a \beta$ for some $a\in \Sigma$, and the current state is $(X, \gamma, k)$, we read $\tau$, and move to the state $(X, \gamma, k)$. If the current string is $\hat{u}\downarrow (a, \oout) \beta$ for some $a\in \Sigma$, and the current state is $(X, \gamma, k)$, we read $a$, output $\oout$, and move to the state $(X, \gamma, k)$. If the current string is $\hat{u}\downarrow\uparrow\beta$, we pop the topmost state from the stack 
  and we move into that state. It is straightforward to see that this run represents exactly the leftmost derivation $S\der{\cG}^* \hat{w}$, and that for each annotated string $\hat{w}\in L(\cG)$ if and only if there is a run of $\cP$ over $w$ that produces $\mu = \ann(\hat{w})$ as output.

This function is also reversible. Consider a run of $\cP$ over a string $w$ which produces $\mu$ as output. This run must start on $q_0$, and then push $q_0$ and move onto a state $(S, \alpha, 0)$ for some rule $S \to \alpha$. Thus, our first two strings in the sequence are $\downarrow S\uparrow$ and $\downarrow \alpha\uparrow\uparrow$. If the current state is $(X, \alpha, k)$ and the next transition is to push $(X, \alpha, k+1)$ onto the stack to move into the state $(Y, \gamma, 0)$, then the current string is of the form $\hat{u}\downarrow X \beta$, so we follow it by the string $\hat{u}\downarrow \gamma\uparrow\beta$. If the next transition is a pop, then the current string is $\hat{u}\downarrow\uparrow\beta$, so we follow it by $\hat{u}\downarrow\beta$. If the current transition is a read, then the current string is $\hat{u}\downarrow a\beta$ for $a\in\Sigma$, so we follow it by $\hat{u}\tau\downarrow\beta$. If the current transition is a read-write, then the current string is $\hat{u}\downarrow (a, \oout)\beta$ for $(a, \oout)\in \Sigma\times\Omega$, so we follow it by $\hat{u}(a, \oout)\downarrow\beta$. It can easily be seen that using the original function over this resulting sequence would give the original sequence back. We point that these two reversible functions mean that there is a one to one correspondence between derivations of $S\der{\cG}^* \hat{w}$ and accepting runs of $\cP$ over $w$ with output $\mu = \ann(\hat{w})$.

Similarly to the observation we made before, we notice that if we start on a sequence in the intermediate model, the profile of the resulting run $\rho$ is fully given by the shape of the sequence (at each step, the size of the stack will be equal to the number of markers $\uparrow$ present in the string).

Now assume that $\cG$ is unambiguous. Seeing that $\cP$ is unambiguous as well comes straightforwardly from the fact that the functions presented above are bijective.

Assume now that $\cG$ is rigid. Let $w$ be an unannotated string and consider two runs $\rho_1$ and $\rho_2$ of $\cP$ over $w$ which output $\mu_1$ and $\mu_2$ respectively. Convert these two runs into sequences $\cS_1$ and $\cS_2$ in the intermediate model. Note that if we convert these two sequences into derivations $S\der{\cG}^*\mu_i(w)$, they will have the same shape. We can apply the functions above to obtain the two runs $\rho_1$ and $\rho_2$ back, and note that they have the same profile. We conclude that if $\cG$ is rigid, then $\cP$ is profiled.
\end{proof}

We then show another direction:

\begin{proposition}
	\label{prp:pdt2g}
	For any PDAnn $\cP$, we can build an equivalent annotated grammar $\cG$ in polynomial
	time. Further, if $\cP$ is unambiguous then so is $\cG$. Moreover, if $\cP$ is profiled, then $\cG$ is rigid.
\end{proposition}

\begin{proof}
	This is again a standard transformation. We first transform the input PDAnn $\cP = (Q, \Sigma, \Omega, \Gamma, \Delta, q_0, F)$ to
	accept by empty stack, i.e., to accept iff the stack is empty. To do this, we build an equivalent PDAnn $\cP' = (Q', \Sigma, \Omega, \Gamma', \Delta', q_0', F')$ where $Q' = Q\cup\{q_0', q_e, q_f\}$, $\Gamma' = \Gamma \cup \{\gamma_0\}$, $F = \{q_f\}$, and we add the following transitions to $\Delta$ to obtain $\Delta'$: A push transition $(q_0', q_0, \gamma_0)$, a pop transition $(q, \gamma_0, q_f)$ for every $q\in F$ (for runs that accept at a point where the stack is already
	empty), plus a pop transition $(q, \gamma, q_e)$ for any other $\gamma\in\Gamma$, and a pop transition $(q_e, q_0, q_f)$.
	
	This clearly ensures that there is a bijection between the accepting runs of
	$\cP$ and those of $\cP'$: given an accepting run~$\rho$ of
	$\cP$, the bijection maps it to an accepting run of $\cP'$ by
	extending it with a push transition at the beginning, and pop transitions at
	the end. Further, all accepting runs in $\cP'$ now finish with an empty
	stack, more specifically a run is accepting iff it finishes with an empty
	stack.
		
	Now, we can perform the transformation. The nonterminals of the grammar are
	triples of the form $(q, \gamma, q')$ for states $q$ and $q'$ and a stack symbol $\gamma$. Intuitively, $(p, \gamma, q')$ will derive the strings that can be
	read by the PDAnn starting from state $p$, reaching some other state $q$ with the same stack, not seeing the stack at all in the process, and then popping $\gamma$ to reach $q'$.
	
	The production rules are the following:
	
	\begin{itemize}
		\item A rule $S \to (q_0, \gamma_0, q_f)$.
		\item A rule $(p, \gamma, q') \rightarrow (q, \gamma', r)
		(r, \gamma, q')$ for every nonterminal $(p, \gamma, q')$, push transition $(p, q, \gamma')\in\Delta$ and state
		$r$.
		\item A rule $(p, \gamma, q)\to \eps$ for each pop transition $(p, \gamma, q)\in \Delta$.
		\item A rule $(p, \gamma, q')\to (a, \oout) (q, \gamma, q')$ for each read-write transition $(p, (a, \oout), q)$, and a rule $(p, \gamma, q')\to a (q, \gamma, q')$ for each read transition $(p, a, q)$, for each nonterminal $(p, \gamma, q')$.
	\end{itemize}

As we did in Proposition~\ref{prp:g2pdt}, we will show a function which receives an accepting run $\rho$ over $w$ in $\cP$ with output $\mu$ and outputs a leftmost derivation $S =\alpha_1\der{\cG}\alpha_2\der{\cG}\ldots\der{\cG}\alpha_m = \mu(w)$. The way we do this is quite straightforward: There is a one-to-one correspondence between snapshots in the run to each $\alpha_i$. Indeed, it can be seen that $\alpha_i = \hat{u}(q_1, \gamma_1, q_2)(q_2, \gamma_2, q_3)\ldots(q_k, \gamma_k, q_f)$ for some string $\hat{u}\in (\Sigma\cup(\Sigma\times\Omega))^*$, some states $q_1,\ldots,q_k$ and stack symbols $\gamma_1, \ldots, \gamma_k$. Moreover, the $i$-th stack in the run is equal to $\gamma_1\gamma_2\ldots\gamma_k$, whereas each state $q_j$ is the first state that is reached after popping the respective $\gamma_{j-1}$. We see that this function is fully reversible, as each production corresponds unequivocally to a transition in particular. This implies that $\cP$ is unambiguous if, and only if $\cG$ is unambiguous.

For the next part of the proof, we bring attention to the fact that there are exactly four possible shapes on the right sides of the rules in $\cG$. Each of these directly map to some type of transition, be it the initial push transition $(q_0', q_0, \gamma_0)$, a different push transition, a pop transition, or a read (or read-write) transition. To be precise, these shapes are the strings $1$, $11$, $\eps$ and $01$ respectively. From here it can be easily seen that, while comparing a run $\rho$ to is respective derivation $S\der{\cG}^*\hat{w}$, each production in the run immediately tells which type of transition was taken, and each transition in the run immediately tells which rule (and therefore, rule shape) was used. Therefore, each derivation shape maps to exactly one stack profile and vice versa, from which we conclude that $\cP$ is profiled if, and only if, $\cG$ is rigid.
\end{proof}

\subsection{Proof of Lemma~\ref{lem:vpaconv}}

We will show a linear-time reduction to enumeration for an I/O-deterministic VPT~\cite{icdt2020nested}.
We will state the necessary preliminaries to use the result presented there. We will also adapt the models slightly to fit our results better while also keeping the results trivially equivalent.

A structured alphabet is a triple $(\opS, \clS, \noS)$ consisting of three disjoint sets $\opS$, $\clS$, and $\noS$ that contain {\it open}, {\it close}, and {\it neutral} symbols respectively.
The set of {\em well-nested strings}
over $\Sigma$, denoted as $\wnS$, is defined as the closure of the following rules: 
$\noS \cup \{\eps\} \subseteq \wnS$,
if $w_1, w_2 \in \wnS\setminus\{\eps\}$ then $w_1 \cdot w_2 \in \wnS$, and if $w \in \wnS$ and $a \in \opS$ and $b \in \clS$ then $a\cdot w\cdot b \in \wnS$. 

A \emph{Visibly Pushdown Transducer} (VPT) is a tuple $\cT = (Q, \hat{\Sigma}, \Gamma, \Omega, \Delta, I, F)$ where $Q$ is a state set, $\hat{\Sigma}$ is a structured alphabet, $\Gamma$ is set of a stack symbols, $\Omega$ is the output alphabet, $I$ is the set of initial states, $F$ is the set of final states, and 
$
\Delta \subseteq  
(Q \times (\opS \cup \opS \times \Omega) \times Q \times \Gamma)  \cup 
(Q \times (\clS \cup \clS \times \Omega) \times \Gamma \times Q)  \cup 
(Q \times (\noS \cup \noS \times \Omega) \times Q)
$
is the transition relation.
A run $\rho$ of $\cT$ over a well-nested string $w = a_1 a_2\cdots a_n \in\wnSann$ %
is a sequence of the form
$
\rho = (q_1, \sigma_1) \xrightarrow{s_1} \ldots  \xrightarrow{s_n} (q_{n+1}, \sigma_{n+1})
$
where $q_i \in Q$, $\sigma_i\in \Gamma^{*}$, $q_1 \in I$, $\sigma_1 = \eps$, each $s_i$ is either equal to $a_i$, or to $a_i/\!\oout_i$ for some $\oout_i\in\Omega$, and for every $i\in[1,n]$ the following holds:
\begin{enumerate}
  \item If $s_{i} = a\in \opS$, then $(q_i, a,q_{i+1},\gamma) \in \Delta$, and if $s_i = a/\!\oout$, for $a\in \opS$, then $(q_i, (a,\oout),q_{i+1},\gamma) \in \Delta$, for some $\gamma\in\Gamma$ with $\sigma_{i+1} = \gamma\sigma_i$,
  \item If $s_{i} = a\in\clS$, then $(q_i, a, \gamma, q_{i+1}) \in \Delta$, and if $s_i = a/\!\oout$, for $a\in \clS$, then $(q_i, (a, \oout), \gamma, q_{i+1}) \in \Delta$, for some $\gamma\in\Gamma$ with $\sigma_i = \gamma\sigma_{i+1}$, and
  \item If $s_i = a\in\noS$, then $(p_i, a,q_{i+1})\in \Delta$, and if $s_i = a/\!\oout$ for $a\in\noS$, then $(p_i, (a,\oout),q_{i+1})\in \Delta$ with $\sigma_i = \sigma_{i+1}$. We say that the run is accepting if $q_{n+1}\in F$.
\end{enumerate}
Given a VPT $\cT$ and a $w \in\wnS$, we define the set $\sem{\cT}(w)$ of all outputs of $\cT$ over $w$ as:
$
\sem{\cT}(w) \ = \ \{ \ann(\rho) \, \mid \, \text{$\rho$ is an accepting run of $\cT$ over $w$}\}.
$
where $\ann$ for runs in VPT is defined analogously to PDAnn. That is, if $\rho = (q_1, \sigma_1) \xrightarrow{s_1} \ldots  \xrightarrow{s_n} (q_{n+1}, \sigma_{n+1})$, then $\ann(\rho) = \omega_1\ldots\omega_n$ where $\omega_i = (\oout, i)$ if $s_i = a/\!\oout$ and $\omega_i = \eps$ otherwise.

We say that $\cT$ is \emph{unambiguous} if for every $w$ and $\mu$ there is at most one accepting run $\rho$ of $\cT$ which produces $\mu$. In~\cite{icdt2020nested} these VPT are called input-output unambiguous.

The theorem we use can be stated as follows:

\begin{theorem}{(\cite{icdt2020nested}, Theorem 3)}\label{eval:prep}
	There is an algorithm that receives an unambiguous VPT $\cT = (Q, \hat{\Sigma}, \Gamma, \Omega, \Delta, q_0, F)$ and an input string $w$, and enumerates the set $\sem{\cT}(w)$ with output-linear delay after a preprocessing phase that takes $\cO(|Q|^2 \cdot \vert\Delta\vert \cdot |w|)$ time.
\end{theorem}

The rest of the proof will consist on showing a linear-time reduction from the problem of enumerating the set $\sem{\cP}_{\pi}(w)$ for an unambiguous PDAnn $\cP$ and input string $w$ to the problem of enumerating the set $\sem{\cT}(w')$ for an unambiguous VPT $\cT$, and input string $w'$.

Let $\cP = (Q, \Sigma, \Omega, \Gamma, \Delta, q_0, F)$ and let $w\in \Sigma^*$ be an input string.
Consider the structured alphabet $\hat{\Sigma} = (\{\texttt{<}\}, \{\texttt{>}\}, \Sigma)$ for some $\texttt{<}, \texttt{>}\not\in\Sigma$. Assume $\pi = \pi_1, \ldots, \pi_m$.
We construct a well-nested string $w' = b_1\cdots b_{m-1}$ where $b_i = 
\texttt{<}$ if $\pi_i > \pi_{i+1}$, $b_i = \texttt{<}$ if $\pi_i < \pi_{i+1}$, and $b_i  = w_j$ otherwise, where $i$ is the $j$-th index in which $\pi_i = \pi_{i+1}$.
We also build a table $\mathsf{Ind}$ such that $\mathsf{Ind}(i) = j$ for each of the indices in the third case.
We build a VPT $\cT = (Q, \hat{\Sigma}, \Gamma, \Omega, \Delta', I, F)$ where $I = \{q_0\}$ and we get $\Delta'$ by replacing every push transition $(p, q, \gamma)\in \Delta$ by $(p, \texttt{<}, q, \gamma)$ and every pop transition $(p,\gamma,q)\in\Delta$ by $(p, \texttt{>}, q, \gamma)$. Note that read and read-write transitions are untouched.

Let $w$ be an input string, and let $\mu$ be an output. Consider the output $\mu'$ which is obtained by shifting the indices in $\mu$ to those that correspond in $w'$. We argue that for each run $\rho$ of $\cP$ over $w$ with profile $\pi$ which produces $\mu$, there is exactly one run $\rho'$ of $\cT$ over $w'$ which produces $\mu'$, and vice versa. We see this by a straightforward induction argument on the size of the run. This immediately implies that for each output $\mu\in\sem{\cT}(w)$ there exists exactly one output $\mu'\in\sem{\cP}_{\pi}(w)$, which has its indices shifted as we mentioned.
The algorithm then consists on simulating the procedure from Theorem~\ref{eval:prep} over $\cT$ and $w'$, and before producing an output $\mu$, we replace the indices to the correct ones following the table  $\mathsf{Ind}$. The time bounds are unchanged since the table $\mathsf{Ind}$ has linear size in $m$, and replacing the index on some output $\mu$ can be done linearly on $|\mu|$. We conclude that there is algorithm that enumerates the set $\sem{\cP}_{\pi}(w)$ with output-linear delay after a preprocessing that takes $\cO(|Q|^2\cdot \vert\Delta\vert\cdot |\pi|)$ time.

\subsection{Proof of Proposition~\ref{prp:basedeterm}}

For this result, we use the notion of PDA (Definition~\ref{def:pda}) and
deterministic PDA (Definition~\ref{def:dpda}) that were omitted from the main
text.

As we have done in previous proofs, the strategy consists on starting with a profiled-deterministic PDAnn $\cP$, and building a PDAnn $\cP'$ by eliminating the output symbols from each transition. This PDAnn behaves almost identically to a pushdown automaton $\cA$ in the sense that if $w\in L(\cA)$,  then $\sem{\cP}(w) = \{\eps\}$, and that if $w\not\in L(\cA)$ then $\sem{\cP}(w) = \emptyset$. Whenever this holds, we say that the PDAnn and the pushdown automaton are \emph{equivalent}. It is simple to see that for this $\cA$ it holds that $L(\cA) = L'$. To conclude the proof, we must show that $\cA$ can be made deterministic. Without loss of generality, we remove from $\cA$ 
all \emph{inaccessible states}, i.e., all states for which there is no run that goes to the state.

First, we will prove that the PDAnn $\cP'$ is profiled-deterministic. Let $w$ be a string in $\Sigma^*$ and let $\rho_1'$ and $\rho_2'$ be two partial runs of $\cP'$ over $w$ with the same profile, and with last configurations $(q, i)$ and $(q', i)$. There clearly exist partial runs $\rho_1$ and $\rho_2$ of $\cP$ over $w$ with the same profile, which can be obtained by replacing each transition by one of the transitions in $\cP$ it was replaced by. Since $\cP$ is profile-deterministic, then one of the following must hold in $\cP$: (1) $\Delta(q) \cup \Delta(q') \subseteq Q \times (\Sigma \cup \Sigma \times
\Omega) \times Q$, i.e., all transitions from~$q$ and $q'$ are read or
read-write transitions; (2) $\Delta(q) \cup \Delta(q') \subseteq Q \times (Q \times \Gamma)$, i.e.,
all transitions from~$q$ and~$q'$ are push transitions; or (3) $\Delta(q) \cup \Delta(q') \subseteq (Q \times \Gamma) \times Q$, i.e., 
all transitions from $q$ and $q'$ are pop transitions. Note that if (2) or (3) hold, then in the new PDAnn $\cP$ the condition holds again in $\cP'$ trivially since none of the transitions in $\Delta(q)$ and $\Delta(q')$ was changed. Moreover, if (1) holds, then it can be seen that all of the transitions that belonged in $Q\times (\Sigma\times\Omega)\times Q$ now belong in $Q\times \Sigma\times Q$, which also leaves the condition unchanged in $\cP'$. We conclude that $\cP'$ is profiled-deterministic.

The next step is to use Lemma~\ref{lem:pdtonechoice} from $\cP'$ to obtain an equivalent PDAnn $\cP''$ which is deterministic-modulo-profile. We will argue that if we start with $\cP'$, which was profiled-deterministic, then the resulting $\cP''$ is equivalent to a pushdown automaton $\cA'$ which is also deterministic. Let $w$ be an input string in $\Sigma^*$ and let $\rho''$ be a partial run of $\cA$ over $w$ with last configuration $(S, i)$ and with topmost symbol on the stack $T$. Let us recall what $\cP''$ being deterministic-modulo-profile entails that the following conditions hold:
  \begin{enumerate}
	\item There is at most one push transition that
	starts on $S$; formally, we have:
        \[\card{\{S', T \in Q'' \times \Gamma'' \mid
                (S, S', T) \in \Delta\}} \leq 1.\]
	\item There is at most
	one pop transition that starts on $S,T$; formally, for each $\gamma$, we
        have:
        \[\card{\{S'
                \in Q \mid (S, \gamma, S') \in \Delta\}} \leq 1.\]
	\item For each letter $a$, and output $\oout \in \Omega$, there is at most one read-write transition that starts on
          $S,a,\oout$; formally, we have \[\card{\{S' \in Q'' \mid (S,(a,\oout),S') \in \Delta''\}}
        \leq 1.\]
	\item For each letter $a$, there is at most one read transition that starts on
          $S,a$; formally, we have: \[\card{\{q \in Q'' \mid (S,a,S') \in \Delta''\}}
        \leq 1.\]
\end{enumerate}
We will show that at most one of these conditions holds. Recall that in the transformation, the states of $\cP''$ are sets which contain pairs of states $(p,q)\in Q'\times Q'$, and the stack symbols are triples $(p, \gamma, q)\in Q'\times\Gamma' \times Q'$. Now, recall the claim that was proven in the lemma, on the backwards direction:

If $\cP''$ has a run $\rho''$ on a string $w$, producing output $\mu$, from its initial state to an instantaneous
description $(S, i), \alpha'$ with $\alpha' =
T_0, \ldots, T_m$ being the sequence of the stack
symbols, then for any choice of elements $(q_0, \gamma_0, p_0) \in T_0$,
$(p_0, \gamma_1, p_1) \in T_1$, ..., $(p_{m-1}, \gamma_m, p_m) \in T_m$
and $(p_m, q) \in S$ it holds that $\cP'$ has a run $\rho'$ on $w$ producing output $\mu$
from some initial state $q_0$ to the instantaneous description $(q, i), \alpha$ with
$\alpha = \gamma_0 q_0, \ldots, \gamma_m q_m$ (writing next to each
stack symbol the state that annotates it), and $\rho''$ and $\rho'$ have
the same profile. 

Since $\cP'$ is profiled-deterministic, then each run $\rho'^+$ which continues $\rho'$ by one step must have the same shape. This implies that exactly one of the following conditions must hold:
\begin{itemize}
	\item The last transition in $\rho'^+$ is a read or read-write transition. Therefore, all transitions from $q$ are either read or read-write transitions.
	\item The last transition in $\rho'^+$ is a push transition. Therefore, all transitions from $q$ are pop transitions.
	\item The last transition in $\rho'^+$ is a pop transition. Therefore, all transitions from $q$ are pop transitions.
\end{itemize}
 Assume bullet point 1 holds. Note that there are no read-write transitions in $\cP'$ so there are only read transitions. From here, we prove that only (4) is true simply by inspecting the transformation in the lemma; if (1) held, then there would be a push transition from $q$ in $\cP'$, if (2) held then there would be a pop transition from $q$ and $\gamma$ in $\cP'$, and (3) never holds. Now, assume bullet point 2 holds. From here, we prove that only (1) can be true; if (2) held, then there would be a pop transition from $q$ and $\gamma$ in $\cP'$, if (4) held, then there would be a read transition from $q$ in $\cP'$, and again, (3) is never true. Lastly, assume bullet point 3 holds. From here, we prove that only (2) can be true; if (1) held, then there would be a push transition from $q$ in $\cP'$, if (4) held, then there would be a read transition from $q$ in $\cP'$, and yet again, (3) is never true. We conclude that from the 4 points, at most one of these can be true at the same time.

Now we prove that the equivalent PDA $\cA$ is deterministic.
Let $q$ be a state of~$\cA$. As all states of~$\cA$ are accessible, pick $\rho''$ to be a run that reaches state~$q$. We have argued that at most one of the points in the list above is true of~$\cP'$, and it cannot be point (3). Now,
we see that (a) is equivalent to (4), that (b) is equivalent to (1) and (c) is equivalent to (2). Since only one of the conditions among (1), (2) or (4) can be true, the same holds for (a), (b) and (c), from which we conclude that $\cA$ is deterministic. This completes the proof.

\subsection{Proof of Proposition~\ref{prp:detlinear}}

Consider a profiled-deterministic PDAnn $\cP$. To prove that it is profiled, consider an input string $w\in \Sigma^*$. We will prove by a simple induction argument that any two runs of $\cP$ over $w$ have the same profile. The base case is trivial since the run is of length 0, and the profile up to now is composed simply of the stack size 0. Assume now that for each pair of runs $\rho$ and $\rho'$ of $\cP$ over $w$ of size $k$, that they have the same profile. We will show that for every pair of runs $\rho_1$ and $\rho_2$ over $w$ of size $k+1$, they have the same profile as well. Note that the runs $\rho_1^-$ and $\rho_2^-$ that are obtained by removing the last step have the same profile, by the hypothesis. From the definition of profiled-deterministic it can be directly seen that if (1) the last transition in $\rho_1$ is a read or read-write transition, then for the runs $\rho_1^-$ and $\rho_2^-$, the only choices are read or read-write transitions, from which we deduce that the last transition in $\rho_2$ is a read or read-write transition as well, if (2) the last transition in $\rho_1$ is a push transition, then for the run $\rho_1^-$ and $\rho_2^-$ the only choice are push transitions, and therefore the last transition in $\rho_2$ has to be a push transition as well, and if (3) the last transition in $\rho_1$ is a pop transition, then for $\rho_1^-$ and $\rho_2^-$ the only choices are pop transitions, so the last transition in $\rho_2$ must be a pop transition as well. We obtain that $\rho_1$ and $\rho_2$ have the same profile, and from the induction argument, we conclude that $\cP$ is profiled.

Now, consider a profiled-deterministic PDAnn $\cP$ and an input string $w$. We will prove that the unique profile of accepting runs of $\cP$ over $w$ can be computed in linear time in $|w|$. The way we do this is by using the pushdown automaton $\cA$ that was constructed in Proposition~\ref{prp:basedeterm}. By inspecting the proof, it can be seen that the unique profile of $\cP$ over $w$ is maintained throughout the construction. Indeed, the first construction simply removes the output symbols, which does not affect the profile, and the second construction has an invariant that keeps the profile intact as well. Therefore, by running the automaton $\cA$ over $w$, and storing the stack sizes at each step, we obtain a profile $\pi$ which is exactly the same profile of the accepting runs of $\cP$ over $w$. To finish the proof, we only need to argue that this profile has linear size on $|w|$ (from a data complexity perspective). This follows from the fact that any run of a deterministic pushdown automaton $\cA$ over a string $w$ has $\cO(f(\cA)\times|w|)$ length, for some computable function $f$. This can be seen from a counting argument: (1) There is a maximum stack size $k$ that can be reached in an accepting run of $\cP$ over $w$ from an empty stack through $\eps$-transitions, which is given by the number of states in $\cA$. Otherwise, there are two configurations which are reachable from one another in a way such the stack, as it was at the first configuration, is not seen. This implies that there is a loop, and since $\cA$ is deterministic, $\cA$ does not accept $w$. (2) From a given stack, the maximum numbers of steps that can be taken without reading from $w$, and without seeing the topmost symbol on the stack is given by the number of possible stacks of size $k$. (3) Between a read (or read-write) transition and the next one, the maximum height difference is $k$, and if we move out of a read (or read-write) transition with a certain stack, from (2) we can see that we can only do a fixed number of steps before consuming some symbol from this stack, and therefore, the number of steps is bounded by a factor depending on $\cA$ multiplied by the size of the stack up until this point, which is linear on the number of symbols in $w$ read so far. We conclude that $w'$ has size linear on $w$, from a data complexity point of view.

\section{Proofs of Section~\ref{sec:spanners}}

\subsection{Formal definitions on extraction grammars}
\label{apx:formalextraction}

We first give formal definitions that were omitted from the main text of the
paper for lack of space. We first formally define \emph{ref-words}, i.e.,
strings
with the special \emph{variable operations}

\begin{definition}[Ref-words]
  \label{def:refwords}
For the set of variables $\var$ we define the \emph{variable operations} of $\var$ by
$\captures_\var = \bigcup_{x \in \var} \{ {\vdash_x}, {\dashv_x} \}$.
A \emph{ref-word} is a string over $\Sigma \cup \captures_\var$, and we let
  $\splain: (\Sigma \cup \captures_\var)^* \rightarrow \Sigma^*$ be the morphism
  over ref-words that removes variable operations
A ref-word is {\em valid} if each variable in $\var$ is opened exactly once and then closed exactly once.
A valid ref-word $r$ then defines a mapping~$\eta_r$ over $\splain(r)$ in the following way:
for each variable $x \in \var$, there is a unique factorization $r = r_x^p \cdot \vdash_x \, \cdot \, r_x \, \cdot\, \dashv_x \cdot r_x^s$,
and we set 
$\eta^r(x) \colonequals \spanc{i}{j}$ for $i \colonequals |\splain(r_x^p)|+1$ and $j \colonequals i + |\splain(r_x)|$. 
\end{definition}

We then define \emph{extraction grammars} from~\cite{Peterfreund21}: they are
simply CFGs over ref-words:

\begin{definition}[Extraction grammar]
  \label{def:extract}
An \emph{extraction grammar} is a tuple
$\cH = (V, \Sigma, \var, P, \allowbreak S)$ where 
$V$ is a finite set of \emph{nonterminals},
$\Sigma$ is the alphabet,
$\var$ is a set of variables, 
$P$ is a finite set of production rules of the form $A \to \alpha$ with $A \in
V$ and $\alpha \in (V \cup \Sigma \cup \captures_\var)^*$, and $S \in V$ is the
\emph{start symbol}.
We assume that $V$, $\Sigma$, and $\var$ are pairwise disjoint.
Note the difference with annotated grammars: the annotations must correspond to
variable operations (in order to capture spans), and they are represented as separate terminals
instead of annotating existing terminals, which we believe makes the design of
enumeration algorithms less convenient.

The semantics of extraction grammars is similar to that for annotated grammars.
It is defined through \emph{derivations}. Specifically, the rules $P$ define the
\emph{(left) derivation relation} $\der{\cH} \ \subseteq \, (V \cup \Sigma \cup \captures_\var)^* \times (V \cup \Sigma \cup  \captures_\var)^*$ such that $u A \beta \der{\cH} u \alpha \beta$ iff $u \in (\Sigma \cup \captures_\var)^*$, $A \in V$, $\alpha, \beta \in (V \cup \Sigma \cup \captures_\var)^*$, and $A \rightarrow \alpha \in P$. We denote by $\der{\cH}^*$ the reflexive and transitive closure of $\der{\cH}$. 
Then the \emph{language} defined by $\cH$ is the set of ref-words $L(\cH) = \{w \in (\Sigma \cup \captures_\var)^* \mid S \der{\cH}^* w\}$.
This language defines a \emph{spanner} $\sem{\cH}$ as follows: for every document $d \in \Sigma^*$,
  \[
\sem{\cH}(d) \ = \ \{ \, \eta^r \ \mid \ r \in L(\cH), \text{ $r$ is valid}, \text{ and } \splain(r) = d\, \}.
\]
Note that ref-words that are not valid are ignored.
An extraction grammar $\cH$ is called \emph{functional} if every $r \in L(\cH)$ is valid, and it is called \emph{unambiguous} if for every $r \in L(\cH)$ there exists exactly one derivation of~$r$ from~$S$. 
\end{definition}

We can now formally define the \emph{equivalence} between an extraction grammar
and an annotated grammar. To do so, we first explain how we can translate
mappings to annotations:

\begin{definition}[Output associated to a mapping]
  \label{def:outputset}
Given a set $\var$ of variables, the corresponding set of annotations $\Omega_\var$ will be the powerset of $\captures_\var$.
Now, given a mapping~$\eta$ on a document~$d$ and variables~$\var$ to an annotation, we 
let $\cI = \bigcup_{x \in \var} \{i, j \mid \eta(x) = \spanc{i}{j}\}$ be the
set of indices which appear in some span of $\eta$.
Further, for each $k \in \cI$, let $S_k = \{\vdash_x \mid \exists j. \, \eta(x) = \spanc{k}{j}\} \cup \{\dashv_x \, \mid \exists i. \, \eta(x) = \spanc{i}{k}\}$. We now define the output $\outf(\eta) = (i_1,S_{i_1})\ldots(i_m,S_{i_m})$ where $\cI = \{i_1,\ldots,i_m\}$ and  $i_1 < \cdots < i_m$, namely, we group the captures for each position as a set and use this set as the annotation.
Note that the largest index that appears in the annotation can be $|d|+1$ because of the range of spans.
\end{definition}

We can now define \emph{equivalence} between extraction grammars and annotation
grammars:

\begin{definition}[Equivalent annotated grammar]
  \label{def:equivalentannotated}
  We say that an extraction grammar $\cH$ over variables $\var$ 
has an \emph{equivalent annotated grammar} $\cG$ if $\cG$ is over the 
set of annotations $\Omega_\var$ and over the
alphabet $\Sigma \cup \{\#\}$
for $\#$ a fresh symbol, and
if for every document $d \in \Sigma^*$ and every mapping $\eta$ of~$d$ over~$\var$, we have
$\eta \in \sem{\cH}(d)$ iff $\outf(\eta) \in \sem{\cG}(d\cdot \#)$.
The $\#$-symbol at the end is used because of the difference in the indexing of spans (from $1$ to $|d|+1$) and annotations (from 1 to~$d$).
\end{definition}

\subsection{Proof of Proposition~\ref{prop:extrac-grammars-general}}

Recall that, in the statement of this result, the formal notion of an
\emph{equivalent annotated grammar} is the one defined above
(Definition~\ref{def:equivalentannotated}). Recall also the formal definition of
\emph{ref-words} (Definition~\ref{def:refwords}).

Let $r$ be a ref-word in $\Sigma \cup \captures_\var$ and let $\hat{w}$ be an annotated string in $\Sigma \cup (\Sigma \times \Omega_\var)$. We say that $r$ and $\hat{w}$ are \emph{equivalent} if $\splain(r\cdot\#) = \str(\hat{w})$ and $\outf(\eta^{r}) = \ann(\hat{w})$. 
For example, the ref-word $r_1 = \ \vdash_x\!\texttt{a}\texttt{a}\!\dashv_x\vdash_y\!\texttt{b}\texttt{b}\!\dashv_y\!\texttt{b}$ is equivalent to $\hat{w}_1 = (\texttt{a},\{\vdash_x\})\,\texttt{a}\,(\texttt{b},\{\dashv_x,\vdash_y\})\,\texttt{b}\,\texttt{b}\,(\texttt{b},\{\dashv_y\})\,\#$.

The overall strategy of this proof is going to be to construct an annotated grammar $\cG$ in a way such that for every ref-word $r\in L(\cH)$ there exists an equivalent annotated string $\hat{w}\in L(\cG)$, and vice versa. It is clear that this implies that $\cG$ and $\cH$ are equivalent.

The way we build $\cG$ will look like we are ``pushing'' the variable operations to the next terminal to the right. 
We will do this process one variable operation at a time.

First, we need to define an intermediate model between those of extraction grammars and annotated grammars. 
We define \emph{extraction grammars with annotations} as a straightforward
extension of extraction grammars which allow annotations on terminals that are
not variable operations. 
For the set of variables $\var$, recall from Definition~\ref{def:refwords} that
we define the variable operations  of $\var$ by $\captures_\var = \{ {\vdash_x},
{\dashv_x} \mid x\in \var\}$,
and recall from Definition~\ref{def:outputset} that we define $\Omega_\var = 2^{\captures_\var}$. An \emph{extraction grammar with annotations} is a tuple
$\cF = (V, \Sigma, \var, P, S)$ where 
$V$ is a finite set of nonterminal symbols, $\Sigma$ is an alphabet and $\var$ is a set of variables, such that $V$, $\Sigma$, $\Sigma\times\Omega_\var$, and $\captures_\var$ are pairwise disjoint, 
$P$ is a finite set of rules of the form $A \to \alpha$ with $A \in
V$ and $\alpha \in (V \cup \Sigma \cup (\Sigma\times\Omega_\var) \cup \captures_\var)^*$, and $S \in V$ is the
start symbol. As in the other models, the semantic of extraction grammars is defined through derivations. Specifically,  the set $P$ defines the (left) derivation relation $\der{\cF} \ \subseteq \, (V \cup \Sigma  \cup (\Sigma\times\Omega_\var) \cup \captures_\var)^* \times (V \cup \Sigma  \cup (\Sigma\times\Omega_\var) \cup  \captures_\var)^*$ such that $\hat{u} A \beta \der{\cF} \hat{u} \alpha \beta$ iff $\hat{u} \in (\Sigma  \cup (\Sigma\times\Omega_\var) \cup \captures_\var)^*$, $A \in V$, $\alpha, \beta \in (V \cup \Sigma \cup (\Sigma\times\Omega_\var) \cup \captures_\var)^*$, and $A \rightarrow \alpha \in P$. We denote by $\der{\cF}^*$ the reflexive and transitive closure of $\der{\cF}$. Then the language defined by $\cF$ is $L(\cF) = \{\hat{w} \in (\Sigma \cup (\Sigma\times\Omega_\var) \cup \captures_\var)^* \mid S \der{\cF}^* \hat{w}\}$. In addition, we assume that no string $\hat{w}$ in $L(\cF)$ has a variable operation as its last symbol.

An extraction grammar with annotations generates
strings over $\Sigma \cup (\Sigma\times\Omega_\var) \cup \captures_\var$, which we now refer to as \emph{annotated ref-words}, and each annotated ref-word defines an output.
We will define the semantics of extraction grammars with annotations recursively by using the semantics of annotated grammars as a starting point, that is, by extending the function $\ann$ to receive strings over $\Sigma \cup (\Sigma\times\Omega_\var) \cup \captures_\var$. In particular, for an annotated ref-word $\hat{r}\in (\Sigma \cup (\Sigma\times\Omega_\var))^*$ the result of $\ann(\hat{r})$ stays the same. For a string $\hat{r} = \hat{u} \kappa a \hat{v} \in (\Sigma \cup (\Sigma\times\Omega_\var) \cup \captures_\var)^*$ where $\kappa\in \captures_{\var}$ and $a\in\Sigma$, we define $\ann(\hat{r}) = \ann(\hat{u} (a,\{\kappa\}) \hat{v})$, and for a string $\hat{r} = \hat{u} \kappa (a,\oout) \hat{v} \in (\Sigma \cup (\Sigma\times\Omega_\var) \cup \captures_\var)^*$ where $\oout\in\Omega_\var$, we define $\ann(\hat{r}) = \ann(\hat{u} (a,\oout\cup\{\kappa\}) \hat{v})$.

Further, we extend the function $\str$ to receive strings over $\Sigma \cup (\Sigma\times\Omega_\var) \cup \captures_\var$ as $\str(\hat{r}) = \str(\splain(\hat{r}))$. Therefore, for an extraction grammar with annotations $\cF$ and a string $w\in\Sigma^*$ we define the function $\sem{\cF}$ as: $\sem{\cF}(w) \ := \  \{\ann(\hat{r}) \mid \hat{r} \in L(\cF) \wedge \str(\hat{r}) = w\}$.
We maintain the notions of equivalency between annotated grammars, extraction grammars, and extraction grammars with annotations. Likewise, we define equality between annotated strings, ref-words and  annotated ref-words in the obvious way. Further, we note that any extraction grammar $\cH = (V, \Sigma, \var, P, S)$ is equivalent to the extraction grammar with annotations $\cF = (V', \Sigma\cup\{\#\}, \var, P', S')$ where $V' = V\cup\{S'\}$ for some $S'\not\in V$, and $P' = P \cup\{S'\to S\# \}$. It is obvious that $\cF$ is unambiguous if and only if $\cH$ is unambiguous.

We proceed as follows. First we convert $\cH$ into a functional extraction grammar (see Definition~\ref{def:extract}).
As detailed in Peterfreund's work
\cite[Propositions 10 and 12]{Peterfreund21}, this takes running time 
$\cO(3^{2k}|\cH|^2)$, and if the initial grammar is unambiguous then so is
the resulting grammar.
Note that this implies that every ref-word $r$ which is derivable from $S$ contains each variable operation at most once. Hence we can build an equivalent extraction grammar with annotations $\cF = (V, \Sigma, \var, P, S)$ using the technique above. We then convert $\cF$ into a version of CNF which is slightly more restrictive than arity-two normal form: We allow rules of the form $X\to YZ$, $X\to \eps$ and $X\to \tau$, for nonterminals $X, Y$ and $Z$ and a terminal $\tau$, but rules of the form $X\to Y$ are not permitted.
Converting to this formalism can be done in linear time in $|\cF|$ while preserving unambiguity,
e.g., by transforming rules of the form $X \rightarrow Y$ to $X \rightarrow E Y$
for some fresh nonterminal $E$ with a rule $E \rightarrow \epsilon$, and
otherwise applying our result on arity-2 normal form (Proposition~\ref{prp:2nf}).
We pick an order over the variable operations in $\captures_\var$ and for each $\kappa\in\captures_\var$ we do the following:

Define a function $\proc_{\kappa}$ that receives an annotated ref-word
$\hat{r}\in (\Sigma \cup (\Sigma\times2^{\captures_\var}) \cup
\captures_\var)^*$ and:
\begin{enumerate}
  \item if $\hat{r}\in (\Sigma \cup (\Sigma\times2^{\captures_\var}) \cup (\captures_\var\setminus\{\kappa\}))^*$, then $\proc_{\kappa}(\hat{r}) = \hat{r}$,
  \item if $\hat{r} = \hat{u}\kappa \beta a\hat{v}$ for some $\hat{u}, \hat{v}\in (\Sigma \cup (\Sigma\times2^{\captures_\var}) \cup (\captures_\var\setminus\{\kappa\}))^*$, $\beta\in (\captures_\var\setminus\{\kappa\})^*$ and $a\in\Sigma$, then $\proc_{\kappa}(\hat{r}) = \hat{u}\kappa \beta (a,\{\kappa\})\hat{v}$,
  \item if $\hat{r} = \hat{u}\kappa \beta (a,T)\hat{v}$, with $T\subseteq\captures_\var\setminus\{\kappa\}$, then  $\proc_{\kappa}(\hat{r}) = \hat{u}\kappa \beta (a,T\cup\{\kappa\})\hat{v}$, and
  \item $\proc_{\kappa}$ is undefined in any other case.
\end{enumerate}
It is straightforward to see that the annotated ref-words $\hat{r}$ and $\hat{t} = \proc_\kappa(\hat{r})$ are equivalent whenever $\proc_\kappa(\hat{r})$ is defined.\\

We build an extraction grammar with annotations $\cF' = (V', \Sigma, \var, P', S')$ where $V' = V_\sfo \cup V_\sfi \cup V_\sfl \cup V_\sfm \cup V_\sfr\cup\{S'\}$, and $V_\mathsf{scr} = \{A^\mathsf{scr} \mid A\in P\}$ for $\mathsf{scr}\in\{\sfo, \sfi, \sfl, \sfm, \sfr\}$, and $P'$ is defined by the following rules:
\begin{itemize}
	\item For each rule $S\to AB$ in $P$, we add the rules $S'\to A^{\sfo}B^{\sfo}$, $S'\to A^{\sfi}B^{\sfo}$ and $S'\to A^{\sfl}B^{\sfr}$ to $P'$.
	\item For each rule $A\to BC$, we add the rules $A^{\sfo} \to B^{\sfo}C^{\sfo}$, $A^{\sfi}\to B^{\sfi}C^{\sfo}$, $A^{\sfi}\to B^{\sfo}C^{\sfi}$, $A^{\sfi}\to B^{\sfl}C^{\sfr}$, $A^{\sfl}\to B^{\sfo}C^{\sfl}$, $A^{\sfl}\to B^{\sfl}C^{\sfm}$, $A^{\sfm}\to B^{\sfm}C^{\sfm}$, $A^{\sfr}\to B^{\sfr}C^{\sfo}$ and $A^{\sfr}\to B^{\sfm}C^{\sfr}$ to $P'$.
	\item For each rule $A\to a$, $a\in \Sigma$, we add $A^{\sfo}\to a$ and $A^{\sfr}\to(a, \{\kappa\})$.
\item For each rule $A\to (a, T)$, $a\in \Sigma$ and $T\subseteq\captures_\var\setminus\{\kappa\}$, we add $A^{\sfo}\to (a, T)$ and $A^{\sfr}\to(a, T\cup\{\kappa\})$.
\item For each rule $A\to \kappa'$, $\kappa'\in\captures_\var\setminus\{\kappa\}$, we add the rules $A^{\sfo}\to\kappa'$ and $A^{\sfm}\to\kappa'$.
\item For each rule $A\to \kappa$, we add the rule $A^{\sfl}\to\eps$.
\item For each rule $A\to\eps$, we add the rules $A^{\sfo}\to\eps$ and $A^{\sfm}\to\eps$.
\end{itemize}

Let $\hat{\Sigma}_{\kappa} = \Sigma \cup (\Sigma\times2^{\captures_\var}) \cup \captures_\var\setminus\{\kappa\}$. For each nonterminal $A\in V$ these hold:
\begin{align*}
	L(A^{\sfo}) &= \{\hat{w}\mid 
	A\der{\cF}^* \hat{w}, 
	\ \text{where } \hat{w} \in  \hat{\Sigma}_{\kappa}^*\}\\
	L(A^{\sfi}) &= \{\proc_{\kappa}(\hat{w}) \mid 
	A\der{\cF}^* \hat{w},
	\ \text{where } \hat{w} = \hat{u}\kappa\beta a \hat{v} \text{ or }\hat{w} = \hat{u}\kappa\beta (a, T) \hat{v},\\ 
	&\ \ \ \ \ \ \ \ \ \ \ \ \ \ \ \ \  \hat{u}, \hat{v} \in \hat{\Sigma}_{\kappa}^*, \beta\in (\captures_\var\setminus\{\kappa\})^*, a\in\Sigma, T\subseteq\captures_\var\setminus\{\kappa\}\}\\
	L(A^{\sfl}) &= \{\hat{w}\beta \mid 
	A\der{\cF}^* \hat{w}\kappa\beta,
	\ \text{where } \hat{w}\in\hat{\Sigma}_{\kappa}^*, \beta\in(\captures_\var\setminus\{\kappa\})^*\}\\
	L(A^{\sfr}) &= \{\beta(a, \{\kappa\})\hat{w} \mid 
	A\der{\cF}^* \beta a \hat{w},
	\ \text{where } \hat{w}\in \hat{\Sigma}_\kappa^*, \beta\in(\captures_\var\setminus\{\kappa\})^*
	\} \ \cup\\ 
	& \!\!\!\{\beta(a, T\cup \{\kappa\})\hat{w} \mid 
	A\der{\cF}^* \beta (a,T) \hat{w},
	\ \text{where }\hat{w}\in\hat{\Sigma}_\kappa^*, \beta\in(\captures_\var\setminus\{\kappa\})^*, T\subseteq\captures_\var\setminus\{\kappa\}\}\\
	L(A^{\sfm}) &= \{\beta\mid A \der{\cF}^*\beta,
	\ \text{where }\beta\in(\captures_\var\setminus\{\kappa\})^*\}
\end{align*}

These equalities are given without proof since they are not used, and are just for illustrating the idea behind the proof.

For the rest of our proof we will represent derivations $X\der{}^* \delta$ as the sequence of productions $X_1\der{}\gamma_1, X_2\der{}\gamma_2, \ldots,X_m\der{}\gamma_m$, where $X_1 = X$ and $\gamma_m = \delta$, which uniquely determines the derivation by doing it in the leftmost way. We use this representation to state exactly how derivations in $\cF$ are translated to derivations in $\cF'$ and vice versa.

Another notion we need to address is how, in a given derivation $X\der{}^* \delta$, instances of nonterminals are located with respect to each other. By an \emph{instance of a nonterminal} (or just \emph{instance}), we mean an $X_i$ along with some specific derivation $X_i\der{} \gamma_i$ in the sequence. For some instances $X_i$ and $X_j$, we say that $X_i$ is a \emph{descendant} of $X_j$ if $X_i = X_j$, or if $Y \der{} X_i Z$, or $Y\der{} Z X_i$ for some $Y$ which is a descendant of $X_j$. We say that $X_i$ is \emph{to the left} of $X_j$ (or $X_j$ is \emph{to the right} of $X_i$) if there is a derivation $X\der{} Y Z$ in the sequence such that $X_i$ is a descendant of $Y$ and $X_j$ is a descendant of $Z$. 

We note a few things in our construction: (1) Each $X\in V_\sfo$ only produces terminals (which do not include $\kappa$) and nonterminals in $V_\sfo$; furthermore, every rule $X\to YZ$ in $P$ is copied into $P'$ as $X^\sfo \to Y^\sfo Z^\sfo$. (2) Nonterminals in $V_\sfi$ do not produce any terminals or $\eps$ directly, so they need to derive into some $X\in V_\sfi$ and some $Y\in V_\sfr$ to derive some string. (3) As with $V_\sfo$, each $X\in V_\sfm$ only produces terminals in $\captures_\var\setminus\{\kappa\}$ and nonterminals in $V_\sfm$. (4) Each $X\in V_\sfl$ (resp. $V_\sfr$) produces exactly one nonterminal $X'\in V_\sfl$ (resp. $V_\sfr$), or $\eps$ (resp. $(a, T)$ for some $a\in \Sigma$ and $T\subseteq\captures_\var$ such that $\kappa\in T$); this, as a consequence, means that on each derivation from $\cF'$ where the first production is not $S' \der{} X^\sfo Y^\sfo$ there is exactly one derivation $X\der{}\eps$ such that $X\in V_\sfl$ (resp., exactly one derivation $X\der{} (a, T)$, such that $X\in V_\sfr$).

From point (1) we see that each annotated ref-word $\hat{r}\in L(\cF)$ such that $\hat{r}\in\hat{\Sigma}_\kappa^*$ (this is, which does not mention $\kappa$ at all) can be derived by $S'$ starting by $S'\der{} A^{\sfo}B^{\sfo}$, $S'\der{} a$, $S'\der{} \eps$ or $S'\der{} \kappa'$. 

On the other hand, each annotated ref-word $\hat{r}\in L(\cF')$ which does not have $\kappa$ on any annotation set was necessarily derived through rules of the form $X^\sfo\to Y^\sfo Z^\sfo$ which correspond to the rule $X\to YZ$ in $P$, so we deduce that $\hat{r}\in L(\cF)$.

We shall now prove that for any string $\hat{r} = \hat{u}\kappa \beta a\hat{v}$, or $\hat{r} = \hat{u}\kappa \beta (a,T)\hat{v}$, where $\hat{u}, \hat{v}\in\hat{\Sigma}_\kappa^*$, $\beta\in\captures_\var\setminus\{\kappa\}$, $a\in\Sigma$ and $T\subseteq\captures_\var\setminus\{\kappa\}$ such that $\hat{r}\in L(\cF)$, it holds that $\proc(\hat{r})\in L(\cF')$. 
W.l.o.g., let $\hat{r} =  \hat{u}\kappa \beta a\hat{v}$ and consider
some leftmost derivation of $\hat{r}$ from $\cF$:
\begin{align*}
	S  &\der{\cF}^* 
	\hat{u}_1 A \delta \\
	&\der{\cF} 
	\hat{u}_1 B C \delta \\
	&\der{\cF}^* 
	\hat{u}_1 \hat{u}_2 \kappa \beta_1 C \delta \\
	&\der{\cF}^*
	\hat{u}_1 \hat{u}_2 \kappa \beta_1 \beta_2 a \hat{v}_1 \delta \\
	&\der{\cF}^* 
	\hat{u}_1 \hat{u}_2 \kappa \beta_1 \beta_2 a \hat{v}_1 \hat{v}_2 = \hat{r},
\end{align*}
where we have that $\beta = \beta_1\beta_2$, $\hat{u} = \hat{u}_1\hat{u}_2$ and $\hat{v} = \hat{v}_1\hat{v}_2$. Note that this is an arbitrary derivation, and we are merely identifying these nonterminals $A$, $B$ and $C$. We also identify the nonterminals $D$, which produces $\kappa$, and $E$, which produces $a$. For the rest of the current part of the proof, we only refer to the \emph{instances} of these nonterminals. Using this, we build a derivation from $\cF'$ step by step:
\begin{enumerate}
	\item We have $S' \der{\cF'}^* \hat{u}_1 A^{\sfi} \delta'$, where $\delta'$ is obtained by replacing each nonterminal $X$ in $\delta$ by $X^{\sfo}$. We get this by starting with the derivation $S \der{\cF}^* \hat{u}_1 A \delta$, and replacing $X\der{\cF} YZ$ by $X^{\sfi}\der{\cF'} Y^{\sfi}Z^{\sfo}$ if $A$ is a descendant of $Y$, by $X^{\sfi}\der{\cF'} Y^{\sfo}Z^{\sfi}$ if $Z$ is, or by $X^{\sfo}\der{\cF'} Y^{\sfo}Z^{\sfo}$ if none is. We also replace each $X\der{\cF} \tau$, for $\tau\in\Sigma\cup(\Sigma\times 2^{\captures_\var})\cup \captures_\var\setminus\{\kappa\}\cup\{\eps\}$, by $X^\sfo\der{\cF'}\tau$. If $A = S$, we replace $A$ it by $S'$.
	\item We have the rule $A^{\sfi} \to B^{\sfl} C^{\sfr}$ which was added to $P'$.
	\item We have $B^\sfl\der{\cF'}^* \hat{u}_2 \beta_1$. We get this by starting from $B\der{\cF}^*\hat{u}_2\kappa\beta_1$, and we replace $X\der{\cF} YZ$ by $X^\sfl\der{\cF'} Y^\sfl Z^\sfm$ if $E$ is a descendant of $D$, by $X^\sfl\der{\cF'} Y^\sfo Z^\sfl$ if $Z$ is, by $X^\sfo\der{\cF'} Y^\sfo Z^\sfo$ if $X$ is to the left of $D$, and by $X^\sfm\der{\cF'} Y^\sfm Z^\sfm$ if it is to the right. We also replace $X\der{\cF} \tau$ by $X^\sfo\der{\cF'}\tau$ if $X$ is to the left of $D$, and by $X^\sfm\der{\cF'}\tau$ if it is to the right. Lastly, we replace $D\der{\cF}\kappa$ by $D^\sfl\der{\cF'}\eps$.
	\item We have $C^\sfr\der{\cF'}^* \beta_2(a,\{\kappa\})\hat{v}_1$. We get this by starting from $C\der{\cF}^*\beta_2 a \hat{v}_1$, and we replace $X\der{\cF} YZ$ by $X^\sfr\der{\cF'} Y^\sfm Z^\sfr$ if $E$ is descendant of $Z$, or by $X^\sfr\der{\cF'} Y^\sfr Z^\sfo$ if $Y$ is, by $X^\sfm\der{\cF'} Y^\sfm Z^\sfm$ if $X$ is to the left of $E$, and by $X^\sfr\der{\cF'} Y^\sfr Z^\sfr$ if it is to the right. We also replace $X\der{\cF} \tau$ by $X^\sfm\der{\cF'}\tau$ if $X$ is to the left of $E$, and by $X^\sfo\der{\cF'}\tau$ if it is to the right. Lastly, we replace $E\der{\cF} a$ by $E^\sfr\der{\cF'}(a, \{\kappa\})$.
	\item We have $\delta'\der{\cF'}^*\hat{v}_2$, which we obtain from $\delta\der{\cF}^*\hat{v}_2$ by replacing each $X \der{\cF} YZ$ by $X^\sfo\der{\cF'} Y^\sfo Z^\sfo$, and each $X\der{\cF}\tau$ by $X^\sfo\der{\cF'}\tau$.
\end{enumerate}
In the end, we get the following leftmost derivation from $\cF'$:
\begin{align*}
	S  &\der{\cF'}^* 
	\hat{u}_1 A^\sfi \delta'\ (\text{or } \hat{u}_1 S' \delta') \\
	&\der{\cF'} 
	\hat{u}_1 B^\sfl C^\sfr \delta' \\
	&\der{\cF'}^* 
	\hat{u}_1 \hat{u}_2 \beta_1 C^\sfr \delta' \\
	&\der{\cF'}^*
	\hat{u}_1 \hat{u}_2 \beta_1 \beta_2 (a, \{\kappa\}) \hat{v}_1 \delta' \\
	&\der{\cF'}^* 
	\hat{u}_1 \hat{u}_2 \beta_1 \beta_2 (a, \{\kappa\}) \hat{v}_1 \hat{v}_2 = \proc_\kappa(\hat{r}),
\end{align*}
which proves that $\proc(\hat{r})\in L(\cF')$. 

We will prove that for every $\hat{s} = \hat{u}(a, T)\hat{v}\in L(\cF')$ where $\kappa\in T$ there is $\hat{r}$ such that $\proc_\kappa(\hat{r}) = \hat{s}$ in a similar way. We argue that any leftmost derivation that produces $\hat{s}$ has the following form:
\begin{align*}
	S'  &\der{\cF'}^* 
	\hat{u}_1 A^\sfi \delta_1\ \  (\text{or } \hat{u}_1 S' \delta_1)\\
	&\der{\cF'} 
	\hat{u}_1 B^\sfl C^\sfr \delta_1\\
	&\der{\cF'}^* 
	\hat{u}_1 \hat{u}_2 D^\sfl \delta_2 C^\sfr \delta_1 \\
	&\der{\cF'}
	\hat{u}_1 \hat{u}_2 \delta_2 C^\sfr \delta_1 \\
	&\der{\cF'}^* 
	\hat{u}_1 \hat{u}_2 \beta_1 C^\sfr \delta_1 \\
	&\der{\cF'}^*
	\hat{u}_1 \hat{u}_2 \beta_1 \beta_2 (a, T) \hat{v}_1 \delta_1 \\
	&\der{\cF'}^* 
	\hat{u}_1 \hat{u}_2 \beta_1 \beta_2 (a, T) \hat{v}_1 \hat{v}_2 = \hat{s},
\end{align*}
where $\delta\in V_\sfo^*$, $\delta'\in V_\sfm^*$, $\beta_1,\beta_2\in (\captures_\var\setminus\{\kappa\})^*$, $\hat{u} = \hat{u}_1\hat{u}_2$ and $\hat{v} = \hat{v}_1\hat{v}_2$. The reasoning goes as follows:
\begin{itemize}
	\item We know that $S'\der{\cF'}^*\hat{u}\beta(a, T)\hat{v}\in L(\cF')$. If $X\der{\cF'}(a, T)$ and $\kappa\in T$, then $X\in V_\sfr$.
	\item From the way $\cF'$ was built, there must be a production $X\der{\cF'} YZ$ in $S'\der{\cF'}^*\hat{s}$ such that $X\in V_\sfi$ (or $X = S'$), $Y\in V_\sfl$ and $Z\in V_\sfr$, as it is the only way to derive a nonterminal in $V_\sfr$. Let $A^\sfi$ (or $S'$), $B^\sfl$ and $C^\sfr$ be these $X$, $Y$ and $Z$ respectively.
	\item Seeing the rules in $P'$ we note that every string of terminals that is derivable from $B^\sfl$ is of the form $\hat{w}\beta$, where $\hat{w}\in\hat{\Sigma}_\kappa$ and $\beta\in (\captures_\var\setminus\{\kappa\})^*$. Furthermore, this string satisfies that there is a production $X\der{\cF'}\eps$ for some $X\in V_\sfl$ such that this $\eps$ is exactly at the left of where $\beta$ begins. Let $\hat{u}_2$ be this $\hat{w}$, let $D^\sfl$ be this $X$, and let $\beta_1$ be this $\beta$.
	\item Likewise, we note that $C^\sfr$ always derives a string of terminals of the form $\beta(a', T')\hat{w}$ for some $\beta\in (\captures_\var\setminus\{\kappa\})^*$ and $\hat{w}\in\hat{\Sigma}_\kappa$. Let $\beta_2$ be this $\beta$ and let $\hat{v}_1$ be this $\hat{w}$.
	\item Lastly, let $S' \der{\cF'}^* \hat{u}_1 A^\sfi \delta_1$ (or $\hat{u}_1 S' \delta_1$) be the one that derives $\hat{s}$. From the rules in $P'$, we note that $\delta_1$ is composed solely of nonterminals in $V_\sfo$.
\end{itemize}

An important point that can be seen from this reasoning is that for each instance $X \neq S'$ that appears in the derivation $S'\der{\cF'}^*\hat{s}$, we can deduce the set $V_{\mathsf{scr}}$ for which $X\in V_{\mathsf{scr}}$, among the options $\mathsf{scr}\in\{\sfo, \sfi, \sfl, \sfm, \sfr\}$, by seeing its position in the derivation. To be precise, this is given from how $X$ relates to the instances $D^\sfl\der{}\eps$, and to $E^\sfr\der{}(a, T)$, for the nonterminal $E^\sfr\in V_\sfr$ that satisfies this. (1) If $X$ is to the left of $D^\sfl$, then $X\in V_\sfo$, (2) if $D^\sfl$ is a descendant of $X$, but $E^\sfr$ is not, then $X\in V_\sfl$, (3) if $X$ is to the right of $D^\sfl$, and is to the left of $E^\sfr$, then $X\in V_\sfm$, (4) if $E^\sfr$ is a descendant of $X$, but $D^\sfl$ is not, then $X\in X_\sfr$, (5) if $X$ is to the right of $E^\sfr$, then $X\in V_\sfo$, and (6) if both $D^\sfl$ and $E^\sfr$ are descendants of $X$, then $X\in V_\sfi$. We bring attention to the fact that in this paragraph we referred only to the instances of $D^\sfl$ and $E^\sfr$ on the derivations mentioned above.

Another, more important point, is this reasoning gives us the derivation presented above. This derivation is translated into the following derivation in $\cF$:
\begin{align*}
	S  &\der{\cF}^* 
	\hat{u}_1 A \delta_1'\ \  (\text{or } \hat{u}_1 S \delta_1')\\
	&\der{\cF} 
	\hat{u}_1 BC\delta_1'\\
	&\der{\cF}^* 
	\hat{u}_1 \hat{u}_2 D \delta_2' C \delta_1' \\
	&\der{\cF}
	\hat{u}_1 \hat{u}_2 \kappa \delta_2' C\delta_1' \\
	&\der{\cF}^* 
	\hat{u}_1 \hat{u}_2 \kappa \beta_1 C \delta_1' \\
	&\der{\cF}
	\hat{u}_1 \hat{u}_2 \kappa \beta_1 \beta_2 (a, T\setminus\{\kappa\}) \hat{v}_1 \delta_1',\ \ \ \ \text{ or}\\
	&\ \ \ \ \ \ \ \ \ \, \, \,
	\hat{u}_1 \hat{u}_2 \kappa \beta_1 \beta_2 a \hat{v}_1 \delta_1',\\
	&\der{\cF}^*
	\hat{u}_1 \hat{u}_2 \kappa \beta_1 \beta_2 (a, T\setminus\{\kappa\}) \hat{v}_1 \hat{v}_2 = \hat{r},\ \ \ \ \text{ or}\\
	&\ \ \ \ \ \ \ \ \ \, \, \,
	\hat{u}_1 \hat{u}_2 \kappa \beta_1 \beta_2 a \hat{v}_1 \hat{v}_2 = \hat{r},
\end{align*}
Where $\delta_1'$ and $\delta_2'$ are obtained by replacing each $X^\sfm$ by $X$ in $\delta_1$ and $\delta_2$, respectively. It is direct to see that this is a valid derivation since for every production $X^\mathsf{x} \der{\cF'} Y^\mathsf{y}  Z^\mathsf{z}$ there exists a valid production  $X \der{\cF} Y Z$, for any $\mathsf{x}, \mathsf{y}, \mathsf{z}\in\{\sfi, \sfo, \sfl, \sfm, \sfr\}$. Furthermore, for the production $D^\sfl\der{\cF'}\eps$ there exists $D\der{\cF}\kappa$, and for $C^\sfr\der{\cF'}(a, T)$ there exists $C\der{\cF} (a, T\setminus\{\kappa\})$ if $T \neq \{\kappa\}$, and $C\der{\cF} a$ if $T = \{\kappa\}$. Further, note that $\proc_\kappa(\hat{r}) = \hat{s}$. We conclude that $\hat{r}\in L(\cF)$ for some $\hat{r}$ such that $\proc_\kappa(\hat{r}) = \hat{s}$.\\

From the arguments above, we obtain that for each annotated ref-word $\hat{r}\in L(\cF)$ there exists an equivalent annotated ref-word $\hat{t}\in L(\cF)$, given by $\hat{t} = \proc_\kappa(\hat{r})$. Furthermore, we showed that for each annotated ref-word $\hat{t}\in L(\cF')$ there exists an equivalent $\hat{r}\in L(\cF)$. This implies that $\cF$ and $\cF'$ are equivalent.\\

Now, assume that $\cF$ is unambiguous. We will prove that $\cF'$ is unambiguous as well. Consider an annotated ref-word $\hat{t}\in L(\cF')$ and consider two sequences $\cS_1$ and $\cS_2$ which define the derivation $S'\der{\cF'}^*\hat{t}$. We showed above how to translate these sequences into sequences $\cS_1'$ and $\cS_2'$ which define the derivation $S\der{\cF}^*	\hat{r}$, for some $\hat{r}$ such that $\hat{t} = \proc_\kappa(\hat{r})$. Since $\cF$ is unambiguous, these sequences are equal. Assume now that $\cS_1$ and $\cS_2$ are not equal, but since their translations into $\cF$ are the same, then it must be that for some production $X\der{\cF}YZ$ or $X\der{\cF}\tau$, there must be two productions $X^\mathsf{x_1} \der{\cF'} Y^\mathsf{y_1}  Z^\mathsf{z_1}$ and $X^\mathsf{x_2} \der{\cF'} Y^\mathsf{y_2}  Z^\mathsf{z_2}$, or $X^\mathsf{x_1} \der{\cF'}\tau$ and $X^\mathsf{x_2} \der{\cF'}\tau$ at the same position, for some $(\mathsf{x_1}, \mathsf{y_1}, \mathsf{z_1}) \neq (\mathsf{x_2}, \mathsf{y_2}, \mathsf{z_2})$. We note that this is not possible since we argued that for a given derivation $S'\der{\cF'}^*\hat{r}$, the set in which each nonterminal instance belongs, among $V_\sfo$, $V_\sfi$, $V_\sfl$, $V_\sfm$, $V_\sfr$, is fixed by its relation to certain instances of $D^\sfl$ and $E^\sfr$. We conclude that $\cF'$ is unambiguous.

Assume $\cF'$ is unambiguous. We will prove that $\cF$ is unambiguous as well. Likewise, consider an annotated ref-word $\hat{r}\in L(\cF)$, and consider two sequences $\cS_1$ and $\cS_2$ which define the derivation $S\der{\cF}^*\hat{t}$. We showed above how to convert these sequences into $\cS_1'$ and $\cS_2'$ which define the derivation $S'\der{\cF'}^*\proc_\kappa(\hat{r})$. Since $\cF'$ is unambiguous, then it must hold that $\cS_1' = \cS_2'$. Note that the translation we showed consisted in replacing productions of the form $X\der{\cF}YZ$ by $X^\mathsf{x} \der{\cF'} Y^\mathsf{y}  Z^\mathsf{z}$, $X\der{\cF}\kappa$ by $X^\sfl\der{\cF'}\eps$, $X\der{\cF}a$ by $X^\sfr\der{\cF}(a, \{\kappa\})$ (or $X\der{\cF}(a, T)$ by $X^\sfr\der{\cF'}(a, T\cup \{\kappa\})$) for a single fixed production, and $X\der{\cF}\tau$ by $X^\mathsf{x}\der{\cF'}\tau$ in any other case, for some $\mathsf{x}, \mathsf{y}, \mathsf{z}\in \{\sfo, \sfi, \sfl, \sfm, \sfr\}$. Therefore, the sequence $S_1$ from which $S_1'$ was obtained is uniquely defined, from which we deduce that $S_1 = S_2$, and we conclude that $\cF$ is unambiguous.\\

At the end of the procedure, we obtain an extraction grammar with annotations $\cF^\dagger = (V^\dagger, \Sigma, \var, P^\dagger, S^\dagger)$ such that there are no rules of the form $X\to\kappa$ in $P^\dagger$, for any $\kappa\in\captures_\var$. From this, we obtain the annotated grammar $\cG = (V^\dagger, \Sigma, \Omega_\var, P^\dagger, S^\dagger)$ which is equivalent to $\cH$. Furthermore, $\cG$ is unambiguous if and only if $\cH$ is unambiguous.

With respect to the running time of building $\cG$, note that in each iteration of the algorithm, by starting on an extraction grammar with annotations $\cF$ with a set of rules $P$, the resulting $\cF'$ has a set of rules $P'$ with a size of $9|P|$. Since this step is repeated twice for each variable $x\in \var$ (once for each variable operation), 
the total running time is $\cO(9^{2|\var|}(3^{2|\var|}|\cH|^2)) = \cO(9^{3|\var|}|\cH|^2)$.

\subsection{Expressiveness examples}
\label{apx:spanexa}

We complete Proposition~\ref{prop:extrac-grammars-general} to give more
intuition about the conciseness of extraction grammars vs annotated grammars,
and the difference in expressiveness.

We first give a simple example to show that, with our notion of equivalence
(Definition~\ref{def:equivalentannotated}), we may indeed need an exponential number of symbols in the annotation set, implying that extraction grammars are in some cases exponentially more concise:

\begin{example}
Consider the following functional extraction grammar $\cH$ with $n$ variables $x_1, \ldots, x_n$ and alphabet $\{a\}$:
  \[
\begin{array}{rrcl}
	\cH: & A_1 & \rightarrow &  \vdash_{x_1} \dashv_{x_1} A_2 \ \ \mid \ \ \vdash_{x_1} A_2 \dashv_{x_1} \\
	&  A_2 & \rightarrow &  \vdash_{x_2} \dashv_{x_2} A_3 \ \ \mid \ \ \vdash_{x_2} A_3 \dashv_{x_2} \\
	&  & \vdots & \\
	&  A_{n} & \rightarrow &  \vdash_{x_n} \dashv_{x_n} a \ \ \mid  \ \ \vdash_{x_n} a \dashv_{x_n} \\
\end{array}
\]
For the document $a$, this extraction grammar will output all possible combinations depending on whether $\dashv_{x_i}$ is at the beginning or end of $a$ for each $i \leq n$. Thus, an equivalent annotated grammar will need to consider all possible subsets of $\{ {\dashv_x} \mid x\in \var\}$ as possible annotations of the character~$a$, which will require an exponential number of rules.
\end{example}

We then illustrate why annotated grammars are in fact \emph{strictly} more
expressive: in addition to capturing all extraction grammars
(Proposition~\ref{prop:extrac-grammars-general}),
annotated grammars can express functions that do not correspond to an annotation grammar.

\begin{example}
  Consider a singleton annotation set $\Omega = \{\oout\}$, a singleton alphabet $\Sigma = \{a\}$, and the annotated grammar with start symbol $S$ and production $S \rightarrow a (a, \oout) S | a | \epsilon$. For each string of $\Sigma^*$, it produces one output where every other character is annotated. This cannot be expressed by an extraction grammar, as such a grammar fixes a finite set $\var$ of variables independently from the input document, and each variable is mapped to only one span.
\end{example}

\end{document}